\documentclass[a4paper,12pt]{article}
\pdfoutput=1
\usepackage[hang,flushmargin,bottom]{footmisc}
\usepackage[colorlinks,citecolor=blue,linkcolor=blue,urlcolor=blue,filecolor=blue,backref=page]{hyperref}
\hypersetup{colorlinks,citecolor=blue,linkcolor=blue,urlcolor=blue,filecolor=blue,backref=page}
\usepackage{footnotebackref}
\usepackage{amsthm}
\usepackage{amsfonts}
\usepackage{amsmath}
\usepackage{authblk, blindtext}
\usepackage[nottoc,numbib]{tocbibind}
\usepackage[round]{natbib}
\usepackage{enumitem}
\usepackage{float}
\usepackage{graphicx}
\usepackage{epstopdf}
\usepackage[margin=0.95in]{geometry}
\usepackage{mathtools}
\usepackage{multirow}
\usepackage{nicefrac}
\usepackage{ragged2e}
\usepackage[normalem]{ulem}
\usepackage{bbm,bm}     
\usepackage{dsfont}  

\providecommand{\keywords}[1]{\textbf{Keywords:} #1}

\DeclareMathOperator{\trace}{tr}
\newcommand{\minsub}[1]{\underset{#1}{\operatorname{min}}\;}
\newcommand{\indicator}[1]{\mathds{1}{{#1}}}
\renewcommand{\tablename}{\bfseries Table}
\def\given{\,|\,}
\newcommand{\argmax}[1]{\underset{#1}{\operatorname{arg}\,\operatorname{max}}\;}
\DeclareMathOperator{\diag}{diag}
\newcommand{\GG}[1]{}

\usepackage[framemethod=TikZ]{mdframed}
\usetikzlibrary{shadows}
\usepackage{environ}
\usepackage{varwidth}
\usepackage{ctable}
\usepackage[labelfont=bf]{caption}
\usepackage{subcaption}
\renewcommand{\thesubtable}{\alph{subtable}}
\usepackage[referable]{threeparttablex}
\newcolumntype{I}{!{\vline width 1.25pt}}
\newcolumntype{K}[1]{>{\centering\arraybackslash}p{#1}}

\newlength{\MyMdframedWidthTweak}%
\NewEnviron{reminder}[1][]{%
	\setlength{\MyMdframedWidthTweak}{\dimexpr%
		+\mdflength{innerleftmargin}
		+\mdflength{innerrightmargin}
		+\mdflength{leftmargin}
		+\mdflength{rightmargin}
	}%
	\savebox0{%
		\begin{varwidth}{\dimexpr0.9\linewidth-\MyMdframedWidthTweak\relax}%
			\BODY
		\end{varwidth}%
	}%
	\begin{mdframed}[
		backgroundcolor=gray!20, 
		shadow=true, 
		align=center,
		shadowsize=4pt,
		roundcorner=5pt,
		userdefinedwidth=\dimexpr\wd0+\MyMdframedWidthTweak\relax, 
		#1]
		\usebox0
	\end{mdframed}
}

\begin{document}
\title{\textbf{Infinite Mixtures of Infinite Factor Analysers}}

\author[1]{\vspace{-2ex} Keefe Murphy}
\author[2]{Cinzia Viroli}
\author[3,4]{Isobel Claire Gormley\vspace{-1ex}}
\affil[1]{{\small Department of Mathematics and Statistics, Maynooth University}}
\affil[2]{{\small Department of Statistical Sciences, University of Bologna}}
\affil[3]{{\small School of Mathematics and Statistics, University College Dublin}}
\affil[4]{{\small Insight Centre for Data Analytics, University College Dublin}}

\affil[ ]{{\small E-mail: \href{mailto:me@keefe.murphy@mu.ie}{\small{keefe.murphy@mu.ie}}}}
\normalsize

\date{}
\vspace*{-3em}
\begin{center}
\begin{reminder}
	{\centering \color{red} \textbf{\textsf{\small This is a preprint. The revised version of this paper is published as}}\\}
	\vspace{3pt}
	{\small \noindent K. Murphy, C. Viroli, and I. C. Gormley (2020) %
		{``Infinite mixtures of infinite factor analysers''.} %
		\textit{Bayesian Analysis}, 15(3): 937--963.\hfill [\href{https://doi.org/10.1214/19-BA1179}{\sf doi: 10.1214/19-BA1179}].}
\end{reminder}
\end{center}
\vspace{-3.5em}
{\let\newpage\relax\maketitle}
\vspace{-3.75em}
\begin{abstract}
	Factor-analytic Gaussian mixtures are often employed as a model-based approach to clustering high-dimensional data. Typically, the numbers of clusters and latent factors must be fixed in advance of model fitting. The pair which optimises some model selection criterion is then chosen. For computational reasons, having the number of factors differ across clusters is rarely considered. 
	
	Here the infinite mixture of infinite factor analysers (\textsf{IMIFA}) model is introduced. \textsf{IMIFA} employs a Pitman-Yor process prior to facilitate automatic inference of the number of clusters using the stick-breaking construction and a slice sampler. Automatic inference of the cluster-specific numbers of factors is achieved using multiplicative gamma process shrinkage priors and an adaptive Gibbs sampler. \textsf{IMIFA} is presented as the flagship of a family of factor-analytic mixtures. 
	
	Applications to benchmark data, metabolomic spectral data, and a handwritten digit example illustrate the \textsf{IMIFA} model's advantageous features. These include obviating the need for model selection criteria, reducing the computational burden associated with the search of the model space, improving clustering performance by allowing cluster-specific numbers of factors, and uncertainty quantification.\bigskip
	
	\noindent \keywords{\small{adaptive Markov chain Monte Carlo, factor analysis, model-based\linebreak clustering, Pitman-Yor process, multiplicative gamma process.}}
\end{abstract}
\small


\section[Introduction]{Introduction}
\label{Section:Intro}

In cases where the number of variables $p$ is comparable to or greater than the number of observations $N$, many clustering techniques tend to perform poorly or be intractable. Factor analysis (\textsf{FA}; \citealp{Knott1999}) is a well-known approach to parsimoniously modelling data. \citet{Bai2012} outline some computational difficulties which arise when $N \ll p$. Model-based clustering methods which rely on latent factor models have long been successfully utilised to cluster high-dimensional data. \citet{Ghahramani1996} propose a mixture of factor analysers model (\textsf{MFA}) with cluster-specific parsimonious covariance matrices and estimate it via an expectation-maximisation algorithm; \citet{McLachlanPeel2000} provide a succinct overview. Estimation of \textsf{MFA} models has also been considered in a Bayesian framework \citep{Diebolt1994, Richardson1997}. \citet{McNicholas2008,McNicholas2010b} develop a suite of similar parsimonious Gaussian mixture models. Other related developments in this area include \citet{Baek2010} and \citet{Viroli2010}, among others. 

Clustering using a \textsf{MFA} model typically requires specifying the number of clusters and factors in advance of model fitting. Generally, a range of \textsf{MFA} models with different numbers of clusters and factors are fitted and then compared through the use~of information criteria, such as the Bayesian Information Criterion (BIC; \citealp{Kass1995}) or the Deviance Information Criterion \citep{Spiegelhalter2002, Spiegelhalter2014}. Within a Bayesian framework \citet{Fokoue2003} use a stochastic model selection approach but do not simultaneously choose the optimal number of clusters and factors. Conducting an exhaustive search of the model space is computationally expensive; the cost is typically reduced by only considering models in which the number of factors is common across clusters. Regardless, even searching the reduced model space can be computationally onerous. The problem of identifying the optimal model is exacerbated by the fraught task of choosing among the range of model selection tools available, which often suggest different optimal models. Moreover, enforcing a common number of factors across clusters may lead to poor clustering performance due to a lack of flexibility. 

The infinite mixture of infinite factor analysers (\textsf{IMIFA}) model is introduced here. It theoretically allows infinitely many components and infinitely many factors within each component. The need to select a model selection criterion is obviated and quantification of the uncertainty in the optimal numbers of non-empty clusters and cluster-specific factors is facilitated. \textsf{IMIFA} relies on an infinite mixture model through the use of a nonparametric Pitman-Yor process (PYP) prior \citep{Perman1992, Pitman1997}, of which the well-known Dirichlet process (DP; \citealp{Ferguson1973}) is a special case. The infinite mixture model framework allows the number of clusters present to be automatically inferred; here the stick-breaking construction \citep{Pitman1996} and an independent slice-efficient sampler \citep{Kalli2011} are employed to facilitate~this. 

By allowing infinitely many factors within each cluster, \textsf{IMIFA} addresses the difficulty in choosing the optimal number of factors. This facilitates fitting factor-analytic models which are more flexible, in the sense that the number of factors may be cluster-specific, thereby potentially improving clustering performance. This is achieved by assuming multiplicative gamma process (MGP) shrinkage priors \citep{Bhattacharya2011,Durante2017} on the cluster-specific factor loading matrices, thus generalising the MGP prior to the mixture setting. Such a prior allows the degree of shrinkage of the factor loadings towards zero to increase as the factor number tends towards infinity. The number of factors with non-negligible loadings can be considered as the `active' number of factors within each cluster. Following \citet{Bhattacharya2011}, a computationally efficient adaptive Gibbs sampling algorithm is employed for estimation. Thus, the choice of the numbers of active factors in different clusters is automated. 

The \textsf{IMIFA} model with its PYP-MGP priors thus offers a single-pass and therefore computationally efficient approach to clustering high-dimensional data. It can be viewed as the most flexible model at the head of a family of Bayesian factor-analytic mixture models. Section \ref{Section:Models} develops the hierarchy of the \textsf{IMIFA} model family, beginning with the \textsf{MFA} model and concluding with the flagship \textsf{IMIFA} model. Between these extremes the novel finite mixture of infinite factor analysers model (\textsf{MIFA}) is introduced. Overfitted factor-analytic mixtures \citep{Papastamoulis2018} also belong to the \textsf{IMIFA} family; the overfitted mixture of infinite factor analysers (\textsf{OMIFA}) model is also introduced here.\enlargethispage{\baselineskip}

Section \ref{Section:Results} considers implementation of the \textsf{IMIFA} family of models. A benchmarking experiment is conducted on the well-known Italian olive oil data set. A real data application follows through the cluster analysis of  spectral metabolomic data from an epilepsy study. Finally an illustrative application is provided through clustering United States Postal Service handwritten digit data, a setting for which fitting sub-models of the \textsf{IMIFA} family is practically infeasible. Comparisons against other clustering methods are provided throughout. Simulation studies demonstrating the performance of \textsf{IMIFA} under different scenarios are deferred to Appendix~\ref{Subsection:SimulationStudies}. Section \ref{Section:Discussion} concludes the article with a discussion of \textsf{IMIFA} and thoughts on future research~directions. 

A software implementation of \textsf{IMIFA} and its family of sub-models is provided by the associated \textsf{R} package \texttt{IMIFA} \citep{IMIFAR2021}, with which all results were generated, which is freely available from \href{https://www.r-project.org}{\texttt{https://www.r-project.org}} \citep{R2021}.

\section[The IMIFA Model Family]{The IMIFA Model Family}\label{Section:Models}

The hierarchy of the \textsf{IMIFA} family of models is delineated herein, including a review of extant methodologies, the introduction of novel sub-models, and concluding with the flagship \textsf{IMIFA} model. Prior specifications, Markov chain Monte Carlo (MCMC) inferential procedures, approaches to posterior predictive model checking, and model-specific implementation issues that arise in practice are addressed. 

\subsection[Mixtures of Factor Analysers]{Mixtures of Factor Analysers}
\label{Subsection:MFA}

Mixtures of factor analysers are Gaussian latent variable models often used for clustering high-dimensional data. For each of $G$ clusters in these finite mixtures, the cluster-specific~\textsf{FA} model in cluster $g$ is given by $\smash{\mathbf{x}_i - \boldsymbol{\mu}_g=\boldsymbol{\Lambda}_g\boldsymbol{\eta}{\mathstrut}_i + \boldsymbol{\varepsilon}_{ig}}$. The observed feature vector $\smash{\mathbf{x}_i=\left(x_{i1},\ldots,x_{ip}\right)^\top}$ with mean $\smash{\boldsymbol{\mu}_g}$ and covariance matrix $\smash{\boldsymbol{\Sigma}_g}$ is assumed to linearly depend on a $q$-vector $\left(q \ll p\right)$ of latent common factor scores $\smash{\boldsymbol{\eta}{\mathstrut}_i}$ and additional sources of variation called specific factors $\smash{\boldsymbol{\varepsilon}_{ig}}$. It is assumed that $\smash{\boldsymbol{\eta}{\mathstrut}_i}$ has a $q$-variate Gaussian distribution $\smash{\textrm{N}_q\negthinspace\left(\mathbf{0}, \bm{\mathcal{I}}_q\right)},$ where $\smash{\bm{\mathcal{I}}_q}$ denotes the $q \times q$ identity matrix, and that $\smash{\boldsymbol{\varepsilon}_{ig} \sim \textrm{N}_p\negthinspace\left(\mathbf{0}, \boldsymbol{\Psi}_g\right)}$, where $\smash{\boldsymbol{\Psi}_g}$ is a diagonal matrix with non-zero elements $\smash{\psi_{1g},\ldots,\psi_{pg}}$ known as uniquenesses. Here, $\smash{\boldsymbol{\Lambda}_g}$ denotes the $p\times q$ factor loadings matrix of cluster $g$ and notably $q=0$ is permitted.

To facilitate estimation, a latent cluster indicator vector $\smash{\mathbf{z}_{i} = (z_{i1}, \ldots, z_{iG})^\top}$ is introduced such that $\smash{z_{ig}=1}$ if observation $i$ belongs to cluster $g$ and $\smash{z_{ig}=0}$ otherwise. Hence, $\smash{\mathbf{z}_i}$ has a $\smash{\textrm{Mult}\negthinspace\left(1, \boldsymbol{\pi}\right)}$ distribution where $\smash{\boldsymbol{\pi} = \left(\pi_1,\ldots,\pi_G\right)^\top}$ are the cluster mixing proportions which sum to $1$. A symmetric uniform Dirichlet prior $\boldsymbol{\pi} \sim \textrm{Dir}_G\negthinspace\left(\boldsymbol{\alpha} = \left(\alpha,\ldots,\alpha\right)=\mathbf{1}\right)$ is assumed. Upon marginalising out $\smash{\mathbf{z}_i}$ and $\smash{\boldsymbol{\eta}{\mathstrut}_i}$, \textsf{MFA} yields a parsimonious finite sum covariance structure for the observed data\vspace{-0.25em}
\begin{equation}
f\negthinspace\left(\mathbf{x}_i\given\boldsymbol{\theta}\right) = \sum_{g=1}^G \pi_g \textrm{N}_p\big(\mathbf{x}_i;\boldsymbol{\mu}_g,\boldsymbol{\Sigma}_g = \boldsymbol{\Lambda}{\mathstrut}_g\boldsymbol{\Lambda}{\mathstrut}_g^\top + \boldsymbol{\Psi}_g\big),\vspace{-0.25em}\label{Equation:MFAmixture}
\end{equation}
where $\textrm{N}_p\negthinspace\left(\mathbf{x}_i;\cdot,\cdot\right)$ denotes the density of a $p$-variate Gaussian distribution evaluated at $\mathbf{x}_i$ and $\smash{\boldsymbol{\theta}_g=\left\{\boldsymbol{\mu}_g, \boldsymbol{\Lambda}_g, \boldsymbol{\Psi}_g\right\}}$ are the cluster-specific \textsf{FA} parameters for which inference is straightforward under a Gibbs sampling scheme. Imposing constraints on $\smash{\mathbf{\Psi}_g}$ \citep{McNicholas2008,McNicholas2010b} and/or fixing $\smash{\pi_g=\nicefrac{1}{G}\:\:\forall\:\:g}$ may be useful in some settings.

\subsubsection[Prior Specification and Practical Issues]{Prior Specification and Practical Issues}\label{Subsubsection:PriorSpec}

The conditionally conjugate nature of the various prior distributions detailed below facilitates MCMC sampling via straightforward Gibbs updates. A multivariate Gaussian prior is assumed for the factor loadings of the variable $j$ across the $q$ factors of cluster $g$:\vspace{-0.25em}
\[\boldsymbol{\Lambda}_{jg} = \left(\lambda_{j1g},\ldots,\lambda_{jqg}\right) \sim \textrm{N}_q\negthinspace\left(\mathbf{0}, \bm{\mathcal{I}}_q\right).\vspace{-0.25em}\]
\noindent Similarly, a diffuse multivariate Gaussian prior is assumed for the component means, \vspace{-0.25em}
\[\boldsymbol{\mu}_g \sim \textrm{N}_p\negthinspace\left(\widetilde{\boldsymbol{\mu}}, \varphi^{-1}\bm{\mathcal{I}}_p\right),\vspace{-0.25em}\]
where $\widetilde{\boldsymbol{\mu}}$ is the overall sample mean and the scalar $\varphi$ controls the level of \mbox{diffusion}.

An inverse gamma prior $\psi_{jg}\sim\textrm{IG}\negthinspace\left(\alpha_0,\beta_j\right)$ is assumed for the uniquenesses of variable $j$ in cluster $g$. Guided by \citet{Fruhwirth-Schnatter2010-II},
hyperparameters are chosen to ensure $\psi_{jg}$ is bounded away from $0$, thereby avoiding Heywood problems. With a sufficiently large shape $\alpha_0$, variable-specific scales are derived from the sample precision matrix $\mathbf{S}^\star=\mathbf{S}^{-1}$ via $\beta_j = \left(\alpha_0-1\right)\negthinspace/S^\star_{jj}$. However, when $\nicefrac{N}{p}$ is close to or less than $1$, or when $\mathbf{S}^{-1}$ is otherwise unavailable, $\mathbf{S}^\star$ is replaced by a ridge-type estimator $\widehat{\mathbf{S}^{-1}} = \big(\beta_0 + \nicefrac{N}{2}\big)\big(\beta_0\bm{\mathcal{I}}_p + 0.5\sum_{i=1}^N \mathbf{x}{\mathstrut}_i\mathbf{x}{\mathstrut}_i^\top\big)^{-1}$, which combines the the inverse Wishart prior $\mathbf{S}^{-1} \sim \textrm{W}_p\negthinspace\left(\beta_0,\beta_0\bm{\mathcal{I}}_p\right)$ with the sample information,  where~$\beta_0$ is a hyperparameter \citep{Fruhwirth-Schnatter2018}. For unstandardised data, this estimator is constructed for the inverse correlation matrix and then appropriately scaled using the diagonal entries of $\mathbf{S}$ \citep{Wang2015}. When the variances are roughly balanced, constraining $\boldsymbol{\Psi}_g$ to $\psi_g\bm{\mathcal{I}}_p\,$, and/or  using $\beta_j = \beta = \left(\alpha_0 - 1\right)\negthinspace/{\max\negthinspace\left(\diag\negthinspace\left(\mathbf{S}^\star\right)\right)}$, provides additional parsimony. Notably, the isotropic constraint provides the link between factor analysis and probabilistic principal component analysis \citep{Tipping1999b}.

The rotational invariance property which makes \textsf{FA} models non-identifiable is well known: most covariance matrices $\boldsymbol{\Sigma}$ cannot be uniquely factored as $\smash{\boldsymbol{\Lambda}\boldsymbol{\Lambda}^\top + \boldsymbol{\Psi}}$ when $q > 1$. Though identifiability of $\boldsymbol{\Lambda}$ is not strictly necessary for the purposes of clustering or inferring $\boldsymbol{\Sigma}$, addressing the identifiability problem offline using the parameter expanded approach of \citet{Ghosh2008} in tandem with Procrustean methods, as in \citet{McParland2014} and \citet{Assman2016}, yields interpretable posterior summaries. Another practical issue is the label switching phenomenon \citep{Fruhwirth-Schnatter2010} which is addressed offline using the cost-minimising permutation given by the square assignment algorithm \citep{CarpToth1980}. Finally, optimal \textsf{FA} and \textsf{MFA} models~are chosen using the BIC-MCMC criterion \citep{Fruhwirth-Schnatter2011} where necessary in what~follows.\vspace{-0.33ex}

\subsection[Mixtures of Infinite Factor Analysers]{Mixtures of Infinite Factor Analysers}\label{Subsection:MIFA}

\begin{figure}[H]
	\centering
	\includegraphics[width=0.9\textwidth, keepaspectratio]{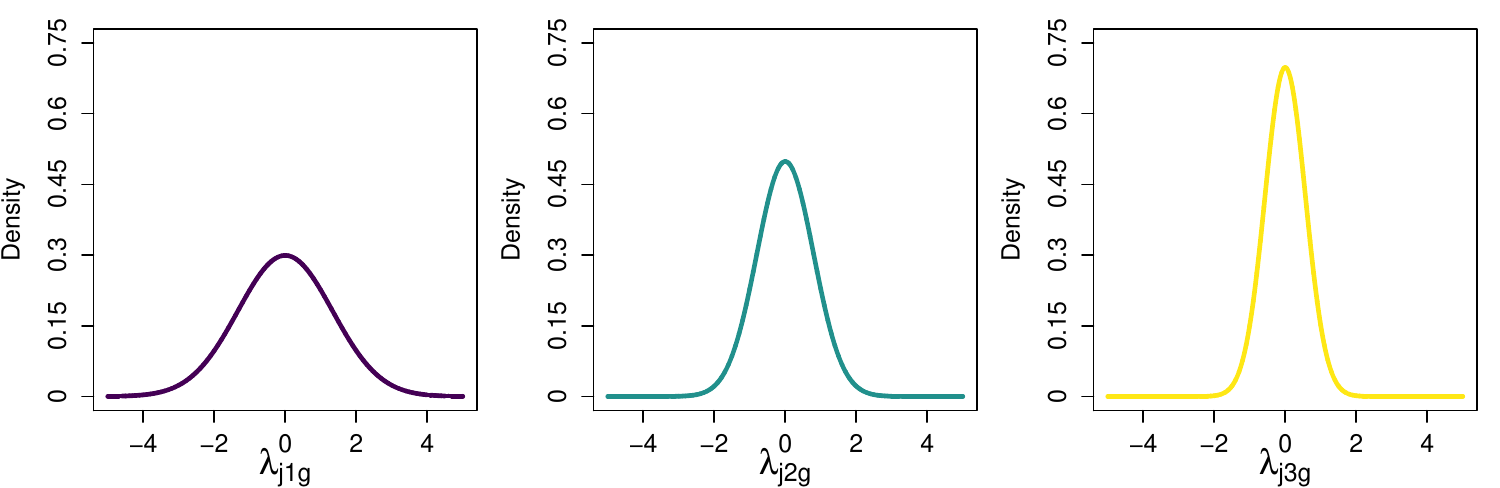}
	\caption[Schematic illustration of the MGP prior.]{Density of a typical element in the first, second, and third columns of a cluster-specific loadings matrix under the MGP shrinkage prior.\label{Plot:MIFA_Prior}}
\end{figure}

To overcome the requirement to specify $q$, infinite factor analysis (\textsf{IFA}) models are employed \citep{Bhattacharya2011}. The \textsf{IFA} model is a factor analysis model which assumes a multiplicative gamma process (MGP) shrinkage prior on the loadings matrix. This prior allows the degree of shrinkage towards zero to increase as the column index $k \rightarrow \infty$, mitigating against the factor splitting phenomenon. Here the \textsf{IFA} model is generalised to the mixture setting, leading to the novel mixture of infinite factor analysers (\textsf{MIFA}) model. Under \textsf{MIFA}, the MGP prior is placed on each parameter expanded $\smash{\boldsymbol{\Lambda}_g}$ matrix with no restrictions on its entries, thereby making the induced prior on $\smash{\boldsymbol{\Sigma}_g}$ invariant to the ordering of the variables. The MGP prior is conditionally conjugate, facilitating block Gibbs updates of the loadings and hence rapid mixing. Thus, the MGP prior in mixture settings is given~by\vspace{-0.33ex}
\begin{equation*}
\vspace{-0.33ex}
\begin{alignedat}{4}
\begin{split}
\lambda_{jkg} \given \phi_{jkg},\tau_{kg},\sigma_g&\sim \textrm{N}_1\negthinspace\left(0,\phi_{jkg}^{-1}\tau_{kg}^{-1}\sigma_g^{-1}\right), &\quad
\phi_{jkg}&\sim \textrm{Ga}\negthinspace\left(\nu_1,\nu_2\right),\label{Equation:MGPPrior}\\[-0.25ex]
\tau_{kg}&= \prod_{h=1}^k \delta_{hg}\,, & \quad\sigma_g & \sim \textrm{Ga}\negthinspace\left(\varrho_1,\varrho_2\right),
\\[-0.25ex]
\delta_{1g}&\sim\textrm{Ga}\negthinspace\left(\alpha_1,\beta_1\right), & \quad\delta_{hg} &\sim \textrm{Ga}\negthinspace\left(\alpha_2,\beta_2\right)\:\:\forall\:\:h\geq 2,
\end{split}
\end{alignedat}
\vspace{-0.33ex}
\end{equation*}
where $\tau_{kg}$ is a column shrinkage parameter for the $k$-th column in the $g$-th cluster's loadings matrix $\boldsymbol{\Lambda}_g\:\:\forall\:\:k = 1,\ldots, \infty$, and $\textrm{Ga}\negthinspace\left(\alpha,\beta\right)$ denotes the gamma distribution with mean $\nicefrac{\alpha}{\beta}$. The role of the local shrinkage parameters $\phi_{1kg}, \ldots, \phi_{pkg}$ for the $p$ elements in column $k$ of $\boldsymbol{\Lambda}_g$ is to favour sparsity while also preserving the signal of non-zero loadings. Lastly, the cluster shrinkage parameter $\sigma_g$ reflects the belief that the degree of shrinkage is cluster-specific. A schematic illustration of the MGP prior is given in Figure \ref{Plot:MIFA_Prior}; note that loadings can shrink arbitrarily close, but not exactly, to zero.

\citet{Bhattacharya2011} fix $\smash{\beta_1 = \beta_2 = 1}$ and recommend that $\smash{\alpha_2 > \beta_2}$. However, \citet{Durante2017} elaborates on the cumulative shrinkage properties and roles played by hyperparameters, showing in particular that $\smash{\alpha_2 > \beta_2 + 1}$ is necessary in order to have column-specific variances $\smash{\tau_{kg}^{-1}}$ that decrease in expectation with growing $k$. It is also recommended that $\smash{\alpha_2}$ be moderately large relative to $\smash{\alpha_1}$ (to ensure that the cumulative shrinkage property for which the prior was developed holds) and to avoid excessively high values for $\smash{\alpha_1}$ (to avoid over-shrinking to increasingly low-dimensional factorisations). While \citet{Bhattacharya2011} assume $\textrm{Ga}\negthinspace\left(\nu, \nu\right)$ priors for the local shrinkage parameters, here more general settings are used to allow control over prior non-informativity. In the spirit of \citet{Durante2017}, the expectation $\smash{\nu_2/\negthinspace\left(\nu_1-1\right)}$ of the induced inverse gamma prior on $\smash{\phi_{jkg}^{-1}}$ is suggested to be $\leq 1$ to induce sparsity on average. Furthermore, following the guidelines of \citet{Durante2017}, it is generally advisable that all MGP hyperparameters are chosen such that the first two moments of the associated hyperprior are defined, as this leads to superior performance in terms of the expected deviation between the true and estimated covariance matrices. In the mixture setting, $\smash{\alpha_1}$ and $\smash{\alpha_2}$ may need to be higher than the values suggested by \citet{Durante2017} to enforce a greater degree of shrinkage in clusters with few units; this aspect is highlighted in simulation studies in Appendix \ref{Subsection:SimulationStudies}.\vspace{-0.5em}

\subsubsection[The Adaptive Gibbs Sampler]{The Adaptive Gibbs Sampler}
\label{Subsubsection:AGS}

An adaptive Gibbs sampler (AGS) is employed when performing inference for \textsf{MIFA}. This dynamically shrinks the loadings matrices (and the infinite scores matrix $\boldsymbol{\eta}$) to have finite numbers of columns, by selecting the number of `active' factors. This practically facilitates posterior computation while closely approximating the \textsf{IFA} model, without requiring specification~of $\smash{\mathbf{Q} = \left(q_1,\ldots,q_G\right)^\top}$. However, a strategy is required for choosing appropriate truncation levels, $\smash{\widehat{q}_g}$, that strike a balance between missing important factors and wasting computational~\mbox{effort}. For computational reasons, a conservatively high upper bound is used, such that $q_g^\star = \min\negthinspace\big(\left\lfloor 3\negthinspace\left(p\right)\right\rceil,N - 1,p - 1\big)\:\:\forall\:\:g$. The number of factors in each $\boldsymbol{\Lambda}_g$ is then adaptively tuned as the MCMC chain progresses. Adaptation can be made to occur only after the burn-in period, in order to ensure the true posterior distribution is being sampled from before truncating the loadings matrices.

At the $t$-th iteration, adaptation occurs with probability $\mathrm{p}\negthinspace\left(t\right) = \exp\negthinspace\left(-b_0 - b_1t\right)$, with $b_0 \geq 0~\mbox{and}~b_1 > 0$ chosen so that adaptation occurs often at the beginning of the chain but then decreases exponentially fast in frequency. Here $b_0 = 0.1$ and $\smash{b_1 = 5 \times 10^{-5}}$ are used. With probability $\mathrm{p}\negthinspace\left(t\right)$, loadings columns having some pre-specified proportion of elements $\varsigma$ in a small neighbourhood $\epsilon$ of zero are monitored. If there are no such columns, an additional column is added by simulation from the MGP prior. Otherwise redundant columns are discarded and the AGS proceeds with all parameters corresponding to non-redundant columns retained. Choice of $\varsigma~\mbox{and}~\epsilon$ can be delicate, as there is an implicit trade-off between these two fixed tuning parameters; smaller $\varsigma$ and larger $\epsilon$ speed up the algorithm by favouring the discarding of factors during the adaptation step, and \emph{vice versa}. Typically, $\varsigma$ should be kept close to $1$ and $\epsilon$ should be kept small, relative to the scale of the data. Here, $\varsigma = \lfloor0.7 \times p\rfloor/p$ and $\epsilon=0.1$ are found to strike an appropriate balance. The dimension of the matrix $\boldsymbol{\eta}$ of factor scores at a given iteration are set to $p \times \overline{q} = p \times \max \left(\mathbf{Q}\left(t\right)\right)$; rows corresponding to observations currently assigned to a cluster with fewer latent factors than $\overline{q}$ are padded with zeros. Notably, here $\smash{\widehat{q}_g}$ may shrink to $0$ thus allowing diagonal covariance structure within a component. If this occurs, the decision to simulate a new column is based on a binary trial with probability $1-\varsigma$ as there are no loadings columns to~monitor.\enlargethispage{\baselineskip}

The numbers of active factors in each cluster for each retained posterior sample can be used to construct a barchart approximation to the posterior distribution of $q_g$. The posterior mode is used to estimate each $q_g$, with credible intervals quantifying uncertainty. Another strategy, which circumvents the need to pre-specify $\varsigma$ and $\epsilon$ is to forego adaptation (provided the computational burden of doing so is tolerable) and estimate $\widehat{q}_g$ from the number of non-redundant columns in the posterior mean loadings matrices. However, this approach is not considered further here.

In any case, the main advantages of \textsf{MIFA} are that different clusters can be modelled by different numbers of factors and that the model search is reduced to one for $G$ only, as $q_g$ is estimated automatically during model fitting. Here, for \textsf{MIFA} models, the optimal $G$ is chosen via the BICM (BIC-Monte (Carlo)) proposed by \citet{Raftery2007}, with
$\smash{\mbox{BICM} = 2\ln\big(\overline{\mathcal{L}}\big) - 2s_l^2 \left(\ln\negthinspace\left(N\right) - 1\right)}$,
where $\overline{\mathcal{L}}$ and $s_l^2$ are the sample mean and sample variance, respectively, of the log-likelihood values calculated for each retained posterior sample. This criterion is particularly useful in the context of nonparametric models where the number of free parameters is difficult to quantify, though we caution that it may be biased in favour of $G=1$ models, under which the log-likelihoods tend to exhibit less variability, and that a large number of posterior samples are required to ensure stable estimation of $s_l^2$.

\subsubsection[Other Infinite Factor Models]{Other Infinite Factor Models}

This work offers an extension of the MGP prior and its related AGS routine to the mixture modelling context. \citet{Wang2016} develop a related model employing a multiplicative exponential process prior. Other nonparametric approaches to inferring the number of factors include \citet{Knowles2007}, in which a two-parameter Indian Buffet Process (IBP) prior is assumed on an infinite binary matrix underlying the factor scores, thus selecting features of interest, with associated standard Gaussian weights. A closely related approach using the beta process (BP) is provided by \citet{Paisley2009}. In \citet{Knowles2011} and \citet{Rockova2016}, an IBP prior is instead assumed for sparsifying the loadings. These models assume a single sparse infinite factor model for the whole data set. However, embedding them in a mixture modelling setting, similar to the \textsf{IMIFA} framework, is intuitively feasible. 

Indeed, \citet{Chen2010} employ the BP prior, coupled with a Dirichlet process prior, to perform clustering in a manifold learning setting. While the BP and IBP priors achieve exact sparsity, which may be advantageous in certain applications, the MGP prior has a weaker notion of sparsity by virtue of cumulatively shrinking an infinite series arbitrarily close to zero, thereby preserving small signals. The block updates of~each row of $\boldsymbol{\Lambda}_g$ facilitated by the MGP prior and parameter expansion mean the AGS approach is a simpler, more computationally efficient alternative to the BP and IBP priors.

\subsection[Overfitted Mixtures of (Infinite) Factor Analysers]{Overfitted Mixtures of (Infinite) Factor Analysers}
\label{Subsection:Overfit}

While \textsf{MIFA} obviates the need to pre-specify $\mathbf{Q}$, the issue of model choice is not yet fully resolved. Overfitted mixtures \citep{Rousseau2011, vanHavre2015} are one means of extending \textsf{MIFA}; indeed \citet{Papastamoulis2018} proposes an overfitted mixture of factor analysers (\textsf{OMFA}), albeit with finite factors. Here, the overfitted mixture of infinite factor analysers (\textsf{OMIFA}) model is introduced.

In overfitted mixtures the symmetric Dirichlet prior on $\boldsymbol{\pi}$ plays an important role. Estimation is approached by initially overfitting the number of clusters expected to be present. Small values of the hyperparameter $\boldsymbol{\alpha}$ encourage emptying out excess components in the posterior distribution; the uniform prior with $\boldsymbol{\alpha}=\mathbf{1}$ is rather indifferent in this respect. The sampler is initialised with a conservatively high number of components: $\smash{G^\star = \max\negthinspace\big(\left\lceil 3\ln\negthinspace\left(N\right)\right\rceil,25, N - 1\big)}$, though this may be too high if it is close to $N$. While $\widetilde{G}=G^\star$ remains fixed throughout the MCMC chain, the number of non-empty clusters is recorded at each iteration of the sampler as $G_0 = \widetilde{G} - \sum_{g=1}^{\widetilde{G}} \indicator{\big(\sum_{i=1}^{N} z_{ig}=0\big)}$ where $\indicator{\left(\cdot\right)}$ is the indicator function. The true $G$ is estimated by $\widehat{G}$, the $G_0$ value visited most often. Cluster-specific inference is conducted only on samples corresponding to those visits. For the \textsf{OMIFA} model, the AGS is modified to handle empty components: the MGP-related parameters are simulated from the relevant priors and each corresponding $\boldsymbol{\Lambda}_g$ matrix is restricted to having $\overline{q}$ factors, i.e. the same number of columns currently in the matrix of factor scores $\boldsymbol{\eta}$, either by truncation or by padding with zeros, as required.

\subsection[Infinite Mixtures of (Infinite) Factor Analysers]{Infinite Mixtures of (Infinite) Factor Analysers}
\label{Subsection:InfiniteMixtures}

Embedding \textsf{MFA} and \textsf{MIFA} in an infinite mixture setting leads, respectively, to the infinite mixture of finite factor analysers model (\textsf{IMFA}) and the flagship infinite mixture of infinite factor analysers model (\textsf{IMIFA}). These models employ a nonparametric Pitman-Yor process (PYP) prior which is easily incorporated into the MCMC sampling~scheme.

The PYP is a stochastic process whose draws are discrete probability measures, whereby $H\sim\textrm{PYP}\negthinspace\left(\alpha, d, H_0\right)$ denotes a PYP probability distribution $H$, with base distribution $H_0$ interpreted as the mean of the PYP, discount parameter $d \in \left[0, 1\right)$, and concentration parameter $\alpha > -d$. For the PYP mixture model \textsf{IMFA} and the PYP-MGP mixture model \textsf{IMIFA} $H_0$ comes from the factor-analytic mixture \eqref{Equation:MFAmixture}, hence\vspace{-0.33em}
\begin{equation}
f\negthinspace\left(\mathbf{x}_i\given\boldsymbol{\theta}\right) = \sum_{g=1}^{\infty}\pi_g\textrm{N}_p\big(\mathbf{x}_i;\boldsymbol{\mu}_g, \boldsymbol{\Lambda}{\mathstrut}_g\boldsymbol{\Lambda}{\mathstrut}_g^\top + \boldsymbol{\Psi}_g\big).\label{Equation:InfiniteMFA}\vspace{-0.33em}
\end{equation}
\indent The stick-breaking representation of the PYP \citep{Pitman1996} is used as a prior process for generating the mixing proportions in \eqref{Equation:InfiniteMFA}. This construction views $\left\{\pi_1,\pi_2, \ldots\right\}$ as pieces of a unit-length stick that is sequentially broken in an infinite process, with stick-breaking proportions $\boldsymbol{\Upsilon} = \left\{\upsilon_1, \upsilon_2, \ldots\right\}$, summarised as
\begin{equation*}
\begin{alignedat}{5}
\begin{split}
\upsilon_g &\sim \textrm{Beta}\negthinspace\left(1 - d, \alpha + gd\right), &\quad\boldsymbol{\theta}_g &\sim H_0\,,\label{Equation:StickBreaking}\\
\pi_g &=\upsilon_g \prod_{l=1}^{g-1}\left(1-\upsilon_l\right), &\quad H &=\sum_{g=1}^\infty \pi_g \delta_{\boldsymbol{\theta}_g} \sim \textrm{PYP}\negthinspace\left(\alpha, d, H_0\right),
\end{split}
\end{alignedat}
\end{equation*}
where $\smash{\delta_{\boldsymbol{\theta}_g}}$ is the Dirac delta centred at $\smash{\boldsymbol{\theta}_g}$, such that draws are composed of a sum of infinitely many point masses. The PYP reduces to the DP when $d=0$, in which case mass shifts to the right with increasing dispersion as $\alpha$ increases, implying an \emph{a priori} larger number of components. However, some important distributional features fundamentally differ when $d \neq 0$ \citep{DeBlasi2015}. The PYP exhibits heavier tail behaviour and allows the stick-breaking distribution to vary according to the component index $g$, without sacrificing much in the way of tractability. In particular, increasing $d$ values have the effect of flattening the prior, controlling its degree of non-informativity.

Slice sampling \citep{Walker2007, Kalli2011} is used here to yield samples from the PYP by adaptively truncating the number of components needed to be sampled at each iteration. By introducing an auxiliary variable $u_i > 0$ which preserves the marginal distribution of the data, and denoting by $\boldsymbol{\xi}=\left\{\xi_1, \xi_2, \ldots\right\}$ a positive sequence of infinite quantities which sum to $1$, the joint density of $\left(\mathbf{x},\mathbf{u}\right)$ is given by $f\negthinspace\left(\mathbf{x}, \mathbf{u} \given \boldsymbol{\theta}, \boldsymbol{\xi}\right) = \sum_{g=1}^\infty \pi_g \textrm{Unif}\negthinspace\left(\mathbf{u}; 0, \xi_g\right) f\negthinspace\left(\mathbf{x}\given \boldsymbol{\theta}_g\right)$. Since only a finite number of $\xi_g$ are greater than $\mathbf{u}$, the conditional density of $\mathbf{x} \given \mathbf{u}$ can be written as a finite mixture with $\widetilde{G} = \max_{i\,\in\,\left\{1,\ldots,N\right\}} \left(\lvert \bm{\mathcal{A}}_{\boldsymbol{\xi}}\negthinspace\left(u_i\right) \rvert\right)$ `active' components at each iteration, where $\vert \cdot \vert$ denotes cardinality and $\bm{\mathcal{A}}_{\boldsymbol{\xi}}\negthinspace\left(\mathbf{u}\right) = \{g\colon\mathbf{u} < \xi_g\}$. Though $G$ is infinite in theory, $\widetilde{G}$ can be at most equal to $N$. Thus, the infinite mixture of (infinite) factor analysers models can be sampled from. 

Typical implementations of the slice sampler arise when $\xi_g = \pi_g$ \citep{Walker2007} but independent slice-efficient sampling \citep{Kalli2011} allows for a deterministic decreasing sequence, e.g. geometric decay, given by $\xi_g = \left(1 - \rho\right)\rho^{g-1}$ where $\rho \in \left[0, 1\right)$ is a fixed value to be chosen with care. Higher values generally lead to better mixing but longer run-times, as the average cardinality of $\bm{\mathcal{A}}_{\boldsymbol{\xi}}\negthinspace\left(\mathbf{u}\right)$ increases, and \emph{vice versa}. Setting $\rho = 0.75$, in line with the recommendations of \citet{Kalli2011}, appears to strike an appropriate balance in the applications considered here.

\subsubsection[Inference for Infinite Mixtures of Factor Analysers Models]{Inference for Infinite Mixtures of Factor Analysers Models}
\label{Subsection:IMIFAInference}

For clarity, what follows focuses on the \textsf{IMIFA} model where inference proceeds via the independent slice-efficient sampler with geometric decay. Inference for other models in the \textsf{IMIFA} family is closely related. The joint density of the \textsf{IMIFA} model~is

\begin{align*}
f\negthinspace\left(\mathbf{X},\boldsymbol{\eta}, \mathbf{Z}, \mathbf{u}, \boldsymbol{\Upsilon}, \boldsymbol{\theta}\right) & \propto f\negthinspace\left(\mathbf{X} \given \boldsymbol{\eta}, \mathbf{Z}, \mathbf{u},\boldsymbol{\Upsilon}, \boldsymbol{\theta}\right) f\negthinspace\left(\boldsymbol{\eta}\right) f\negthinspace\left(\mathbf{Z}, \mathbf{u} \given \boldsymbol{\Upsilon},\boldsymbol{\pi}\right) f\negthinspace\left(\boldsymbol{\Upsilon} \given \alpha, d\right) f\negthinspace\left(\boldsymbol{\theta}\right) \\[-0.5ex]
& = \Bigg\{\negthinspace\prod_{i=1}^N \prod_{g\,\in\,\bm{\mathcal{A}}_{\boldsymbol{\xi}}\negthinspace\left(u_i\right)}\negthinspace\textrm{N}_p\big(\mathbf{x}_i; \boldsymbol{\mu}_g + \boldsymbol{\Lambda}_g \boldsymbol{\eta}{\mathstrut}_{i}, \boldsymbol{\Psi}_g\big)^{z_{ig}}\negthinspace\Bigg\}\Bigg\{\negthinspace\prod_{i=1}^N \textrm{N}_q\big(\boldsymbol{\eta}{\mathstrut}_{i}; \mathbf{0}, \bm{\mathcal{I}}_q\big)\negthinspace\Bigg\} \\[-0.5ex]
& \phantom{=\,\,}\left\{ \prod_{i=1}^N \prod_{g=1}^\infty\negthinspace \left(\dfrac{\pi_g}{\xi_g} \indicator{\big(u_i < \xi_g\big)}\negthinspace\right)^{z_{ig}}\negthinspace\right\}\negmedspace\left\{ \prod_{g=1}^\infty \dfrac{\left(1 - \upsilon_g\right)^{\alpha + gd-1}}{\upsilon_g^{d}~\mathrm{B}\negthinspace\left(1 - d,\alpha + gd\right)} \right\}\negthinspace f\negthinspace\left(\boldsymbol{\theta}\right)\negthinspace, \label{Equation:IMIFAjoint}
\end{align*}
where $\mathrm{B}\negthinspace\left(\cdot\right)$ is the Beta function and $f(\boldsymbol{\theta})$ is the product of the previously defined collection of conditionally conjugate priors with additional layers for hyperparameters. Only the parameters of the~$\widetilde{G}$ active components are sampled at each iteration. The algorithm is initialised with the same $G^\star$ value detailed in Section \ref{Subsection:Overfit}, typically above~the anticipated number to which the algorithm will converge, in the spirit of \citet{Hastie2014}.  Here, however, $\widetilde{G}$ can theoretically exceed this value. For computational reasons, a finite upper limit is placed on $\widetilde{G}$ with $\max\negthinspace\left(G^\star, \min\negthinspace\left(N - 1, 50\right)\right)$ found~to be sufficiently large. However, $\widetilde{G}$ is only regarded as a set of proposals as to where to allocate observations; as in Section \ref{Subsection:Overfit}, it is the subset of non-empty clusters $G_0$ that is of inferential interest.

Bayesian approaches to clustering are known to be sensitive to initial cluster allocations. While starting values for $\mathbf{z}_i$ can be obtained by any means, model-based agglomerative hierarchical clustering \citep{Scrucca2016} is used here. Though this is fast~and intuitive given that \textsf{IMIFA} models are initialised at a conservatively high number~of~components, which are then merged as the sampler proceeds, heavily imbalanced initial cluster sizes are cautioned against. By extension, initial cluster means and mixing proportions are computed empirically. Other parameter starting values are simulated from~their relevant prior distributions. The adaptive inferential algorithm for \textsf{IMIFA}~then proceeds mostly via Gibbs updates. For those which are multivariate Gaussian, using the Cholesky factor of the covariance matrices and employing block updates speeds up the algorithm \citep{GMRFbook}. The allocations $\mathbf{z}_i$ are sampled in a fast, numerically stable fashion, using the Gumbel-Max~trick \citep{Yellott1977}. Finally, state spaces for applications of \textsf{IMIFA} to real data can be highly multimodal with well-separated regions of high posterior probability coexisting, corres\-ponding to clusterings with different numbers of components. Thus,~label switching~moves \citep{Papaspiliopoulos2008} are incorporated in order to improve~mixing. Details of the Gibbs updates, Gumbel-Max trick, and label switching moves are provided in Appendix \ref{Subsection:PosteriorConditional}.

\subsubsection[Assessing Model Fit]{Assessing Model Fit and Mixing}

As is good statistical practice, posterior predictive model checking \citep{Gelman2004} is employed. Sampled model parameters from the MCMC chain are used to generate replicate data from the posterior predictive distribution. Valid posterior samples, after conditioning on $\smash{\widehat{G}}$, are those for which $\max\negthinspace\left(\mathbf{Q}\left(t\right)\right)\negthinspace\geq \max\negthinspace\left(\widehat{q}_1,\ldots,\widehat{q}_{\widehat{G}}\right)$ such that the dimension of the estimated scores matrix $\smash{\widehat{\boldsymbol{\eta}}}$ is preserved. To assess model fit, histograms of the modelled data $\mathbf{X}$ are compared to histograms of the replicate data in a global sense using the Posterior Predictive Reconstruction Error (PPRE), calculated as follows:

\begin{enumerate}[listparindent=\parindent]
	\itemsep 1em
	\item Gather the histogram bin counts of each variable in $\mathbf{X}$ into the $h\times p$ matrix $\bm{\mathcal{H}}$, where~$h$ is the maximum number of bins across all variables and $\bm{\mathcal{H}}$ is padded with zeros as required.
	\item Generate $r \in \left\{1,\ldots,R\right\}$ data sets $\bm{\mathcal{X}}^{\left(r\right)}$ from the posterior predictive distribution.
	\item Create a similar matrix of histogram bin counts $\bm{\mathcal{H}}^{\left(r\right)}$ for each $\bm{\mathcal{X}}^{\left(r\right)}$ using the same break-points with which $\bm{\mathcal{H}}$ was constructed (with endpoint bins extended to $\pm\,\infty$).
	\item Compute the Frobenius norm ${\left\lVert\cdot\right\rVert}_\mathcal{F}$ 
	between $\bm{\mathcal{H}}$ and $\bm{\mathcal{H}}^{\left(r\right)}$, standardising to the 0-1 scale using the triangle inequality: 
	\[\Big\lvert {\big\lVert\bm{\mathcal{H}}\big\rVert}_\mathcal{F} - {\big\lVert\bm{\mathcal{H}}^{\left(r\right)}\big\rVert}_\mathcal{F} \Big\rvert \leq {\big\lVert\bm{\mathcal{H}} - \bm{\mathcal{H}}^{\left(r\right)}\big\rVert}_\mathcal{F} \leq {\big\lVert\bm{\mathcal{H}}\big\rVert}_\mathcal{F} + {\big\lVert\bm{\mathcal{H}}^{\left(r\right)}\big\rVert}_\mathcal{F}\,.\]
\end{enumerate} 
The distribution of PPRE values can be visualised using boxplots and summarised by the median, with credible intervals quantifying uncertainty. This discrepancy measure is well-suited to assessing model adequacy for mixtures of multivariate data: it accounts for inherent multimodality and gives a global quantitative measure of agreement between the distributions of the observed variables and their posterior predictive counterparts. 

Convergence of the MCMC chains is assessed using the potential scale reduction factor (PSRF; \citealp{Brooks1998, Coda2006}). Random allocations of the initial cluster labels, resulting in different draws from the relevant priors for parameter initialisation, are used to construct the multiple overdispersed chains required. The MAP labels of each chain are matched to the main chain prior to computing the diagnostics; $\boldsymbol{\Lambda}_g$ matrices are also rotated to a common template for each cluster. Good convergence is indicated by upper PSRF $95\%$ confidence interval limits close to $1$; this is a stricter requirement than the PSRF values themselves being near $1$.

\subsubsection[Comparing the IMIFA Family Models]{Comparing the IMIFA Family Models}

Though \textsf{IMIFA} and \textsf{OMIFA} come with the computational complexities inherent in nonparametric methods, diminishing adaptation, and extra tuning parameters, their advan\-tages over other models in the \textsf{IMIFA} family are numerous:  i) flexibility, in the sense~that models where $q_g \neq q_g^\prime$ can be fitted, ii) computational efficiency, in the sense that~the burden is reduced relative to searching over a range of fitted \textsf{MFA}~or \textsf{MIFA} models, iii) removing the need for model selection criteria, and iv) the ability to quantify the uncertainty in $\smash{\widehat{G}}$ and $\smash{\widehat{q}_g}$. Both methods offer simpler alternatives to 
reversible jump MCMC \citep{Richardson1997} and birth-death MCMC \citep{Stephens2000}. Hence, among the \textsf{IMIFA} family, the infinite factor models are recommended over the finite factor models and the infinite and overfitted mixtures are recommended over the finite mixtures. However, the \textsf{MIFA} model is appropriate if one wishes to fix $G$ but infer $q_g$.

While infinite mixtures are often used for density estimation, they are also employed to infer the number of components in cluster analyses (e.g. \citealt{
Kim2006, Xing2006, Yerebakan2014}). However, \citet{Miller2013, Miller2014} raise concerns about the guarantee of posterior consistency for the number of non-empty clusters, showing the number uncovered is typically greater than or equal to the truth, often with several vanishingly small clusters inferred. These concerns highlight the need~for practi\-tioners to pay due consideration to the uncertainty in the number of clusters offered~by \textsf{IMIFA} models. Relatedly, \citet{FSMW2019} compare infinite mixtures to overfitted (`sparse finite') mixtures. They highlight that overfitted mixtures are useful for applications in which the data arise from a moderate number~of clusters, even as the sample size increases, whereas infinite mixtures are suited to cases where the number of clusters also increases. However, they show that clustering results are driven less by the assumption of whether the data arose~from a finite or infinite mixture, but by the hyperprior on the DP parameters or the sparseness of the Dirichlet prior in the overfitted setting. Indeed, they show that overfitted and infinite mixtures yield comparable clustering performance on the observed data when these hyperpriors are matched. This matching leads to `sparse' infinite mixtures~that avoid overfitting~the number of clusters. Similar behaviour is observed for the PYP prior in the applications in Section \ref{Section:Results}, where the \textsf{IMIFA} and \textsf{OMIFA} models, with matched hyperpriors, give comparable results.

The issue of choosing $\alpha$ can make implementing overfitted models challenging. With fixed $\alpha = \nicefrac{\gamma}{G^\star}$, the prior approximates a DP with concentration parameter $\gamma$ as $G^\star$ tends to infinity \citep{Green2001}. Here, following \citet{FSMW2019}, a $\textrm{Ga}\negthinspace\left(a, b G^\star\right)$ hyperprior is assumed for $\alpha$. This favours small values and allows $\alpha$ to be updated via Metropolis-Hastings. In the infinite mixture setting, learning the PYP parameters (which also requires Metropolis-Hastings steps) and adopting the label-switching moves enables accurate inference on $G_0$. A joint hyperprior $\mathrm{p}\negthinspace\left(\alpha,d\right) = \mathrm{p}\negthinspace\left(\alpha\given d\right)\mathrm{p}\negthinspace\left(d\right)$ is assumed \citep{Carmona2019} where $\mathrm{p}\negthinspace\left(\alpha \given d\right) = \textrm{Ga}\negthinspace\left(\alpha + d; a, b\right)$; choosing a large $b$ encourages clustering \citep{Muller2013}. A spike-and-slab hyperprior $d \sim \kappa\delta_0 + \left(1-\kappa\right)\textrm{Beta}\negthinspace\left(a^\prime, b^\prime\right)$ is assumed. The estimated proportion $\smash{\widehat{\kappa}}$ can then be used to assess whether the data arose from a DP or a PYP at little extra computational cost. See Appendix \ref{Subsection:PosteriorConditional} for further details.

\section[Illustrative Applications]{Illustrative Applications}
\label{Section:Results}

The flexibility and performance of the \textsf{IMIFA} model and its related model family are demonstrated below through application to benchmark and real data sets. All results are obtained through the \texttt{IMIFA} \textsf{R} package; code to reproduce many of the results is available in the associated vignette\protect\footnote{\href{https://cran.r-project.org/web/packages/IMIFA/vignettes/IMIFA.html}{\,\texttt{https://cran.r-project.org/web/packages/IMIFA/vignettes/IMIFA.html}}.}. Appendix \ref{Subsection:SimulationStudies} reports on simulation studies demonstrating the performance of \textsf{IMIFA} under different scenarios, including effects of the $\nicefrac{N}{p}$ ratio, the PYP parameters, imbalanced cluster sizes, uncommon $q_g$, and the degree of loadings sparsity. Appendix \ref{Subsection:Robustness} explores the robustness of \textsf{IMIFA}.

MCMC chains were run for $50,000$ iterations, except for Section \ref{Subsection:Digits} in which $20,000$ were run. Every 2\textsuperscript{nd} sample was thinned and the first $20\%$ of iterations were discarded as burn-in. All computations were performed on a Dell Latitude 5491 laptop, equipped with a 6-core 2.60 GHz Intel Core i7-8850H processor and 16 GB of RAM. Where necessary, the optimal finite and infinite factor models are chosen by the BIC-MCMC and BICM criteria, respectively.  Throughout, $\smash{\widehat{\cdot}}$ denotes the posterior mode, posterior mean, or relevant optimal value. Unless otherwise stated, data were mean-centred and unit-scaled and no constraints were imposed on the uniquenesses. Hyperprior specifications are detailed in Table \ref{Table:Hyperparameters}. While there are many hyperparameters to select, the choices are all reasonably standard. However, poor settings may introduce additional factors or clusters to maintain flexibility and so care in specifying hyperparameters is advised.
\begin{table}[H]
	\caption[Hyperparameter specifications for the \textsf{IMIFA} model.]{Hyperparameter specifications for the \textsf{IMIFA} model. Note that the specification of the beta distribution in the prior for $d$ amounts to a standard uniform.}
	\label{Table:Hyperparameters}
	\centering
	\scriptsize
	\extrarowheight 2.5pt
	
	\begin{tabular}[pos=center]{ c | c | c }
		Parameter(s) & Hyperparameter(s) & Value(s) \\\hline
		$\boldsymbol{\mu}_g$ & $\varphi$ & $0.01$\\
		$\boldsymbol{\Psi}_g$ & $\left(\alpha_0,\beta_0\right)$ & $\left(2.5, 3\right)$\\
		$\phi_{jkg}$ & $\left(\nu_1, \nu_2\right)$ & $\left(3, 2\right)$\\
		$\delta_{1g}$ & $\left(\alpha_1, \beta_1\right)$ & $\left(2.1, 1\right)$\\
		$\delta_{kg}$ & $\left(\alpha_2, \beta_2\right)$ & $\left(3.1, 1\right)$\\
		$\sigma_{g}$ & $\left(\varrho_1, \varrho_2\right)$ & $\left(3, 2\right)$\\
		$\alpha$ & $\left(a, b\right)$ & $\left(2, 4\right)$\\
		$d$ & $\left(a^\prime, b^\prime, \kappa\right)$ & $\left(1, 1, 0.5\right)$
	\end{tabular}
\end{table}

\subsection[Benchmark Data: Italian Olive Oils]{Benchmark Data: Italian Olive Oils}\label{Subsection:OliveOil}

The Italian olive oil data \citep{Forina1982,Forina1983} is often clustered using factor-analytic models, e.g. \citet{McNicholas2010}. The data detail the percentage composition of $8$ fatty acids in $572$ Italian olive oils, known to originate from three areas: southern and northern Italy and Sardinia. Each area is composed of different regions: southern Italy comprises north Apulia, Calabria, south Apulia, and Sicily; Sardinia is divided into inland and coastal Sardinia; and northern Italy comprises Umbria and east and west Liguria. Hence, the true number of clusters is hypothesised to correspond to either $3$ areas or $9$ regions. 

The full family of \textsf{IMIFA} models is fitted to the olive oil data with results detailed in Table \ref{Table:OliveResults}. Models relying on pre-specification of finite ranges of $G$ and/or $q$ are based on $G = 1,\ldots,9$ and $q = 0,\ldots,6$. Clustering performance is evaluated using the adjusted Rand index (ARI; \citealp{Hubert1985}) and the misclassification rate, compared to the $3$ area labels. The $\alpha$ parameter is reported as its fixed value or posterior mean, as appropriate. Table \ref{Table:OliveResults} shows the flexibility and accuracy of the developed model family, and of the \textsf{IMIFA} model in particular which has the best clustering performance. Additionally, \textsf{IMIFA} is the most computationally efficient model considered, among those in the \textsf{IMIFA} family achieving clustering, as it requires only one run. This speed improvement would be exacerbated with larger data sets. However, methods requiring fitting of multiple models were run here in series; parallel implementations would reduce runtimes. Finally, models with different numbers of cluster-specific factors show improved clustering performance compared to the corresponding finite factor model in every case.

\begin{figure}[H]
	\centering
	\begin{minipage}[t]{.49\textwidth}
		\centering
		\includegraphics[width=\textwidth, keepaspectratio]{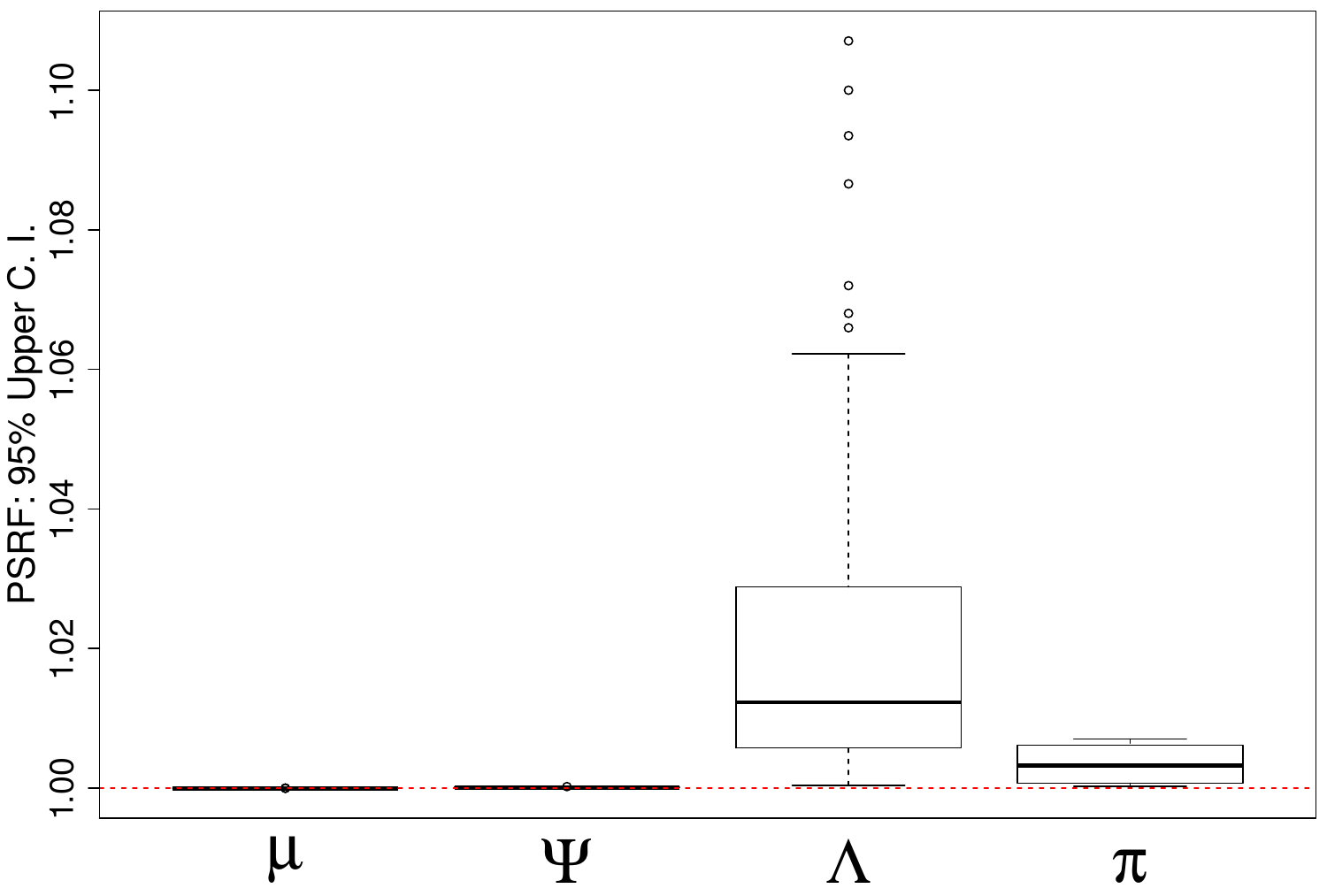}
		\caption[Boxplots of the upper PSRF limits for all cluster-specific parameters in the overdispersed IMIFA chains fit to the olive oil data.]{Boxplots of the upper PSRF limits for all cluster means, uniquenesses, loadings, and mixing proportions in the overdispersed \textsf{IMIFA} chains fit to the olive oil data, with red reference line at $1$.}
		\label{Plot:OlivePSRF}
	\end{minipage}\hfill%
	\begin{minipage}[t]{.49\textwidth}
		\centering
		\includegraphics[width=\textwidth, keepaspectratio]{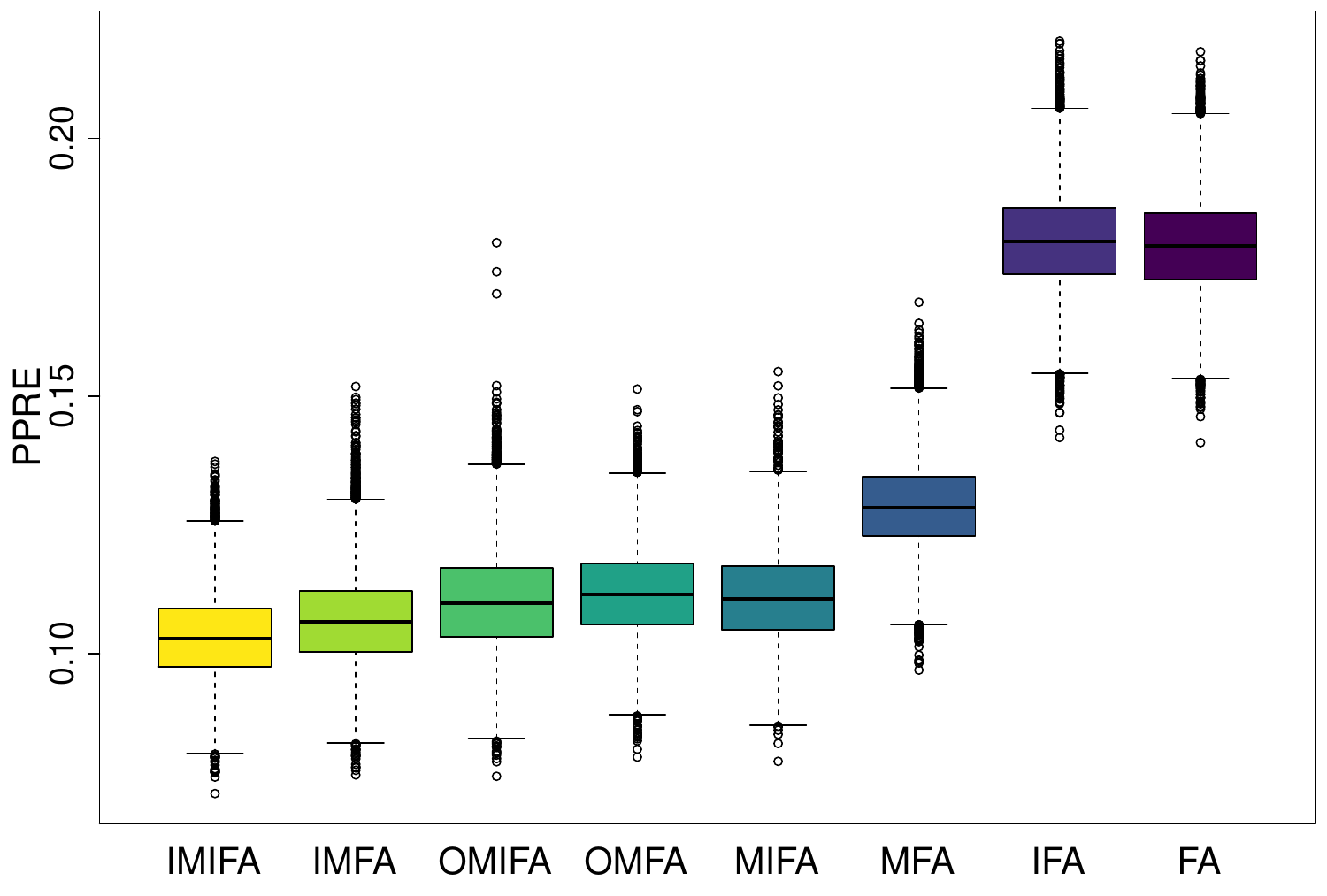}
		\caption[Boxplots of the PPRE values for the full family of IMIFA models fit to the olive oil data.]{Boxplots of the PPRE values for the full family of \textsf{IMIFA} models fit to the olive oil data. Values close to zero indicate good model fit.}
		\label{Plot:OlivePPRE}
	\end{minipage}
\end{figure}

Table \ref{Table:OliveResults} also shows that the performance of the \textsf{IMIFA} model compares favourably to the best parsimonious Gaussian mixture model, fit via the \texttt{pgmm} \textsf{R} package \citep{pgmm2019} and the best mixture of factor mixture analysers (\textsf{MFMA}) model \citep{Viroli2010}, evaluated with $1,\ldots, 5$ components in both layers. Models with zero factors were not considered~in either case. \textsf{IMIFA} also outperforms the best constrained Gaussian mixture model fitted using \texttt{mclust} \citep{Scrucca2016}. These finite mixtures are fit via maximum likelihood and use the BIC for model selection after fitting a large number of candidate models.
\begin{table}[H]
	\caption[Results of fitting the full family of IMIFA models and other methods to the olive oil data.]{Results of fitting a range of models, including the full \textsf{IMIFA} family, to the Italian olive oil data, detailing the number of candidate models explored, the run-time relative to the \textsf{IMIFA} run (approx. $782$ seconds), the posterior mean or fixed value of $\alpha$, the posterior mean of $d$, modal estimates of $G$ and $\mathbf{Q}$, and the ARI and misclassification rate as evaluated against the known area labels, under the optimal or modal model as appropriate.}
	\label{Table:OliveResults}
	\centering
	\scriptsize
	\extrarowheight 5pt
	
	\begin{threeparttable}
	\begin{tabular}[pos=center]{ c | c | c | c | c | c | c | c | c}
		{{\centering}Model} & {{\centering}\#~Models} &  {{\centering}Relative~Time} &
		{{\centering}$\alpha$} &
		{{\centering}$d$} & {{\centering}$G$} & {{\centering}$\mathbf{Q}$} & {{\centering}ARI} & {{\centering}Error~(\%)}\\\hline
		\textsf{IMIFA} & $1$ & $1.00$ & $0.48$ & $0.01$ & $4$ & $6$, $3$, $6$, $2$ & $0.94$ & $8.39$\\
		\textsf{IMFA} & $7$ & $4.14$ & $0.62$ & $0.01$ & $5$ & $6$, $6$, $6$, $6$, $6$ & $0.91$ & $14.86$\\
		\textsf{OMIFA} & $1$ & $1.19$ & $0.02$ & --- & $4$ & $6$, $3$, $6$, $4$ & $0.93$ & $9.97$\\
		\textsf{OMFA} & $7$ & $5.11$ & $0.03$ & --- & $5$ & $6$, $6$, $6$, $6$, $6$ & $0.85$ & $15.56$\\
		\textsf{MIFA} & $9$ & $3.41$ & $1$ & --- & $5$ & $6$, $3$, $6$, $6$, $4$ & $0.92$ & $10.31$\\
		\textsf{MFA} & $63$ & $13.86$ & $1$ & --- & $2$ & $5$, $5$ & $0.82$ & $17.13$\\
		\textsf{IFA} & $1$ & $0.11$ & --- & --- & $1$ & $6$ & --- & ---\\
		\textsf{FA} & $7$ & $0.37$ & --- & --- & $1$ & $6$ & --- & ---\\
		\hline
		\centering{\texttt{mclust}\tnotex{tnA}} & $115$ & $0.01$ & --- & --- & $6$ & --- & $0.56$ & $38.64$\\
		\centering{\textsf{MFMA}\tnotex{tnA}} & $1,350$ & $4.68$ & --- & --- & $4$ & $5$, $5$, $5$, $5$ & $0.68$ & $20.28$\\
		\centering{\texttt{pgmm}\tnotex{tnA} \textsuperscript{~,}\tnotex{tnB}} & $588$ & $4.46$ & --- & --- & $5$ & $6$, $6$, $6$, $6$, $6$ & $0.53$ & $35.84$
	\end{tabular}
	\vspace{1.5em}
	\begin{tablenotes}[flushleft]
	\itemsep 1em
	\item[$\dagger$] \label{tnA} Due to the various covariance matrix decompositions considered, the results for \texttt{mclust}, \textsf{MFMA}, and \texttt{pgmm} are reported for the unstandardised data, for which superior clustering performance in terms of the ARI was achieved in each case. 
	\item[$\ddagger$] \label{tnB} The optimal \texttt{pgmm} model uses the \textsf{UCU} constraints on the uniquenesses (i.e. $\boldsymbol{\Psi}_g=\boldsymbol{\Psi}$). Among the more directly comparable unconstrained \textsf{UUU} models, the optimal one according to BIC has $G=6$ components, each with $5$ factors, and achieves an ARI of $0.43$. Notably, the \texttt{pgmm} models chosen by BIC both have more components than the IMIFA model.
	\end{tablenotes}
	\end{threeparttable}
\end{table}
It is also notable that within the set of \textsf{IMIFA} models relying on information criteria, those deemed optimal were not necessarily optimal in a clustering sense. For instance, the $4$-cluster \textsf{MIFA} model yields an ARI of $0.94$ and a misclassification rate of $6.99\%$, with respect to the $3$ area labels, despite its sub-optimal BICM. Similarly, the BICM and BIC-MCMC criteria suggest different optimal \textsf{MFA} models. For the \textsf{IMIFA} model $\smash{\widehat{\kappa}\approx0.89}$, suggesting similar inference would have resulted under a DP prior. Indeed, the results obtained by the \textsf{OMIFA} and \textsf{OMFA} models are similar to those of their infinite mixture counterparts, though the latter provide a better fit to the data (see Figure \ref{Plot:OlivePPRE}).\clearpage

\begin{figure}[H]
	\centering
	\begin{minipage}[t]{.49\textwidth}
		\centering
		\includegraphics[width=\textwidth, keepaspectratio]{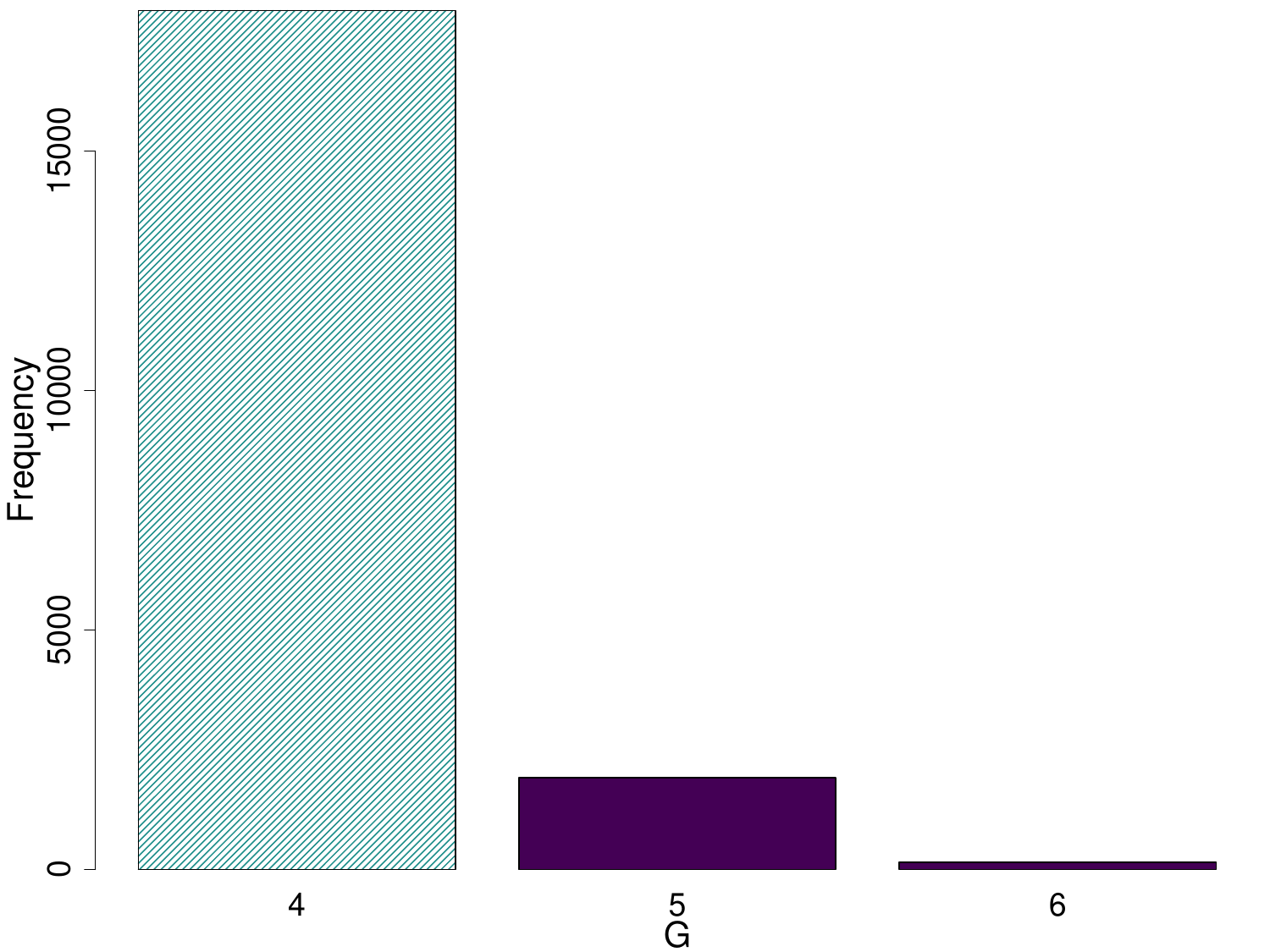}
		\caption[Posterior distribution of $G$ under the \textsf{IMIFA} model fit to the olive oil data.]{Posterior distribution of $G$ under the \textsf{IMIFA} model fit to the olive oil data. The number of clusters is estimated by the modal value, $\widehat{G}=4$.}
		\label{Plot:Gdist}
	\end{minipage}\hfill%
	\begin{minipage}[t]{.49\textwidth}
		\centering
		\includegraphics[width=\textwidth, keepaspectratio]{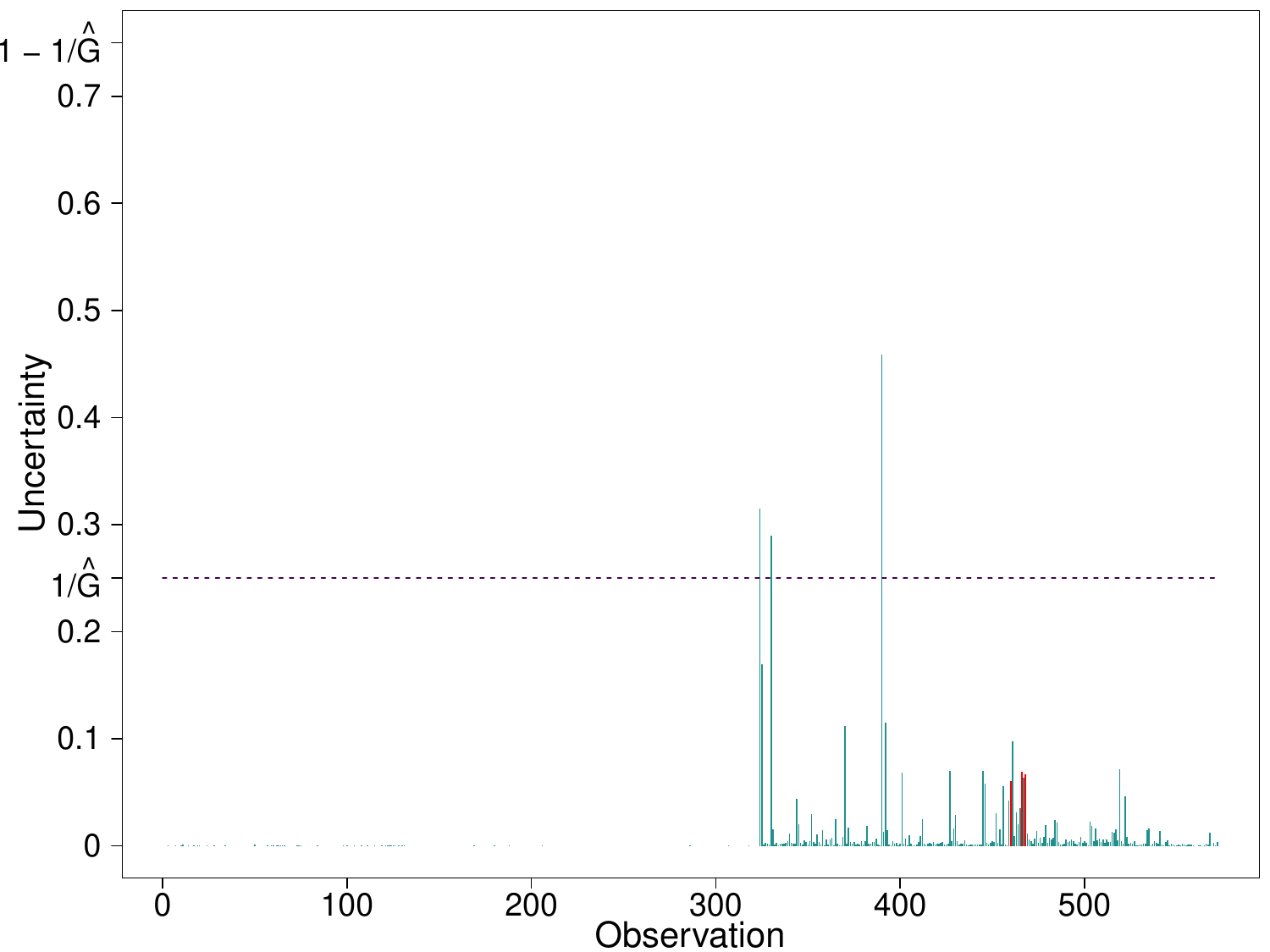}
		\caption[Clustering uncertainties for the \textsf{IMIFA} model fit to the olive oil data.]{Clustering uncertainties for the \textsf{IMIFA} model fit to the olive oil data. Oils misclassified according to the labels in Table \ref{Table:OliveFour} are highlighted in red.}
		\label{Plot:Uncertainty}
	\end{minipage}
\end{figure}

Figure \ref{Plot:Gdist} shows a barchart approximation to the posterior distribution of $G$ under the \textsf{IMIFA} model. The modal value of $4$, visited in $\approx90\%$ of posterior samples, is used as the estimate of the true number of clusters (with $95\%$ credible interval $[4,5]$). Table \ref{Table:OliveArea} tabulates the MAP clustering against the $3$ area labels and suggests this solution makes geographic sense, in that northern oils are cleanly split into two sub-clusters. Cluster 1 contains all of the $323$ southern Italy oils: this large cluster requires the largest number of factors ($\widehat{q}_1 = 6~[5, 6]$, with $95\%$ credible intervals in brackets). Some of the other clusters require notably fewer ($\widehat{q}_2 = 3~[1,6]$, $\widehat{q}_3 = 6~[3,6]$, and $\widehat{q}_4 = 2~[1,4]$). Table \ref{Table:OliveFour} gives the confusion matrix with oils from the north labelled by their associated region(s), yielding an ARI of $0.994$ and a misclassification rate of $0.52\%$. Figure \ref{Plot:Uncertainty} shows the uncertainty in the allocations to these clusters. Only three oils have large probability of belonging to a cluster other than the one to which they were assigned by the \textsf{IMIFA} model.

\begin{table}[H]
	\caption[Confusion matrices of the MAP IMIFA clustering of the olive oil data.]{Confusion matrices of the MAP \textsf{IMIFA} clustering of the Italian olive oils against (\subref{Table:OliveArea}) the known $3$ area labels and (\subref{Table:OliveFour}) the new labelling in which northern Italy is split into its constituent sub-regions.}
	\begin{subtable}[t]{.455\linewidth}
		\centering
		\scriptsize
		\extrarowheight 5pt
		
		\caption{3 area cross tabulation}		
		\label{Table:OliveArea}
		\begin{tabular}{l | c c c c}
			& 1 & 2 & 3 & 4 \\\hline
			Southern Italy& $323$ & $0$ & $0$ & $0$\\
			Sardinia & $0$ & $98$ & $0$ & $0$\\
			Northern Italy& $0$ & $0$ & $103$ & $48$
		\end{tabular}
	\end{subtable}\hfill%
	\begin{subtable}[t]{.545\linewidth}
		\centering
		\scriptsize
		\extrarowheight 2.5pt
		
		\caption{4 area cross tabulation}
		\label{Table:OliveFour}
		\begin{tabular}{l | c c c c}
			& 1 & 2 & 3 & 4 \\\hline
			Southern Italy & $323$ & $0$ & $0$ & $0$\\
			Sardinia & $0$ & $98$ & $0$ & $0$\\
			East Liguria\,\&\,Umbria & $0$ & $0$ & $100$ & $0$\\
			West Liguria & $0$ & $0$ & $3$ & $48$
		\end{tabular}		
	\end{subtable} 
\end{table}

To assess sensitivity to starting values, the \textsf{IMIFA} model was re-fitted using multiple random initial allocations, implying also different random draws from the priors for parameter starting values. These runs led to identical inference about $\widehat{G}$ and $\widehat{\mathbf{Q}}$ and equivalent clustering performance. These overdispersed chains were used to compute the upper $95\%$ PSRF confidence limits depicted in Figure \ref{Plot:OlivePSRF}, which indicate good convergence. The PPRE boxplots in Figure \ref{Plot:OlivePPRE} demonstrate the superior fit of the \textsf{IMIFA} model (with a median PPRE of $0.10$) to the olive oil data, compared to the other \textsf{IMIFA} family models. Histograms comparing the bin counts between the modelled and replicate data sets for each variable, under the \textsf{IMIFA} model, are given in Appendix~\ref{Subsection:AdditionalResults}. 

\subsection[Spectral Metabolomic Data]{Spectral Metabolomic Data}\label{Subsection:SpectralData}
\begin{figure}[H]
	\centering
	\includegraphics[width=0.7125\textwidth, keepaspectratio]{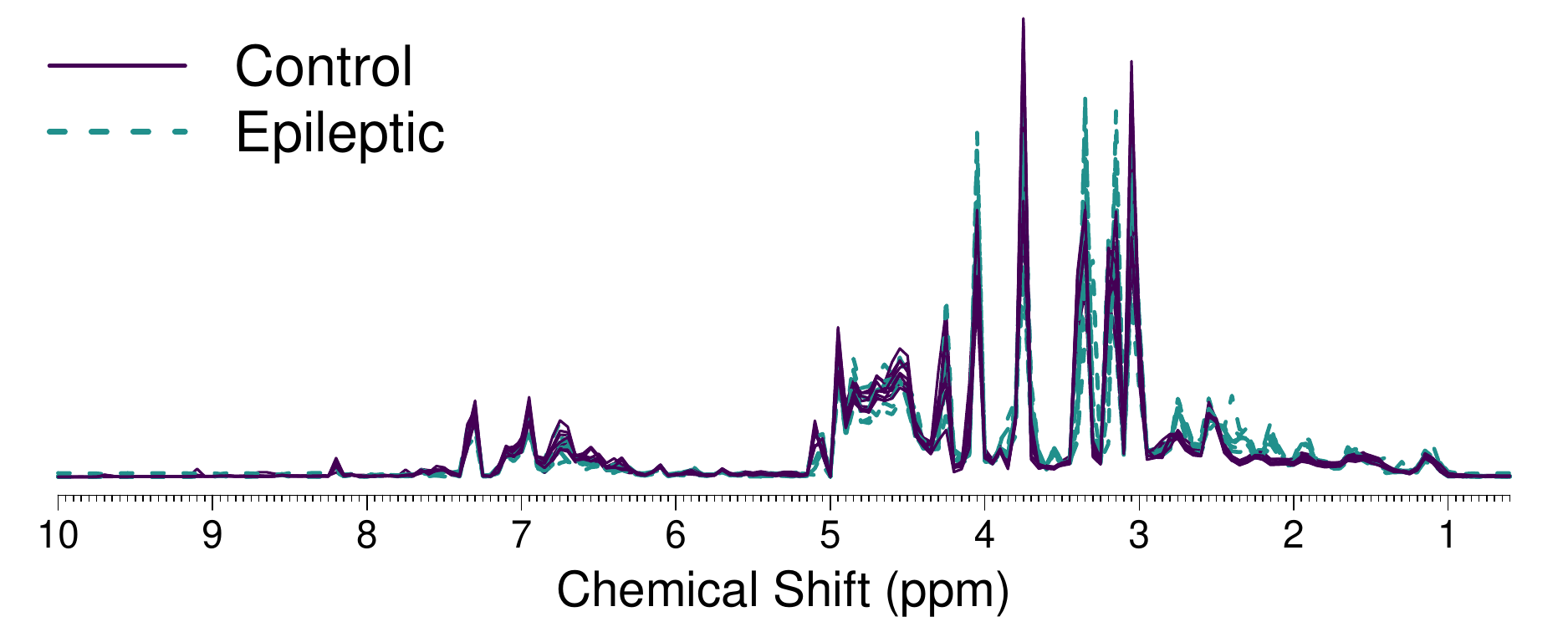}
	\caption[Raw spectral metabolomic data.]{Raw spectral metabolomic data.}
	\label{Plot:UrineData}
\end{figure}

\textsf{IMIFA} is employed to cluster spectral metabolomic data for which $N \ll p$ (Figure \ref{Plot:UrineData}). The data are nuclear magnetic resonance spectra consisting of $p = 189$ spectral peaks from urine samples of $N = 18$ participants, half of which are known to have epilepsy \citep{Carmody2010, Gift2010}. Interest lies in uncovering any underlying clustering structure given the $N \ll p$ setting.\vspace{-0.25em}
\vspace{-0.25em}
Data were mean-centred and Pareto scaled \citep{vandenBerg2006}. Although $N \ll p$, no restrictions are imposed on the uniquenesses as the sample variances are quite imbalanced. Fitting \textsf{MIFA} models for $G = 1, \ldots, 5$ is feasible as $N$ is small. The BICM criterion chooses $\smash{\widehat{G} = 2}$ as optimal and one participant is misclassified. \textsf{IMIFA}, however, unanimously visits a 2-cluster model and perfectly uncovers the group structure.

\begin{figure}[H]
	\centering
	\includegraphics[width=0.8\textwidth, keepaspectratio]{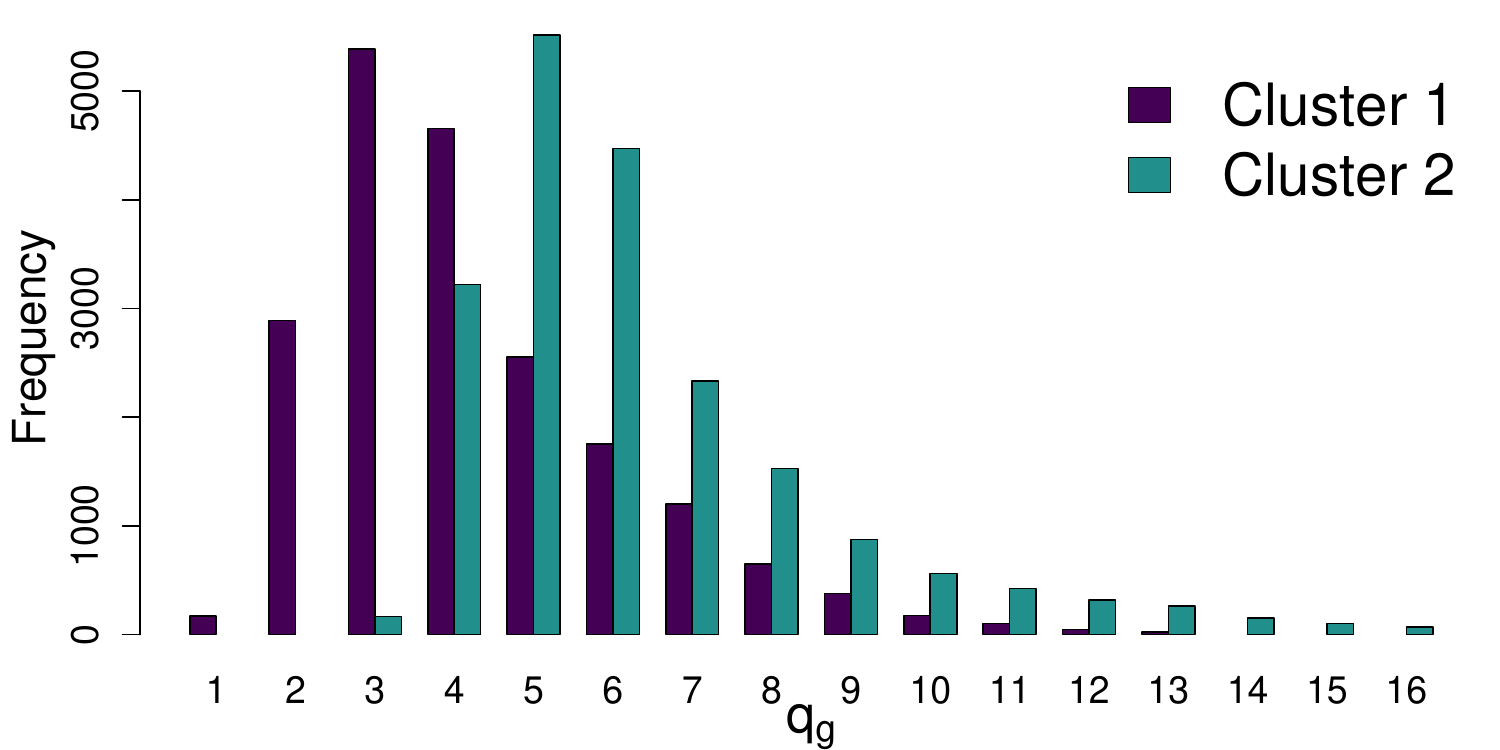}
	\caption[Posterior distribution of $q_g$ under the \textsf{IMIFA} model fit to the metabolomic data.]{Posterior distribution of $q_g$ under the \textsf{IMIFA} model fit to the metabolomic data.} 
	\label{Plot:MIFA_Qdist}
\end{figure}

\begin{figure}[H]
	\centering
	\includegraphics[width=\textwidth, keepaspectratio]{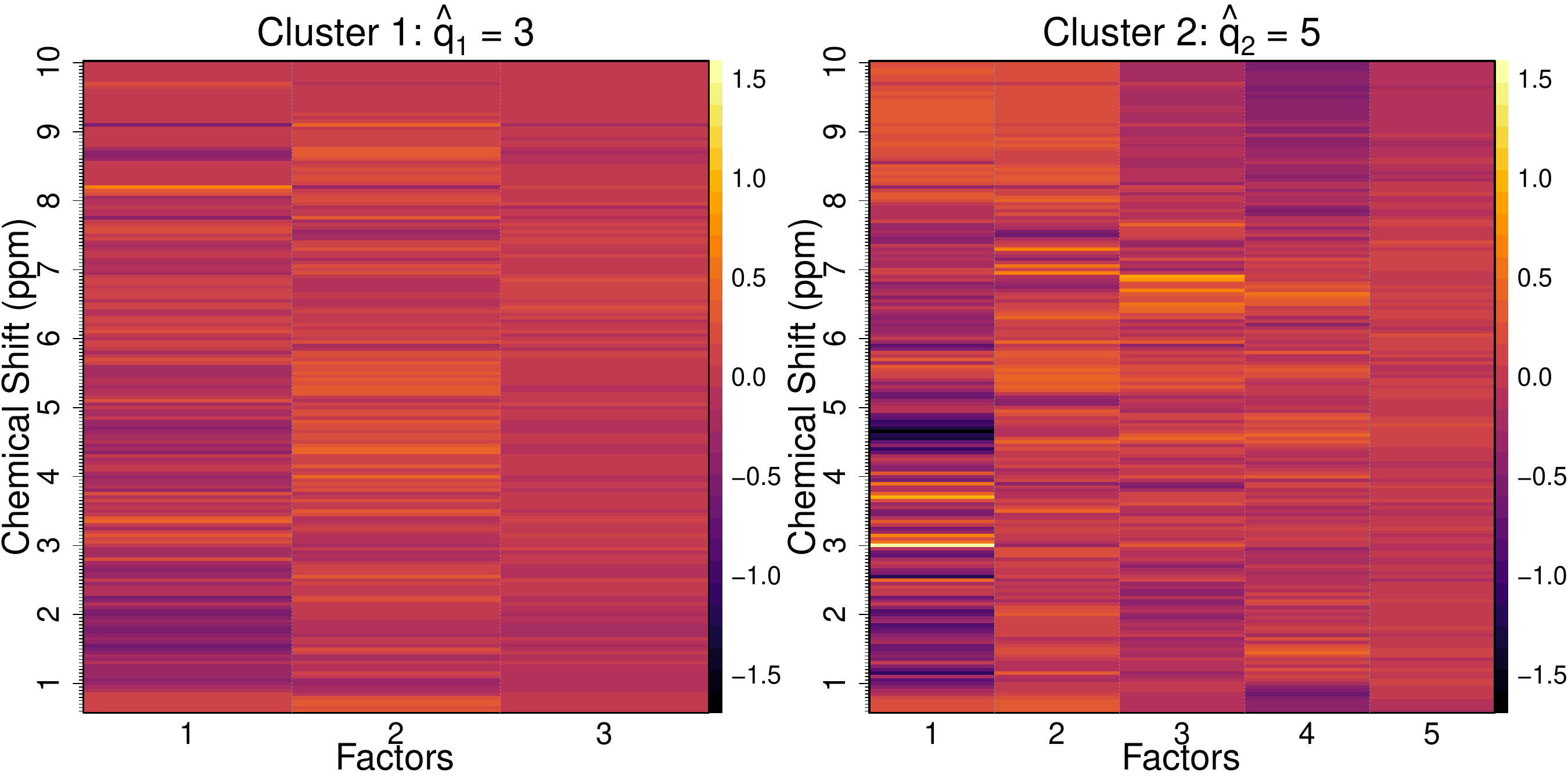}
	\caption[Heat maps of cluster-specific posterior mean loadings matrices in the IMIFA model fit to the metabolomic data.]{Heat maps, calibrated to a common colour scale, of posterior mean loadings matrices in the clusters uncovered by fitting \textsf{IMIFA} to the spectral metabolomic data. }
	\label{Plot:MIFA_Loadings}
\end{figure}

The modal estimates of the number of factors in each \textsf{IMIFA} cluster are $\widehat{q}_1 = 3~[2, 9]$ and $\widehat{q}_2 = 5~[4, 13]$ (see Figure \ref{Plot:MIFA_Qdist}). Cluster 1 corresponds to the control group and Cluster 2 to the epileptic participants. Figure \ref{Plot:MIFA_Loadings} illustrates the $\smash{p\times \widehat{q}_g}$ posterior mean loadings matrices, based on retained samples with $\smash{\widehat{q}_g}$ or more factors, after Procrustes rotation to a common template for both clusters. The sparsity and shrinkage induced by the MGP prior is apparent, as is the greater complexity in Cluster 2, given the greater variation in colour and larger number of factors. For instance, many elevated loadings are visible for chemical shift values between $8$ and $10$ for the first two factors in Cluster 2; this activity is not present for other factors in either cluster. In general, the distributions of the loadings within a factor exhibit narrow spread around zero, particularly for the cluster of control participants, with the exception of the regions of the spectrum corresponding to the large peaks between chemical shifts of $3$ and $5$ in Figure \ref{Plot:UrineData}.
\vspace{-1em}
\textsf{IMIFA} outperforms the optimal $\smash{\widehat{G}=3}$ \texttt{mclust} model and the optimal $\smash{\widehat{G}=2}$, $\smash{\widehat{q}=5}$ \texttt{pgmm} model, with respective ARI values of $0.73$ and $0.27$. The clustering performance~of the optimal \textsf{MFMA} model is identical to the optimal \textsf{MIFA} model described above. Given the $N \ll p$ nature of the data, spectral clustering with the Gaussian kernel \citep{Ng2001} is also considered. The eigengap heuristic suggests $\smash{\widehat{G}=2}$ and a~perfect clustering is achieved almost instantaneously. However, the approach  does not characterise the uncovered clusters in an interpretable manner, nor provide estimates of cluster membership uncertainty as given by model-based clustering approaches such as \textsf{IMIFA}. 

The median PPRE for the \textsf{IMIFA} model of $0.21~[0.18,0.24]$ shows~good model fit, given the size and dimensionality of the data. The median PSRF upper~$95\%$ confidence limits, using three randomly initialised auxiliary chains, for the cluster means, uniquenesses, loadings, and mixing proportions of $1.01\,(0.01)$, $1.00\,(<0.01)$, $1.01\,(0.08)$,~and $1.00\,(<0.01)$ respectively, show good mixing also (standard deviations in parentheses). Notably, all chains yield the same inference about~$\smash{\widehat{G}}$ and $\smash{\widehat{\mathbf{Q}}}$. So too, again, does the \textsf{OMIFA} model, although its model fit is inferior (median PPRE=$0.26$).

\subsection[Handwritten Digit Data]{Handwritten Digit Data}\label{Subsection:Digits}

A final illustration of \textsf{IMIFA} is given through its application to handwritten digit data from the United States Postal Service (USPS; \citealp{LeCun1990,Hastie2001}). Here $N=7,291$ images of the digits $0,\ldots,9$ are considered, taken from handwritten zip codes. The data are~not balanced in terms of digit labels. Each image is a $16\times16$ grayscale grid concatenated into~a $p=256$-dimensional vector; data were mean-centred but not scaled.  Such data are often considered in the context of manifold learning, positing that the data dimensionality is artificially~\mbox{high. Given} $N$ and $p$, fitting a range of \textsf{MFA} or \textsf{MIFA} models is practically infeasible. Results of a single \textsf{IMIFA} run are presented here. For these data, it is reasonable to expect the number of components to grow as the sample size grows. It is anticipated that the flexibility afforded by having cluster-specific numbers of factors will help characterise digits with different geometric features.

\begin{figure}[H]
	\centering
	\includegraphics[width=0.9375\textwidth, keepaspectratio]{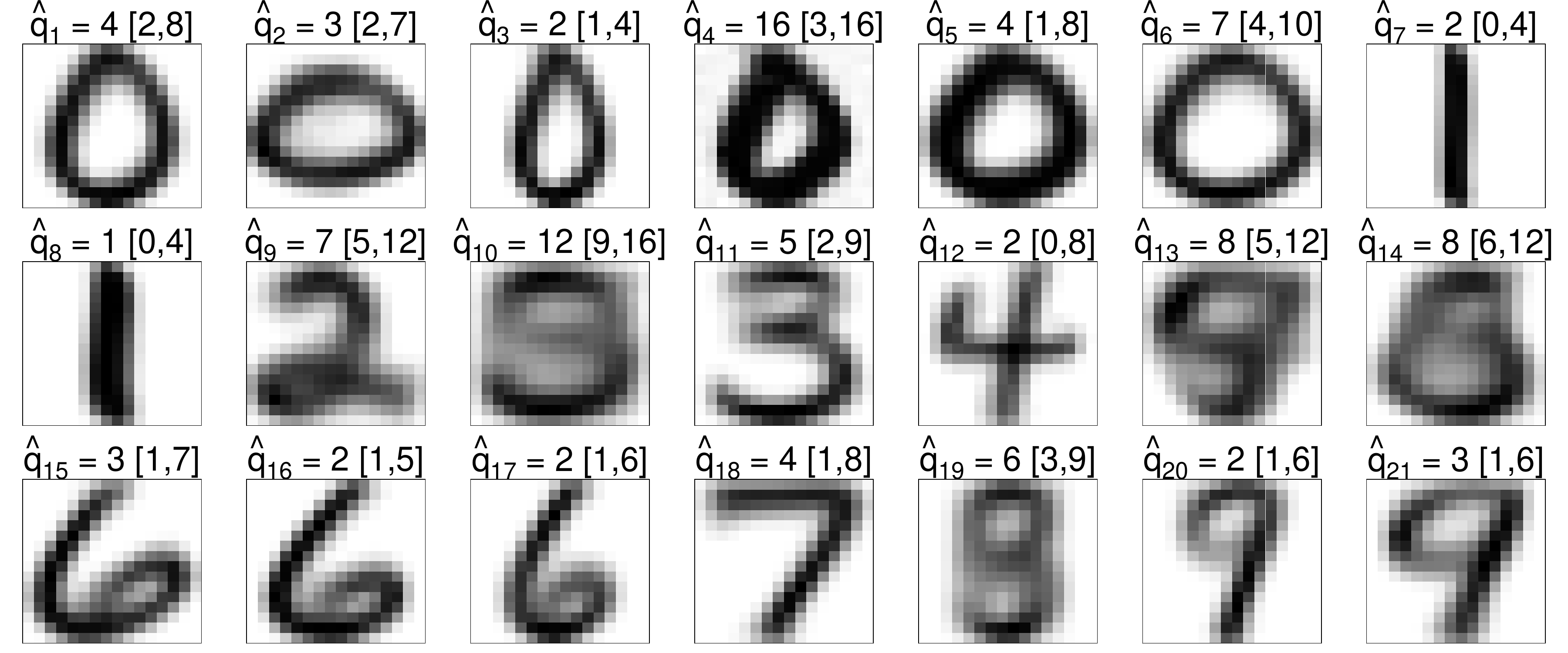}
	\caption[Posterior mean images for clusters uncovered by fitting IMIFA to the USPS data.]{Posterior mean images for clusters uncovered by fitting \textsf{IMIFA} to the USPS data. Plots are ordered according to Table \ref{Table:DigitConfusion} and labelled with the modal $\smash{\widehat{q}_g}$.}
	\label{Plot:USPSDigits}
\end{figure}

The \textsf{IMIFA} model visited a $\widehat{G}=21$ cluster solution in all posterior samples; Table \ref{Table:DigitConfusion} cross-tabulates the MAP clustering against the known digit labels and 
achieves an ARI of $0.33$. The median PPRE of $0.05$ $[0.04,0.06]$ indicates good model fit. The overdispersed chains used to~compute the PSRF diagnostics lead to identical inference about~the number of clusters but slightly different inference about the modal numbers of cluster-specific factors. The ARI values between each resulting pair of MAP partitions all exceed $0.93$. As before, good mixing is indicated by median PSRF upper $95\%$ confidence limits for the cluster means, uniquenesses, and mixing proportions of $1.01\, (0.01)$, $1.01\,(0.01)$, and $1.01\,(<0.01)$, respectively. In computing the diagnostic for the loadings ($1.14\,(0.35)$), only the first factor (common to all loadings matrices across all clusters in all chains) was considered, for reasons of fairness and computational resource constraints.\vspace{-0.125em}
\begin{table}[H]
	\caption[Cross tabulation of the IMIFA model's MAP clustering against the true digit labels for the USPS data.]{Cross tabulation of the \textsf{IMIFA} model's MAP clustering (rows) against true digit labels (columns) for the USPS data. Cells that are $0$ are blank for clarity. Posterior means $\smash{\widehat{\pi}_g}$ and modal estimates $\smash{\widehat{q}_g}$, with associated $95\%$ credible intervals, are also given.}
	\label{Table:DigitConfusion}
	\centering
	\scriptsize
	
	\tabcolsep 7.5pt
	\extrarowheight 1.25pt
	\begin{tabular}[pos=center]{l | c c c c c c c c c c | c | l}
		\centering
		& 0 & 1 & 2 & 3 & 4 & 5 & 6 & 7 & 8 & 9 & $\widehat{\pi}_g$ & $\widehat{q}_g$\\\hline
		1 & $359$ & & & & & & & & & & $0.05$ &$4~[2, 8]$\\
		2 & $58$ & & $12$ & & & $3$ & $2$ & & & & $0.01$ & $3~[2, 7]$\\
		3 & $108$ & & & & & & & & & & $0.01$ & $2~[1, 4]$\\
		4 & $9$ & & & & & & & & & & $0.00$ & $16~[3, 16]$\\
		5 & $95$ & & & & & & & & & & $0.01$ & $4~[1, 8]$\\
		6 & $308$ & & & & & $3$ & & & & & $0.04$ & $7~[4, 10]$\\
		7 & & $844$ & & & $2$ & & & & & & $0.12$ &$2~[0, 4]$\\
		8 & & $133$ & & & & & & & $1$ & & $0.02$ & $1~[0, 4]$\\
		9 & & $2$ & $392$ & $10$ & & $1$ & & & & & $0.05$ & $7~[5, 12]$\\
		10 & $59$ & & $121$ & $93$ & $19$ & $91$ & $13$ & $2$ & $25$ & $4$ & $0.06$ & $12~[9, 16]$\\
		11 & & & & $136$ & & $64$ & & & & & $0.03$ & $5~[2, 9]$\\
		12 & & & & & $38$ & $1$ & & $1$ & & & $0.01$ & $2~[0, 8]$\\
		13 & $25$ & & $3$ & $7$ & $98$ & $51$ & $2$ & $36$ & $59$ & $28$ & $0.04$ & $8~[5, 12]$\\
		14 & $48$ & & $73$ & $61$ & $62$ & $135$ & $32$ & $1$ & $16$ & $6$ & $0.06$ & $8~[6, 12]$\\
		15 & $1$ & & & & & & $83$ & & & & $0.01$ & $3~[1, 7]$\\
		16 & $1$ & & & & & & $74$ & & & & $0.01$ & $2~[1, 5]$\\
		17 & & $2$ & & & $4$ & $19$ & $381$ & & $2$ & & $0.06$ & $2~[1, 6]$\\
		18 & & & & & & & & $207$ & & & $0.03$ & $4~[1, 8]$\\
		19 & $123$ & $8$ & $129$ & $348$ & $247$ & $184$ & $77$ & 26 & $420$ & $84$ & $0.23$ & $6~[3, 9]$\\
		20 & & $16$ & $1$ & $3$ & $120$ & $1$ & & $338$ & $19$ & $451$ & $0.13$ & $2~[1, 6]$\\
		21 & & & & & $62$ & $3$ & & $34$ & & $71$ & $0.02$ & $3~[1, 6]$
	\end{tabular}
\end{table}
\vspace{-0.25em}
\indent Generally, \textsf{IMIFA} assigns images of the same digit, albeit written differently, to different clusters. Posterior mean images for each cluster are shown in Figure \ref{Plot:USPSDigits}, ordered,~as is Table \ref{Table:DigitConfusion}, from~0 to 9 according to the digit most frequently assigned to the related cluster. Cluster 7 and the smaller cluster 8 capture the digit 1 written in a straight and slanted fashion, respectively. \mbox{Clusters} 15--17 represent the digit 6 written with extended, medium, and compact loop curvature, respectively. Notably, cluster 15 requires more factors than clusters 16 and 17. A similar interpretation follows for clusters 20 and 21 ($\smash{\widehat{q}_{20} = 2, \widehat{q}_{21} = 3}$), mostly capturing the digit~9 with small and large loops, respectively. Cluster 19 appears to mostly represent the digit 8 and has a large number of factors ($\smash{\widehat{q}_{19} = 6}$) in comparison, say, to clusters 7 and 8 ($\smash{\widehat{q}_7 = 2, \widehat{q}_{8} = 1}$) which capture the digit 1. This is intuitive, as 8 is a more geometrically complex digit than 1. However, some clusters appear to be diluted by the confusion of the so-called `closed' 4, in contrast to the `open' 4 in cluster 12, with the digits 3, 5, and 8 (cluster 19) and the digits 7 (written with a horizontal bar) and 9 (clusters 20 and 21). Many clusters capture the most common digit 0, with differing degrees of elongation and border thickness. Of concern here is cluster 4, containing just 9 units; the fact that $\smash{\widehat{q}_{4}=16}$, the upper AGS limit, suggests that the model struggles to~shrink the number of factors in poorly populated clusters. This difficulty is highlighted further in the simulation studies in Appendix \ref{Subsection:SimulationStudies}. Finally, Table \ref{Table:DigitConfusion} indicates that clusters 10, 13, and 14 also capture several other digits, all of which are reflected in the blurriness of the resulting posterior mean images and in $\smash{\widehat{q}_{10}}$, $\smash{\widehat{q}_{13}}$, and $\smash{\widehat{q}_{14}}$ being quite large. The cluster-membership uncertainties are visualised in Appendix~\ref{Subsection:AdditionalResults}.\vspace{-0.33em}
\vspace{-0.5em}
It is computationally infeasible to run \texttt{mclust}, \texttt{pgmm}, or \textsf{MFMA} on these large data, as an exhaustive model search would be too vast. For comparative purposes, a DP-BP model \citep{Chen2010} is fitted; this approach also simultaneously assumes infinitely many components and factors. It finds $43$ clusters, each with around $14$ factors, and achieves an ARI of $0.32$. Cross tabulating this clustering against the~$21$ clusters of the  \textsf{IMIFA} model shows that some of the DP-BP clusters are encapsulated by the larger \textsf{IMIFA} clusters. \textsf{IMIFA} is thus the more parsimonious approach and affords greater cluster-specific factor flexibility. Additionally, a finite mixture of matrix-normal distributions \citep{Viroli2011} is also fitted. This approach accounts for the grid nature of~the~data, but is computationally infeasible for $G > 15$ and requires a model selection strategy. The optimal model according to BIC yields $\smash{\widehat{G}=12}$ and $\mbox{ARI} = 0.38$. While neither \textsf{IMIFA} nor the DP-BP model account for the spatial structure in the data, they demonstrate comparative performance without the need for a computationally expensive model search.\vspace{-0.25em}

\section[Discussion]{Discussion}
\label{Section:Discussion}

The proposed \textsf{IMIFA} model is a Bayesian nonparametric approach  clustering high-dimensional data using factor-analytic mixture models. By extending the MGP prior \citep{Bhattacharya2011} to the PYP-MGP setting, the model sidesteps the fraught and computationally intensive task of determining the optimal number of clusters and factors using  model selection criteria. Thus, the \text{IMIFA} model is recommended when fitting factor-analytic mixtures in settings where an exhaustive model search is computationally infeasible. Though \textsf{IMIFA} is not entirely choice-free, it achieves improved clustering results by allowing factor-analytic models of different dimensions in different clusters. If small clusters are inferred, one may wish to prune or merge small clusters with the larger clusters \citep{West1994} or assess whether the small clusters are in fact of domain-specific interest. While comparative performance can be achieved by the \textsf{IMIFA} and \textsf{OMIFA} models, one may wish to fit a \textsf{MIFA} or \textsf{OMIFA} model when the expectation is that the number of clusters is fixed or unlikely to grow with $N$, respectively.

Future research directions are varied and plentiful. Incorporating covariates, in the spirit of Bayesian factor regression models \citep{West2003, Carvalho2008}, would allow for direct inclusion of the weight and urine pH covariates available with the metabolomic data, for example. Furthermore, the models could be extended to the (semi-)supervised model-based classification setting where all (or some) of the data are labelled. While constraints on the uniquenesses across variables and/or clusters are allowed, there is scope for also constraining the loadings across clusters. Though the number of factors would no longer be cluster-specific, the common number of loadings columns would be estimated in a similarly automatic fashion. However, incorporating covariance matrix constraints in the \textsf{IMIFA} model family problematically reintroduces the need for model selection strategies, in order to choose between them, though the BICM criterion could feasibly be used for this purpose also.

As proposed by \citet{Bhattacharya2011}, the MGP hyperparameters could be learned via Metropolis-Hastings, and thus also be made cluster-specific. This could help combat some difficulties identified in the simulation studies in Appendix \ref{Subsection:SimulationStudies}. For example, learning those related to local shrinkage may help when loadings are notably dense. Learning those related to column shrinkage may help in settings with many small clusters, where \textsf{IMIFA} struggles to adaptively truncate loadings columns. In principle,~a further global shrinkage parameter $\varpi$ could be added to the MGP prior to borrow information across clusters, i.e. $\smash{\lambda_{jkg} \given \ldots \sim \textrm{N}_1\big(0, \phi_{jkg}^{-1}\tau_{kg}^{-1}\sigma_g^{-1}\varpi^{-1}\big)}$. Alternatively, the infinite factor prior of \citet{Legramanti2020} could be employed, which decouples control over the cumulative shrinkage rate and the active loadings terms. The IBP prior \citep{Knowles2007,Knowles2011,Rockova2016} could be also extended to the infinite mixture setting, as per the DP-BP model of \citet{Chen2010}. Thus, the \textsf{IMIFA} family can in fact be considered to be wider than the range of models presented here, with potential for further expansion.

For applied problems, a mismatch between the assumed model and the data distribution will impact inference. \citet{Miller2013,Miller2014} highlight~that posterior consistency for the number of non-empty clusters in infinite mixtures is contingent on correct specification of the component distributions. While they do not discourage the use of infinite mixtures for clustering, they show that a few tiny extra clusters are typically fitted and suggest robustifying inference.
If the data distribution~is close to but not exactly a finite mixture of Gaussians, an infinite Gaussian mixture will introduce more components as the amount of data increases. Potential avenues of exploration thus include considering the \textsf{IMIFA} model with the heavy tailed multivariate $t$-distribution \citep{Peel2000}.
Similarly, modelling of complex component distributions can be achieved by considering the MFMA approach in the context of infinite factor models. Defining robust inference functions as in \citet{Lee2014} or using nonparametric unimodal component distributions as in \citet{Rodriguez2014} may also prove fruitful. Another means of
robustifying inference is to explicitly include a noise component with zero factors to capture outliers which depart from the component multivariate normality assumption. 
Finally, a `coarsened' posterior \citep{Miller2018} could be used for addressing misspecification, by conditioning on~the event that the model generates data close to the observed data in a distributional sense.
\subsection*{Acknowledgements}
\footnotesize{This research was supported by Science Foundation Ireland (SFI/12/RC/2289\_P2). The authors thank the members of the UCD Working Group in Statistical Learning and Prof. Adrian Raftery's Working Group in Model-based Clustering and Prof. David Dunson for helpful discussion. The authors also thank Prof. Lorraine Brennan (UCD), for the metabolomic data, and the anonymous reviewers for constructive feedback from which this work greatly benefitted.}
\small

\section*{Appendices}
\renewcommand{\thesubsection}{\Alph{subsection}}
\renewcommand\thetable{\thesubsection.\arabic{table}}
\renewcommand\thefigure{\thesubsection.\arabic{figure}}
\renewcommand{\thesubtable}{\alph{subtable}}
\renewcommand{\tablename}{\bfseries Table}

\subsection[Posterior Conditional Distributions: \\ Technical details for sampling from the \textrm{IMIFA} model]{Posterior Conditional Distributions: \\ Technical details for sampling from the \textrm{IMIFA} model}
\label{Subsection:PosteriorConditional}
\setcounter{table}{0}
\setcounter{figure}{0}

The structure of the Metropolis-within-Gibbs sampler to conduct inference for the \textsf{IMIFA}~model and the exact forms of the required conditional distributions are detailed below. Note that $\textrm{Ga}\negthinspace\left(\alpha,\beta\right)$ refers throughout to the gamma distribution with mean $\nicefrac{\alpha}{\beta}$. The number of observations in a component is denoted by $n_g$, where $\smash{\mathbf{n} = \left(n_1,\ldots, n_{\widetilde{G}}\right)^\top}$ sums to $N$, where $\widetilde{G}$ is the current number of active components (of which some may be empty), and $\widetilde{q}_g$ is the current sample of the cluster-specific number of active factors. Algorithms for sampling other models in the \textsf{IMIFA} family can all be considered as special cases of what follows.~The algorithm is implemented in the associated \textsf{R} package \texttt{IMIFA} \citep{IMIFAR2021}.

For $g = 1, \ldots, \widetilde{G}$:
\begin{alignat*}{2}
\boldsymbol{\mu}_g \given \ldots & \sim \textrm{N}_p\Bigg(\boldsymbol{\Omega}_{\boldsymbol{\mu}}^{-1}\bigg(\boldsymbol{\Psi}_g^{-1}\Big(\sum_{i\colon z_{ig} = 1}\mathbf{x}_i - \negmedspace\sum_{i\colon z_{ig} = 1}\boldsymbol{\Lambda}_g\boldsymbol{\eta}{\mathstrut}_i\Big) + \varphi\bm{\mathcal{I}}_p\widetilde{\boldsymbol{\mu}}\bigg), \boldsymbol{\Omega}_{\boldsymbol{\mu}}^{-1}\Bigg),\\[3ex]
\boldsymbol{\eta}{\mathstrut}_i\given z_{ig} = 1, \ldots&\sim\textrm{N}_{\widetilde{q}_g}\bigg(\boldsymbol{\Omega}_{\boldsymbol{\eta}}^{-1}\boldsymbol{\Lambda}_g^\top\boldsymbol{\Psi}_g^{-1}\Big(\mathbf{x}_{i\colon z_{ig} = 1} -\boldsymbol{\mu}_g\Big), \boldsymbol{\Omega}_{\boldsymbol{\eta}}^{-1}\bigg)\quad\mbox{for}~i = 1, \ldots, n_g\,,&\\[3ex]
\psi_{jg}\given \ldots &\sim\textrm{IG}\negthinspace\left(\alpha_0 + \frac{n_g}{2}, \beta_j + \frac{\mathcal{S}_{jg}}{2}\right)\quad\mbox{for}~j = 1, \ldots, p,&\\[3ex]
\boldsymbol{\Lambda}_{jg}\given \ldots &\sim\textrm{N}_{\widetilde{q}_g}\bigg(\boldsymbol{\Omega}_{\boldsymbol{\lambda}}^{-1}\boldsymbol{\eta}{\mathstrut}_{i\colon z_{ig} = 1}^\top\psi_{jg}^{-1}\Big(\mathbf{x}_{i\colon z_{ig} = 1}^{(j)} -\mu_{jg}\Big), \boldsymbol{\Omega}_{\boldsymbol{\lambda}}^{-1}\bigg)\quad\mbox{for}~j = 1, \ldots, p,&\\[3ex]
\phi_{jkg}\given \ldots &\sim \textrm{Ga}\negthinspace\left(\nu_1 + \frac{1}{2},\nu_2 +\frac{\sigma_g\tau_{kg}\lambda_{jkg}^{2}}{2}\right) \quad\mbox{for}~j = 1, \ldots, p\:\:\mbox{and}\:\:k = 1, \ldots, \widetilde{q}_g\,,&\\[3ex]
\delta_{1g}\given \ldots&\sim\textrm{Ga}\negthinspace\left(\alpha_1 + \frac{p\widetilde{q}_g}{2},\beta_1 + \frac{\sigma_{g}}{2}\sum_{h=1}^{\widetilde{q}_g}\tau_{hg}^{\left(1\right)}\sum_{j=1}^p\phi_{jhg}\lambda_{jhg}^{2}\right),\\[3ex]
\delta_{kg} \given \ldots&\sim\textrm{Ga}\negthinspace\left(\alpha_2 + \frac{p}{2}\Big(\widetilde{q}_g-k+1\Big), \beta_2 + \frac{\sigma_{g}}{2}\sum_{h=k}^{q_g}\tau_{hg}^{\left(k\right)}\sum_{j=1}^p\phi_{jhg}\lambda_{jhg}^{2}\right)\quad\mbox{for}~k = 2, \ldots, \widetilde{q}_g\,,&\\[3ex]
\sigma_{g} \given \ldots &\sim \textrm{Ga}\negthinspace\left(\varrho_1 + \frac{p\widetilde{q}_g}{2},\varrho_2 + \frac{\sum_{k=1}^{\widetilde{q}_g}\tau_{kg}\sum_{j=1}^p\phi_{jkg}\lambda_{jkg}^2}{2}\right),\\[3ex]
\upsilon_g \given \ldots &\sim\textrm{Beta}\bigg(1 - d + n_g, \alpha + gd + N -\sum_{l=1}^gn_l\bigg),\\[3ex]
u_i\given z_{ig} = 1, \ldots &\sim\textrm{Unif}\big(0, \xi_g\big)\quad\mbox{for}~i = 1, \ldots, N,&\\[0.5ex]
\intertext{where}
\boldsymbol{\Omega}_{\boldsymbol{\mu}} &= \varphi\bm{\mathcal{I}}_p + n_g\boldsymbol{\Psi}_g^{-1},\\[3ex]
\boldsymbol{\Omega}_{\boldsymbol{\eta}}&=\bm{\mathcal{I}}_{\widetilde{q}_g} + \boldsymbol{\Lambda}{\mathstrut}_g^\top\boldsymbol{\Psi}_g^{-1}\boldsymbol{\Lambda}{\mathstrut}_g\,,\\[3ex]
\boldsymbol{\Omega}_{\boldsymbol{\lambda}} &=\diag\Big(\phi_{j1g}\tau_{1g}\sigma_g,\ldots,\phi_{j\widetilde{q}_gg}\tau_{\widetilde{q}_gg}\sigma_g\Big) + \psi_{jg}^{-1}\boldsymbol{\eta}{\mathstrut}_{i\colon z_{ig} = 1}^\top\boldsymbol{\eta}{\mathstrut}_{i\colon z_{ig} = 1}^{\phantom{^\top}}\,,\\[3ex]
\tau_{hg}^{\left(k\right)} &= \prod_{t=1}^{h}\frac{\delta_{tg}}{\delta_{kg}}\,,\\[3ex]
\pi_g &=\upsilon_g\prod_{l=1}^{g-1}\left(1-\upsilon_l\right),\\
\intertext{and}
\mathcal{S}_{jg} &= \sum_{i\colon z_{ig} = 1}\Big(x_{ij} - \mu_{jg} - \boldsymbol{\Lambda}_{jg}\boldsymbol{\eta}{\mathstrut}_{i}\Big)^\top\Big(x_{ij} - \mu_{jg} - \boldsymbol{\Lambda}_{jg}\boldsymbol{\eta}{\mathstrut}_{i}\Big).
\end{alignat*}
Here $\mathbf{x}^{(j)}$ denotes the $j$-th column of the data matrix, $
\lambda_{jkg}^2$ denotes a single squared loading, and $\smash{\tau_{kg}=\prod_{h=1}^{k}\delta_{hg}}$ is updated after every update of $\delta_{hg}$.

Parsimonious parameterisations of the component covariance matrices are easily incorporated. Uniquenesses can be constrained to be isotropic, with $\boldsymbol{\Psi}_g = \diag\negthinspace\left(\psi_g,\ldots,\psi_g\right)$, leading to a model that corresponds to an infinite mixture and infinite-dimensional extension of probabilistic principal components analysers \citep{Tipping1999b}. Uniquenesses can also be constrained across clusters, with or without the isotropic constraint across variables. These restrictions define the models in the \texttt{pgmm} family \citep{McNicholas2008} named \textsf{UUC}, \textsf{UCU},~and \textsf{UCC}, respectively, to which the following Gibbs updates are Bayesian analogues
\begin{align*}
\psi_g\given \ldots &\sim \textrm{IG}\negthinspace\left(\alpha_0 + \frac{pn_g}{2}, \beta + \frac{\trace{\left(\bm{\mathcal{S}}_g\right)}}{2}\right),\\[3ex]
\psi_j\given \ldots &\sim\textrm{IG}\negthinspace\left(\alpha_0 + \frac{N}{2},\beta_j + \frac{\sum_{g=1}^G\mathcal{S}_{jg}}{2}\right),\\[3ex]
\psi\given \ldots &\sim\textrm{IG}\negthinspace\left(\alpha_0 + \frac{pN}{2},\beta + \frac{\sum_{g=1}^G\trace{\left(\bm{\mathcal{S}}_g\right)}}{2}\right).
\end{align*}
\indent In the contexts of finite and overfitted mixtures (i.e. \textsf{MFA}, \textsf{MIFA}, \textsf{OMFA}, and \textsf{OMIFA}) $\mathbf{z}_i \given \mathbf{x}_i,\ldots \sim \textrm{Mult}\negthinspace
\left(1,p_{i1},\ldots,p_{i\widetilde{G}}\right)$, with
\[p_{ig} =\Pr\negthinspace\left(z_{ig}=1\given\mathbf{x}_i,\ldots\right) = \frac{\pi_g \textrm{N}_p\big(\mathbf{x}_i; \boldsymbol{\mu}_g,\boldsymbol{\Lambda}{\mathstrut}_g\boldsymbol{\Lambda}{\mathstrut}_g^\top+\boldsymbol{\Psi}_g\big)}{\sum_{g=1}^{\widetilde{G}}\pi_g \textrm{N}_p\big(\mathbf{x}_i; \boldsymbol{\mu}_g,\boldsymbol{\Lambda}{\mathstrut}_g\boldsymbol{\Lambda}{\mathstrut}_g^\top+\boldsymbol{\Psi}_g\big)},\]
whereas under the \textsf{IMIFA} and \textsf{IMFA} models
\begin{equation}
p_{ig}\propto \textrm{N}_p\big(\mathbf{x}_i;\boldsymbol{\mu}_g, \boldsymbol{\Lambda}{\mathstrut}_g\boldsymbol{\Lambda}{\mathstrut}_g^\top + \boldsymbol{\Psi}_g\big)\frac{\pi_g}{\xi_g}\indicator{\big(u_i < \xi_g\big)}.\label{eq:IMIFApig}
\end{equation}
The allocations $\mathbf{z}_i$ are sampled in a fast, numerically stable fashion, using the unnormalised log-probabilities and independent draws from the standard Gumbel distribution \citep{Yellott1977} via $s_{ig}=-\ln\negthinspace\left(m_{ig}\right)$, with $m_{ig} \sim \textrm{Exp}\negthinspace\left(\lambda=1\right)$. Observation $i$ is assigned the label $g$ satisfying
\[\argmax{g\,\in\, \left\{1,\ldots,\widetilde{G}\right\}}\big(\negthinspace\ln\negthinspace\left(p_{ig}\right) + s_{ig}\big).\]
For the \textsf{IMIFA} and \textsf{IMFA} models, the sampler need only find the maximum over, and only draw Gumbel noise for, log-probabilities for which the indicator function in \eqref{eq:IMIFApig} evaluates to $1$.

Sampling the parameters of the PYP for non-zero $d$ values requires the introduction of Metropolis-Hastings steps within the Gibbs sampler. A joint hyperprior of the form $\mathrm{p}\negthinspace\left(\alpha,d\right) = \mathrm{p}\negthinspace\left(d\right)\mathrm{p}\negthinspace\left(\alpha\given d\right)$ is assumed, as per \citet{Jara2010}. Firstly, the hyperprior for the discount parameter $d$ is similar to the one assumed by \citet{Carmona2019}; a mixture of~a point-mass at zero and a continuous beta distribution, in order to consider the DP special case~with $d=0$ with positive probability, i.e. $d \sim \kappa\delta_0 + \left(1-\kappa\right)\textrm{Beta}\negthinspace\left(a^\prime, b^\prime\right)$. This facilitates explicit comparison between DP models and encompassing PYP alternatives. Secondly, the hyperprior for $\alpha$ is given conditionally on $d$, s.t. $\left(\alpha \given d\right) \sim \textrm{Ga}\negthinspace\left(\alpha + d; a, b\right)$, and includes the constraint $\alpha>-d$ by shifting the support of the gamma density to the interval $\left(-d, \infty\right)$; choosing a large $b$ value is particularly relevant as it encourages clustering \citep{Muller2013}.

The likelihood for $\alpha$ and $d$ is given by the exchangeable partition probability function induced by the PYP \citep{Pitman1995}. Thus, the required conditional posterior distributions~are
\begin{align}
\alpha \given d, \ldots &\propto \frac{\Gamma\left(\alpha + 1\right)}{\Gamma\left(\alpha + N\right)}\Bigg\{\prod_{g=1}^{G_0-1}\left(\alpha + gd\right)\Bigg\}p\left(\alpha \given d\right),\label{eq:alphaconditional}\\[1.5ex]
d \given \alpha, \ldots &\propto \Bigg\{\prod_{g=1}^{G_0-1}\left(\alpha + gd\right)\Bigg\}\Bigg\{\prod_{g=1}^{G_0}\frac{\Gamma\left(n_g - \alpha\right)}{\Gamma\left(1-\alpha\right)}\Bigg\}p\left(d\right).\label{eq:discountconditional}
\end{align}
Sampling from the distributions in \eqref{eq:alphaconditional} and \eqref{eq:discountconditional}, while always considering the support $\alpha > -d$, proceeds~as per \citet{Carmona2019}; a Metropolis-Hastings step is implemented for the discount parameter with independent proposal distribution $0.5\delta_0 + 0.5\textrm{Beta}\negthinspace\left(d; 1, 1\right)$, and a random walk Metropolis-Hastings step with proposal distribution given by $\alpha^\star\given\alpha \sim \textrm{Unif}\negthinspace\left(\alpha - \zeta,\alpha + \zeta\right)$ is implemented for the concentration parameter, where $\zeta$ (= $2$ in our implementation) is used~to control the acceptance rate. For $d$, the mutation rate is considered rather than the acceptance rate, whereby a move is only considered accepted if the proposal differs from the current value.

However, when the DP prior is assumed, or when the sampled value of $d$ is exactly zero under the PYP prior, $\alpha$ is updated according to the auxiliary variable routine of \citet{West1992}, with Gibbs updates by simulation from a weighted mixture of two gamma distributions, via
\[\alpha \given G_0, \chi,\ldots \sim \omega_\chi \textrm{Ga}\big(a + G_0, b - \ln\negthinspace\left(\chi\right)\big) + \left(1 - \omega_\chi\right)\textrm{Ga}\big(a + G_0 - 1, b - \ln\negthinspace\left(\chi\right)\big),\vspace{-0.25ex}\]
where $G_0$ denotes the current number of non-empty clusters,
\begin{align*}
\left(\chi \given \alpha, G_0\right) &\sim \textrm{Beta}\negthinspace\left(\alpha + 1, N\right),
\intertext{and the mixing weights $\omega_\chi$ are defined by}
\dfrac{\omega_\chi}{1 - \omega_\chi} &=\dfrac{\left(a + G_0 - 1\right)}{N\big(b - \ln\negthinspace\left(\chi\right)\big)}.
\end{align*}
\indent The complementary label switching moves of \citet{Papaspiliopoulos2008}, which are effective at swapping similar and unequal clusters, respectively, are also incorporated. Firstly,~the labels of two randomly chosen non-empty clusters $g$ and $h$ are swapped with probability
\[\min \big(1, \left(\pi_h / \pi_g\right)^{n_g - n_h}\negmedspace\big).\] 
Secondly, the labels of neighbouring active components $l$ and $l+1$ are swapped with probability 
\[\min\big(1, \left(1-\upsilon_{l+1}\right)^{n_l}\negmedspace/ \left(1-\upsilon_{l}\right)^{n_{l+1}}\negmedspace\big);\]
if accepted, $\upsilon_l$ and $\upsilon_{l+1}$ are also swapped. Cluster-specific parameters are reordered accordingly after each accepted move. Finally, for updating $\alpha$ under the sparse finite \textsf{OMIFA} or \textsf{OMFA} models, a random walk Metropolis-Hastings step is implemented, with a Gaussian proposal distribution, where
\[\alpha \given \mathbf{Z}, \widetilde{G}, \ldots \propto \frac{\Gamma\big(\alpha \widetilde{G}\big)}{\Gamma\big(N + \alpha \widetilde{G}\big)}\Bigg\{\prod_{g\colon n_g > 0}\frac{\Gamma\left(n_g + \alpha\right)}{\Gamma\left(\alpha\right)}\Bigg\}p\left(\alpha\right).\]
Further details of this update can be found in \citet{FSMW2016} and \citet{FSMW2019}.


\subsection[Simulation Studies]{Simulation Studies}
\label{Subsection:SimulationStudies}
\setcounter{table}{0}
\setcounter{figure}{0}

The performance of the novel \textsf{IMIFA} model with its PYP-MGP priors, in terms of inferring both the number of clusters and the cluster-specific numbers of factors, is assessed here through simulation studies. Section \ref{Subsection:SimulationStudy1} explores sensitivity to the PYP parameters in a range of dimensionality scenarios, with balanced cluster sizes and a common number of factors. The simulation study in Section \ref{Subsection:SimulationStudy2} is more challenging; a larger number of clusters (many of which are small) are simulated for $N < p$ data, with different numbers of cluster-specific factors (some of which are large). The final simulation study in Section \ref{Subsection:SimulationStudy3} mirrors the design in Section \ref{Subsection:SimulationStudy2}, only here the true $\boldsymbol{\Lambda}_g$ matrices used to generate the data are sparse.

\subsubsection[Simulation Study 1]{Simulation Study 1}
\label{Subsection:SimulationStudy1}

Firstly, data with $G = 3$ clusters and $p = 50$ variables are simulated with $q_g = 4\:\:\forall\:\:g$, and~with $\boldsymbol{\pi} = \left(\nicefrac{1}{3},\nicefrac{1}{3},\nicefrac{1}{3}\right)$ so that clusters are roughly equally sized. Other model parameters are simulated, with $\boldsymbol{\eta}{\mathstrut}_i \sim \textrm{N}_q\negthinspace\left(\mathbf{0}, \bm{\mathcal{I}}_q\right)$, $\psi_{jg} \sim \textrm{IG}\negthinspace\left(2, 1\right)$, and $\boldsymbol{\Lambda}_{jg} \sim \textrm{N}_q\negthinspace\left(\mathbf{0},\bm{\mathcal{I}}_q\right)$. Notably, loadings are not drawn from the MGP prior \citep{Bhattacharya2011} underpinning the \textsf{IMIFA} model. To ensure clusters are reasonably closely located, $\boldsymbol{\mu}_g \sim\textrm{N}_p\negthinspace\left(\left(2g - G - 1\right)\mathbf{1},\bm{\mathcal{I}}_p\right)$. The data are then simulated according to the conditional mixture model
\begin{equation*}
f\big(\mathbf{x}_i\given \boldsymbol{\eta}{\mathstrut}_i,\boldsymbol{\theta}\big) = \sum_{g=1}^G\pi_g\textrm{N}_p\big(\mathbf{x}_i;\boldsymbol{\mu}_g + \boldsymbol{\Lambda}_g\boldsymbol{\eta{\mathstrut}}_i, \boldsymbol{\Psi}_g\big).
\end{equation*}
\indent To evaluate performance in different settings, sample sizes less than, equal to, and greater than $p$ are considered, i.e. $N = 25, 50$, and $300$. Sensitivity to the PYP and DP parameters is explored by firstly assuming a DP prior with various values of $\alpha$ less than, equal to, and greater than $1$, and by allowing $\alpha$ to be learned as per \citet{West1992}, and secondly by incorporating Metropolis-Hastings steps to learn both $\alpha$ and $d$, assuming a PYP prior.

\begin{figure}[H]
	\centering
	\includegraphics[width=\textwidth, keepaspectratio]{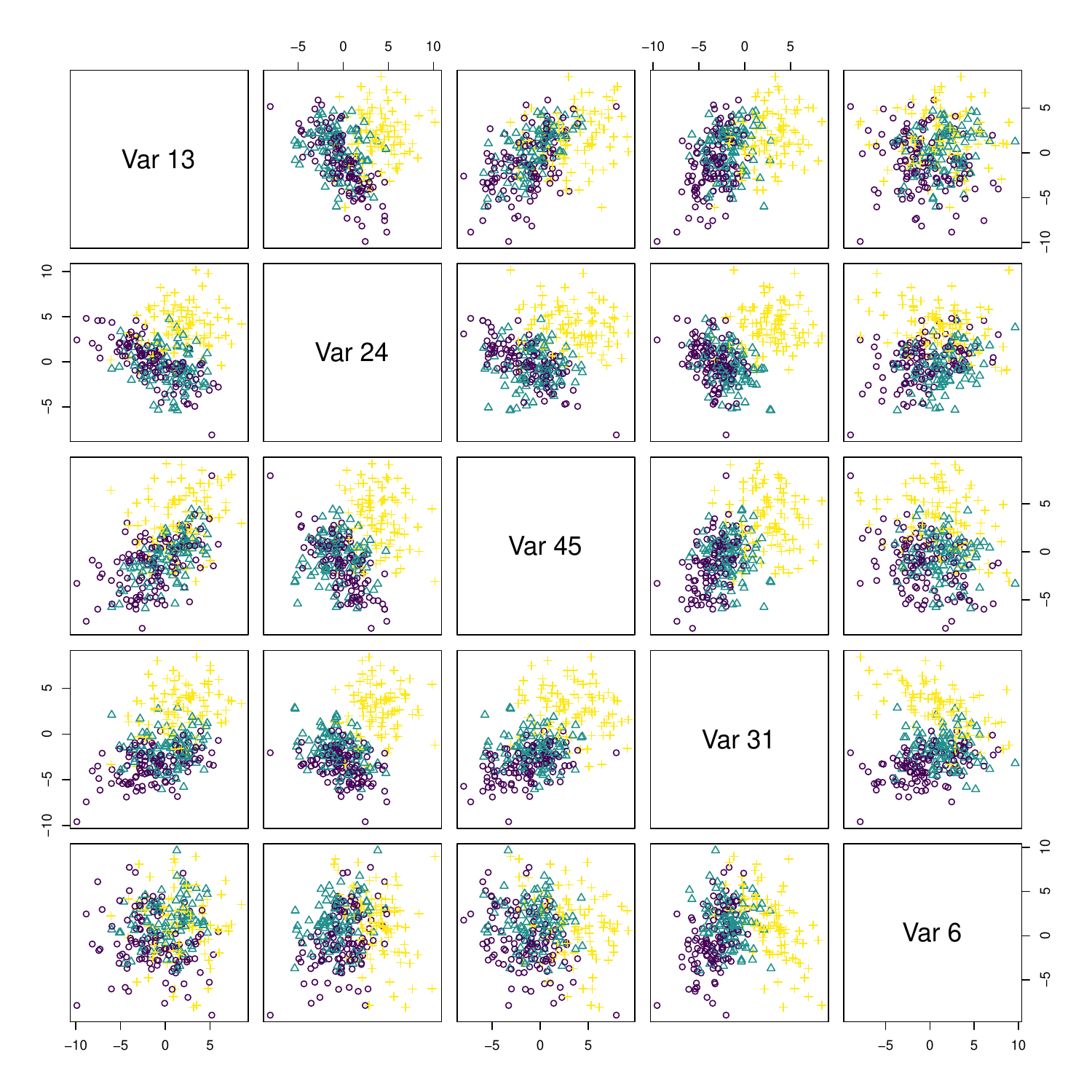}
	\caption[Pairwise scatterplots of a subset of variables for a replicate data set under Simulation Study 1.]{Pairwise scatterplots of $5$ randomly chosen variables from one of the raw replicate data sets under the $N>p$ scenario in Table \ref{Table:SimulationStudy}, demonstrating the overlap between the $3$ clusters.}
	\label{Plot:SimulatedPairs}
\end{figure}

Results, provided in Table \ref{Table:SimulationStudy}, are based on $10$ replicate data sets, standardised prior to model fitting, for each scenario. MCMC chains were run for $25,000$ iterations, with every 2\textsuperscript{nd} sample thinned and the first $20\%$ of iterations discarded as burn-in. Cluster labels were initialised using \texttt{mclust} \citep{Scrucca2016}, as hierarchical clustering gave poor, heavily imbalanced starting values. As the cluster-specific $\boldsymbol{\Lambda}_g$ and $\boldsymbol{\Psi}_g$ parameters could still induce separation among clusters, pairwise scatterplots from one randomly chosen raw replicate data set under the $N > p$ scenario are shown in Figure \ref{Plot:SimulatedPairs} to demonstrate the extent of the overlap; for visual clarity, only $5$ randomly chosen variables are depicted. 
\begin{table}[H]
	\caption[Aggregated results of Simulation Study 1.]{Aggregated simulation study results for the \textsf{IMIFA} model under different dimensionality scenarios and settings of the concentration and discount parameters $\alpha$ and $d$ (posterior mean estimates thereof in parentheses where appropriate). The modal estimates of $G$ and associated estimates of $q_g\:\:\forall\:\:g$ are reported (with $95\%$ credible intervals in brackets). Clustering performance is assessed through the average percentage error rate against the known cluster labels.}
	\label{Table:SimulationStudy}
	\centering
	\scriptsize
	\extrarowheight 2.5pt
	
	\begin{tabular}[pos=center]{c | c | c | c | c | c | c | c}
		{{\centering}Dimension} & {{\centering}$\alpha$} & {{\centering}$d$} & {{\centering}$G$} & {{\centering}$q_1$} & {{\centering}$q_2$} & {{\centering}$q_3$} & {{\centering}Error~(\%)}\\\hline
		\multirow{5}{0.1\linewidth}{\centering $N=25$\\$\left(N<p\right)$}&$0.5$&$0$&$3\,[3,3]$&$5\,[3,9]$&$5\,[3,9]$&$5\,[3,9]$&$0$\\
		&$1$&$0$&$3\,[3,3]$&$5\,[3,9]$&$5\,[3,9]$&$5\,[3,9]$&$0$\\
		&$5$&$0$&$3\,[3,4]$&$5\,[3,9]$&$5\,[3,9]$&$5\,[3,9]$&$6.4$\\
		&$(0.57)$&$0$&$3\,[3,3]$&$5\,[3,9]$&$5\,[3,9]$&$5\,[3,9]$&$0$\\
		&$(0.51)$&$(0.05)$&$3\,[3,3]$&$5\,[3,9]$ &$5\,[3,9]$&$5\,[3,9]$&$0$\\
		\hline
		\multirow{5}{0.1\linewidth}{\centering $N=50$\\$\left(N=p\right)$}&$0.5$&$0$&$3\,[3,3]$&$5\,[4,7]$&$5\,[4,7]$&$5\,[4,7]$&$0$\\
		&$1$&$0$&$3\,[3,3]$&$5\,[4,7]$&$5\,[4,7]$&$5\,[4,7]$&$0$\\
		&$5$&$0$&$3\,[3,3]$&$5\,[4,7]$&$5\,[4,7]$&$5\,[4,7]$&$0$\\
		&$(0.52)$&$0$&$3\,[3,3]$&$5\,[4,7]$&$5\,[4,7]$&$5\,[4,7]$&$0$\\
		&$(0.48)$&$(0.03)$&$3\,[3,3]$&$5\,[4,7]$&$5\,[4,7]$&$5\,[4,7]$&$0$\\
		\hline
		\multirow{5}{0.1\linewidth}{\centering $N=300$\\$\left(N>p\right)$}&$0.5$&$0$&$3\,[3,3]$&$5\,[4,6]$&$5\,[4,6]$&$5\,[4,6]$&$0$\\
		&$1$&$0$&$3\,[3,3]$&$5\,[4,6]$&$5\,[4,6]$&$5\,[4,6]$&$0$\\
		&$5$&$0$&$3\,[3,3]$&$5\,[4,6]$&$5\,[4,6]$&$5\,[4,6]$&$0$\\
		&$(0.42)$&$0$&$3\,[3,3]$&$5\,[4,6]$&$5\,[4,6]$&$5\,[4,6]$&$0$\\
		&$(0.39)$&$(0.02)$&$3\,[3,3]$&$5\,[4,6]$&$5\,[4,6]$&$5\,[4,6]$&$0$
	\end{tabular}
\end{table}
Table \ref{Table:SimulationStudy} clearly demonstrates that the \textsf{IMIFA} model performs well overall for these data, exhibiting capability to uncover the structure within the simulated data sets regardless of dimensionality. The modal estimate of $G$ is equal to the truth in all cases, with only the $\smash{N < p}$, $\smash{\alpha = 5}$ scenario showing some deviation in the $95\%$ credible interval. Perhaps surprisingly, given the closeness of the cluster means, and the equality of the clusters in terms of their mixing proportions and numbers of factors, $G$ is never underestimated. Indeed, clustering performance is mostly perfect. Furthermore, in every case, the true value of $\smash{q_g=4}$ is within the limits of the associated credible intervals, which intuitively become narrower as more data accumulates. While the modal estimates $\smash{\widehat{q}_g}$ are consistently greater than the truth throughout Table \ref{Table:SimulationStudy}, overestimation should be preferred to underestimation; a less parsimonious model which nevertheless fits well and uncovers the true clustering structure is better than one which loses information and fits poorly due to having too few factors. Recall that the loadings were drawn from a standard multivariate Gaussian, rather than the MGP prior underpinning the \textsf{IMIFA} model, i.e. entries in the true $\smash{\boldsymbol{\Lambda}_g}$ matrices did not shrink with the column index, nor were the loadings sparse. Thus, there is evidence to suggest the model is liable to overestimate the number of factors when the $\smash{\boldsymbol{\Lambda}_g}$ matrices, and by extension the cluster-specific marginal covariance matrices, are dense. This is explored further in the subsequent simulation studies.\vspace{-0.5em}

\subsubsection[Simulation Study 2]{Simulation Study 2}
\label{Subsection:SimulationStudy2}

\begin{figure}[H]
	\centering
	\begin{subfigure}[t]{\textwidth}
		\centering
		\includegraphics[width=1.015\textwidth, height=1.3125in]{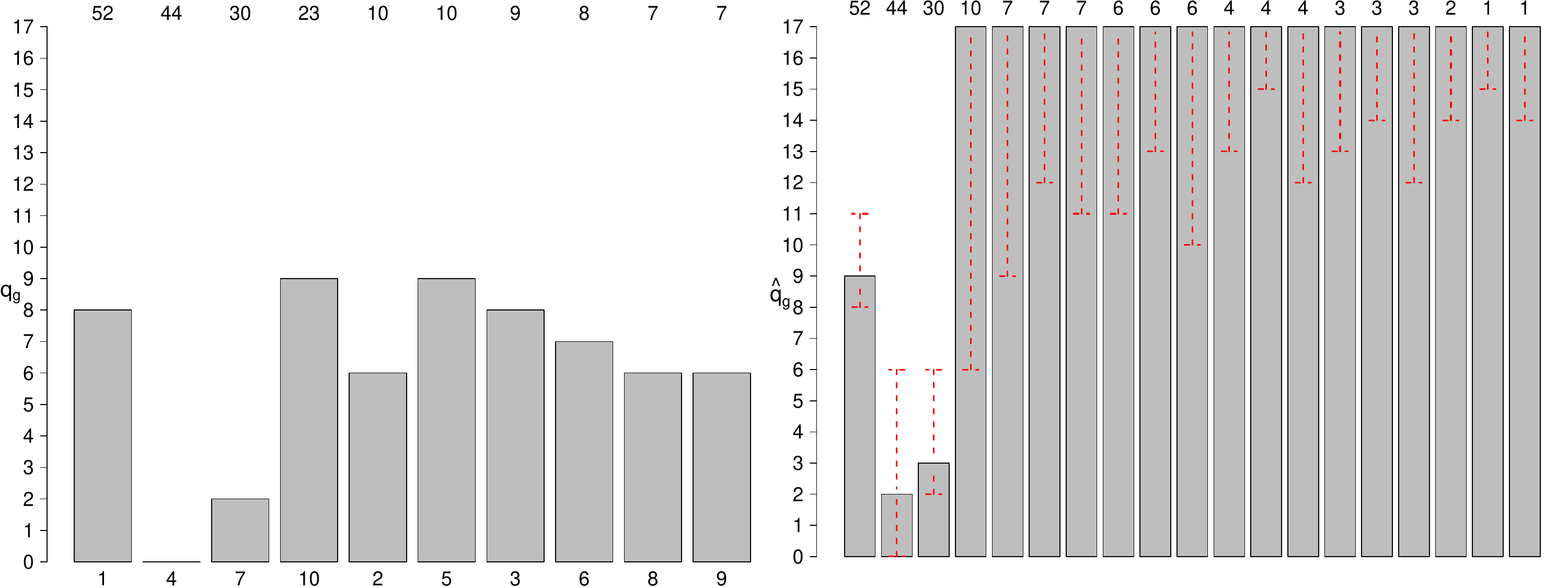}
		\caption{Replicate 1: $\smash{\widehat{G}=19~[19,19]}$, $\mbox{ARI}=0.94$, $\smash{\widehat{\alpha}=0.80}$, $\smash{\widehat{d}=0.26}$.}
		\label{Plot:SimStudy2a}
	\end{subfigure}
	\begin{subfigure}[t]{\textwidth}
		\setlength{\abovecaptionskip}{1pt plus 0pt minus 0pt}
		\setlength{\belowcaptionskip}{1pt plus 0pt minus 0pt}
		\centering
		\includegraphics[width=1.015\textwidth, height=1.3125in]{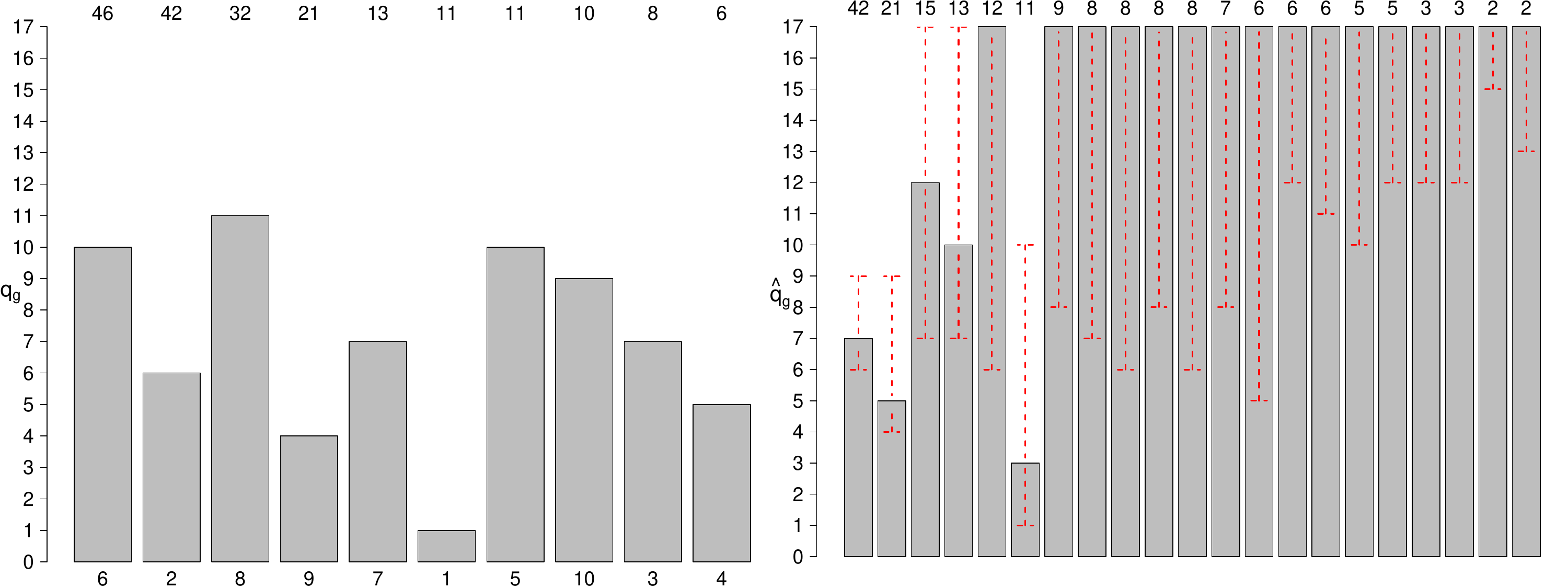}
		\caption{Replicate 2: $\smash{\widehat{G}=21~[21,22]}$, $\mbox{ARI}=0.66$, $\smash{\widehat{\alpha}=1.20}$, $\smash{\widehat{d}=0.20}$.}
		\label{Plot:SimStudy2b}
	\end{subfigure}
	\begin{subfigure}[t]{\textwidth}
		\setlength{\abovecaptionskip}{1pt plus 0pt minus 0pt}
		\setlength{\belowcaptionskip}{1pt plus 0pt minus 0pt}
		\centering
		\includegraphics[width=1.015\textwidth, height=1.3125in]{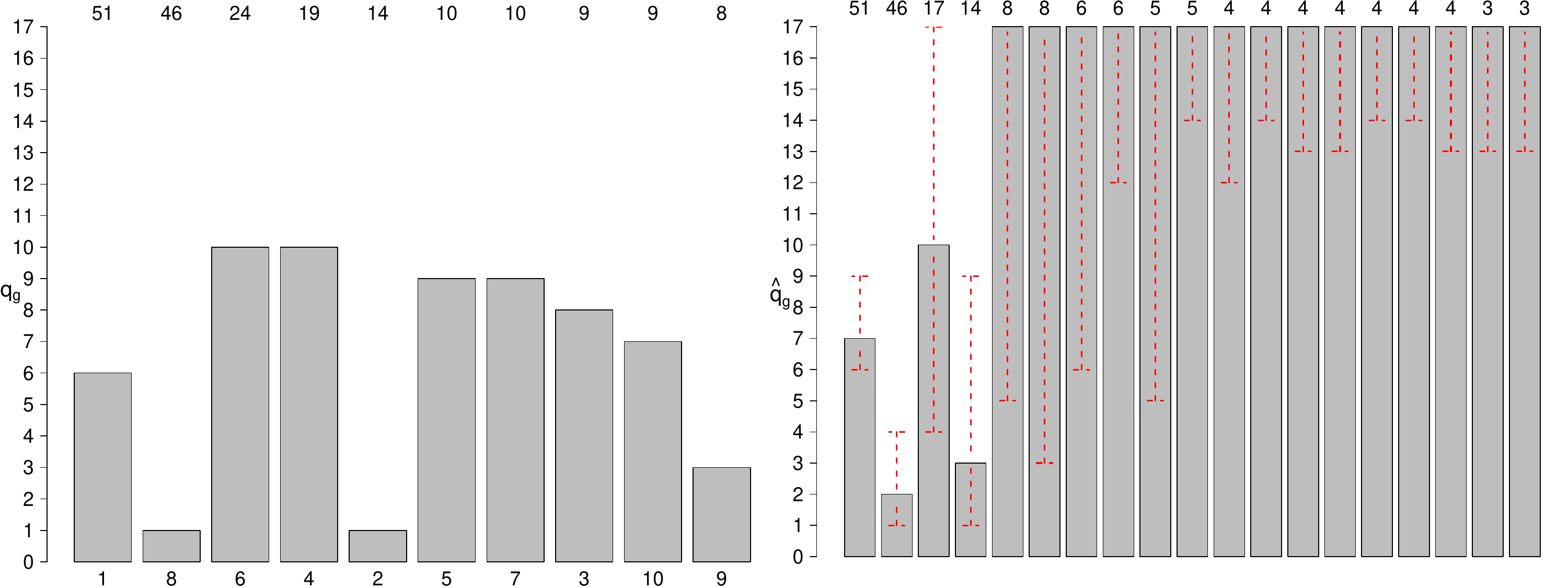}
		\caption{Replicate 3: $\smash{\widehat{G}=19~[19,24]}$, $\mbox{ARI}=0.90$, $\smash{\widehat{\alpha}=1.06}$, $\smash{\widehat{d}=0.21}$.}
		\label{Plot:SimStudy2c}
	\end{subfigure}
	\begin{subfigure}[t]{\textwidth}
		\setlength{\abovecaptionskip}{1pt plus 0pt minus 0pt}
		\setlength{\belowcaptionskip}{1pt plus 0pt minus 0pt}
		\centering
		\includegraphics[width=1.015\textwidth, height=1.3125in]{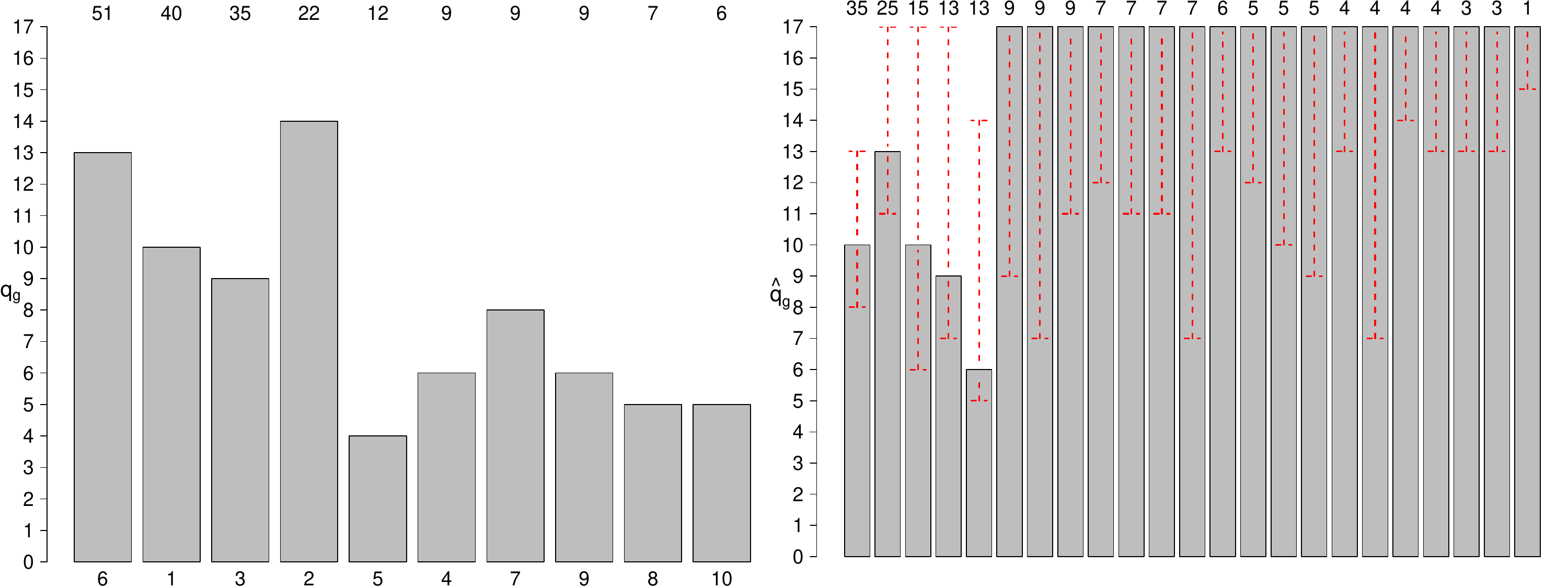}
		\caption{Replicate 4: $\smash{\widehat{G}=23~[20,25]}$, $\mbox{ARI}=0.56$, $\smash{\widehat{\alpha}=1.16}$, $\smash{\widehat{d}=0.23}$.}
		\label{Plot:SimStudy2d}
	\end{subfigure}
	\begin{subfigure}[t]{\textwidth}
		\setlength{\abovecaptionskip}{1pt plus 0pt minus 0pt}
		\setlength{\belowcaptionskip}{1pt plus 0pt minus 0pt}
		\centering
		\includegraphics[width=1.015\textwidth, height=1.3125in]{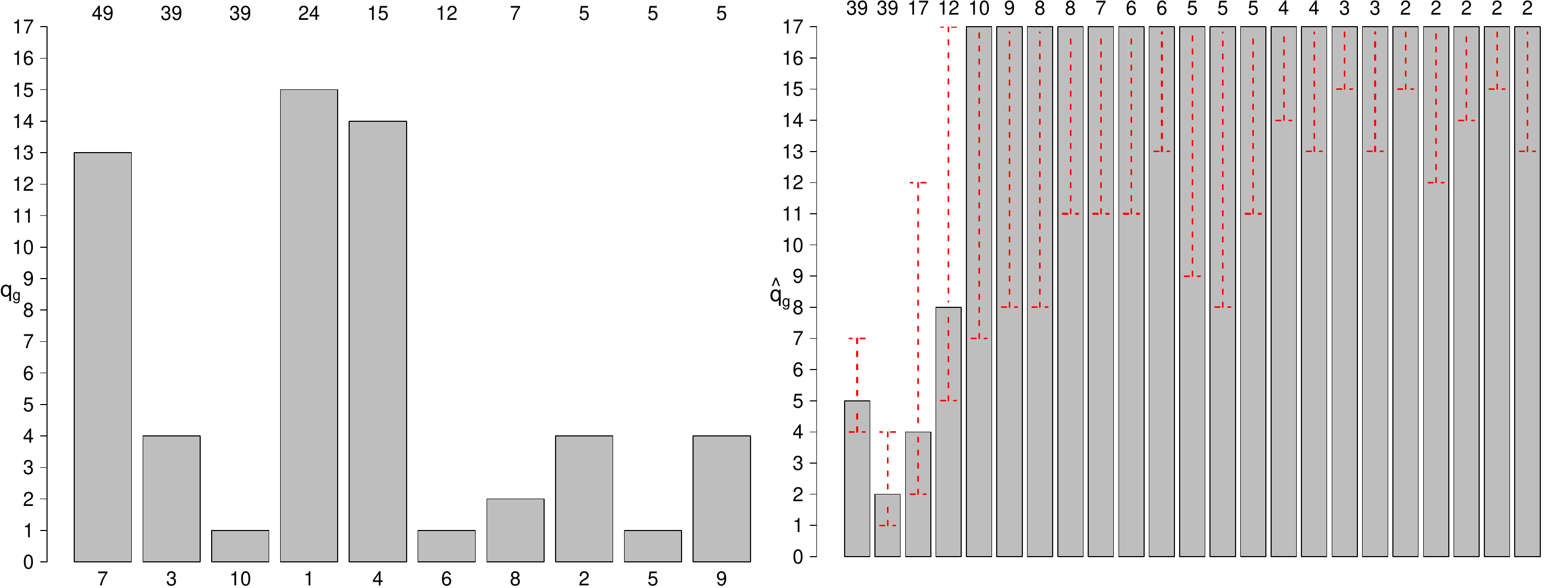}
		\caption{Replicate 5: $\smash{\widehat{G}=23~[23,23]}$, $\mbox{ARI}=0.68$, $\smash{\widehat{\alpha}=0.92}$, $\smash{\widehat{d}=0.26}$.}
		\label{Plot:SimStudy2e}
	\end{subfigure}
	\caption[Barplots of the true and estimated numbers of factors for each replicate data set comprising Simulation Study 2.]{Barplots of the true number of cluster-specific factors $q_g$ (left) and estimates $\smash{\widehat{q}_g}$ (right)~for each replicate data set and corresponding fitted \textsf{IMIFA} model comprising Simulation Study 2.~Bars~are sorted in descending order of $n_g$ and $\smash{\widehat{n}_g}$, respectively, and labelled above with these true and estimated cluster sizes. The plots on the left are also labelled below with the cluster indices. Vertical red lines~in the plots on the right show $95\%$ credible intervals for $\smash{\widehat{q}_g}$. Modal $\smash{\widehat{G}}$ estimates (with $95\%$ credible intervals in brackets), ARI values, and posterior mean estimates $\smash{\widehat{\alpha}}$ and $\smash{\widehat{d}}$ are given for each~replicate.}
	\label{Plot:SimulationStudy2}
\end{figure}

Results of a more challenging simulation study are presented in Figure \ref{Plot:SimulationStudy2}; here, $\smash{N < p}$ data ($\smash{N=200, p=250}$) are simulated with a large number of clusters and uncommon numbers of cluster-specific factors. In particular, many of the $G=10$ clusters are small (a setting often studied in Bayesian nonparametric modelling), with $\smash{\boldsymbol{\pi} = \left(0.25, 0.2, 0.15, 0.1, 0.05, \ldots, 0.05\right)^\top}$. The numbers of factors $\smash{q_1,\ldots,q_g}$ are drawn randomly from $\smash{0,\ldots,\min\left(15, n_g - 1\right)}$, where the upper limit ensures that no cluster has more factors than observations.  Otherwise, the same parameter settings as Simulation Study 1 above (Section \ref{Subsection:SimulationStudy1}) were used to generate the data.

\begin{figure}[H]
	\centering
	\includegraphics[width=\textwidth, keepaspectratio]{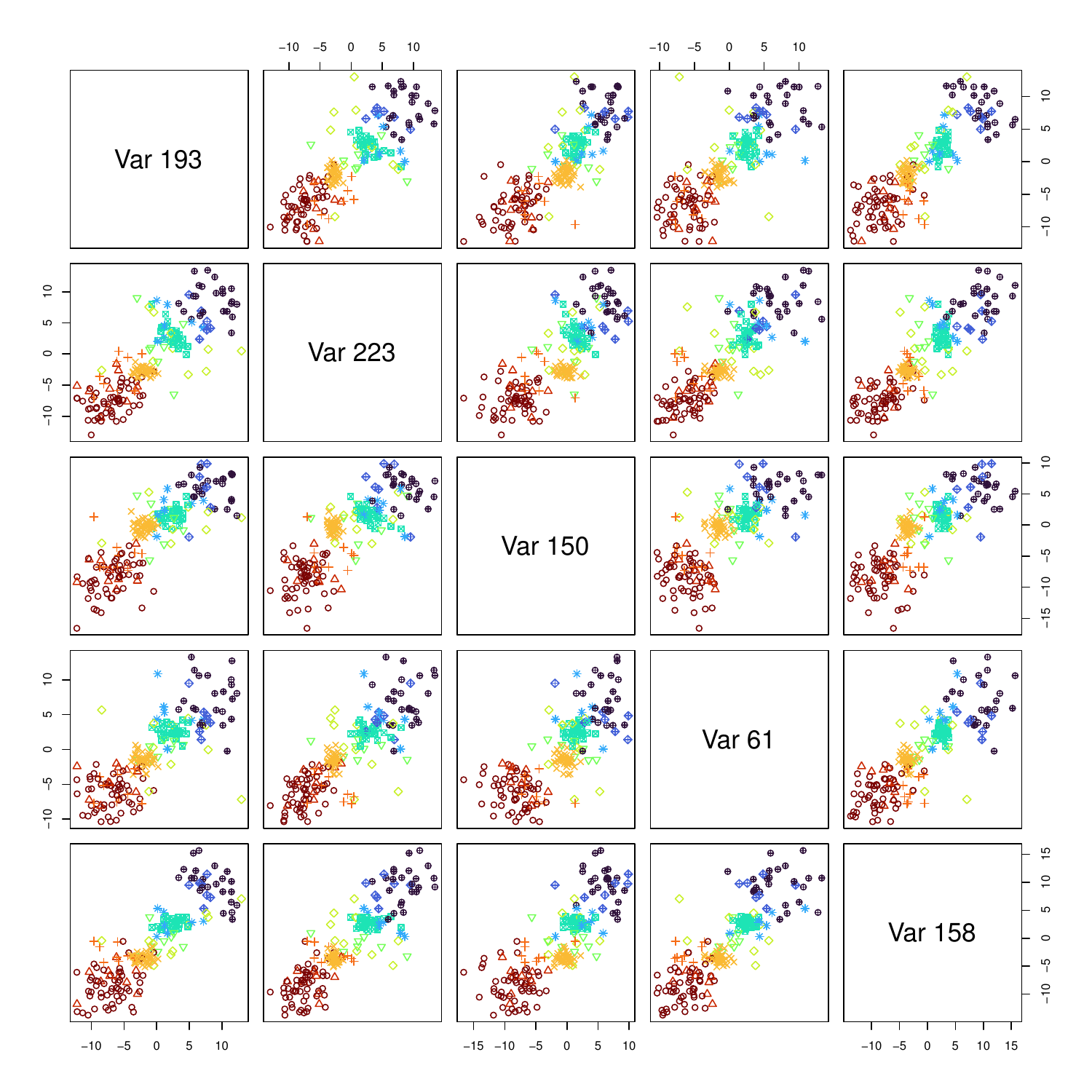}
	\caption[Pairwise scatterplots of a subset of variables for a replicate data set under Simulation Study 2.]{Pairwise scatterplots of $5$ randomly chosen variables from the first raw replicate data set in Simulation Study 2 (Section \ref{Subsection:SimulationStudy2}), demonstrating the overlap between the $10$ clusters.}
	\label{Plot:DensePairs}
\end{figure}

Results of fitting an \textsf{IMIFA} model assuming a PYP prior, allowing both $\alpha$ and $d$ to be learned, and otherwise using the same sampler settings as in Section \ref{Subsection:SimulationStudy1} above, are given for $5$ replicates of this scenario, with the $\boldsymbol{\pi}$ vector ordered randomly for each data set. To demonstrate the extent of the challenge these settings represent, pairwise scatterplots are again shown for $5$ randomly chosen variables for the first replicate data set in Figure \ref{Plot:DensePairs}.\vspace{-0.5em}
Figure \ref{Plot:SimulationStudy2} shows that the model over-estimates the number of clusters, though in some cases the ARI values are nonetheless quite good, as the larger clusters are generally uncovered well. However, the smaller clusters are further divided, albeit cleanly, into smaller sub-clusters with, in some cases, just $1$ or $2$ units inside. In these cases, the modal $\smash{\widehat{q}_g}$ estimates are close~or equal to the upper limit of the adaptive Gibbs sampler ($3\ln\negthinspace\left(p\right)$), and hence or otherwise greater than the corresponding estimated cluster sizes $\smash{\widehat{n}_g}$. Thus, there is evidence that the model has difficulty in adaptively shrinking the $\smash{\boldsymbol{\Lambda}_g}$ matrices when there are many clusters with few units.

\subsubsection[Simulation Study 3]{Simulation Study 3}
\label{Subsection:SimulationStudy3}

In both previous simulation studies, the true loadings were dense, having been drawn from~a standard multivariate Gaussian, rather than the MGP prior underpinning the model. The design of this final simulation study exactly mirrors the parameter and sampler settings used~in Section \ref{Subsection:SimulationStudy2} with the sole exception that, as per the simulation study design in \citet{Bhattacharya2011}, the true loadings matrices used to generate the data are sparse.

\begin{figure}[H]
	\centering
	\setlength{\abovecaptionskip}{1pt plus 0pt minus 0pt}
	\setlength{\belowcaptionskip}{1pt plus 0pt minus 0pt}
	\includegraphics[width=\textwidth, keepaspectratio]{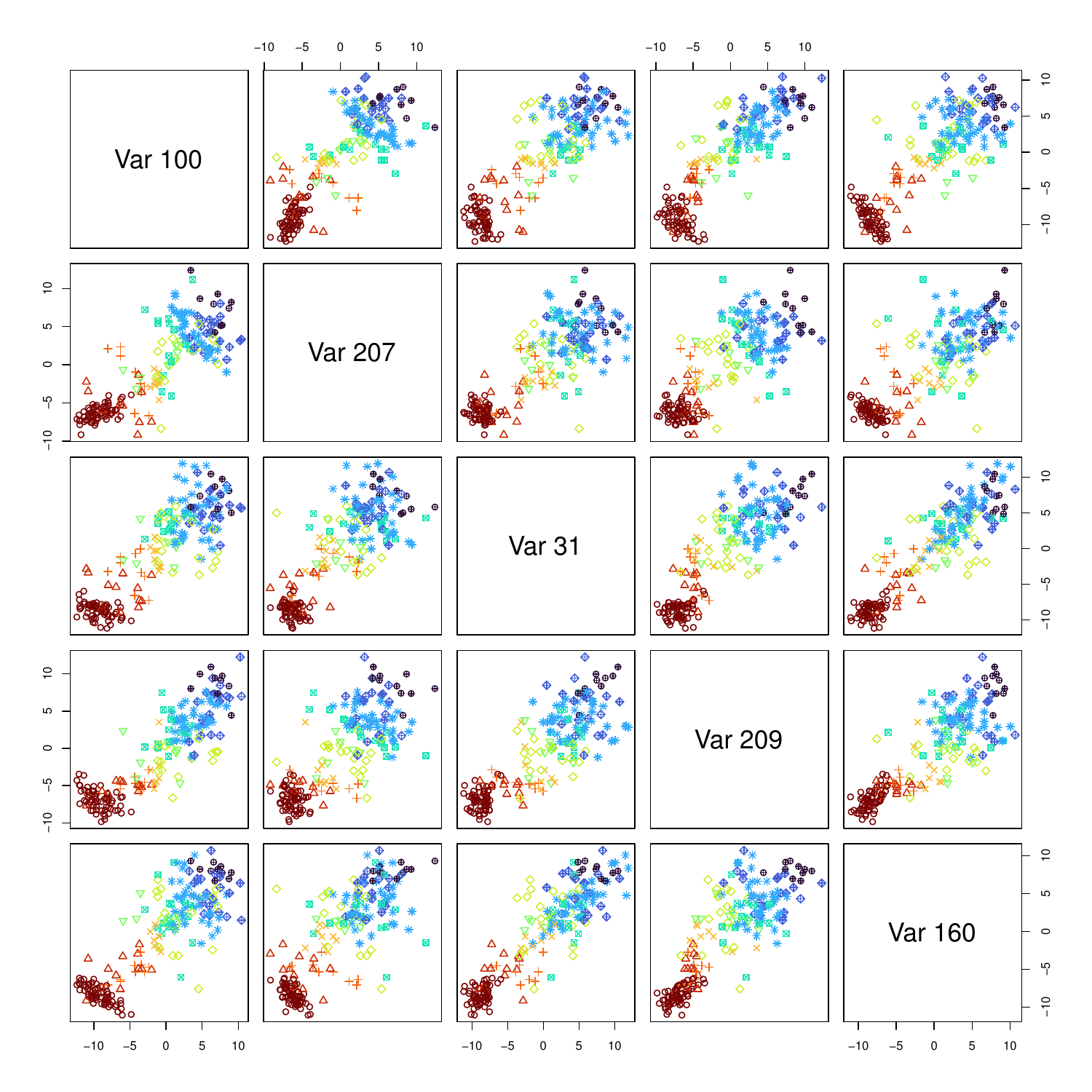}
	\caption[Pairwise scatterplots of a subset of variables for a replicate data set under Simulation Study 3.]{Pairwise scatterplots of $5$ randomly chosen variables from the first raw replicate data set in Simulation Study 3 (Section \ref{Subsection:SimulationStudy3}), demonstrating the overlap between the $10$ clusters.}
	\label{Plot:SparsePairs}
\end{figure}

Specifically, the number of non-zero loadings in each $\boldsymbol{\Lambda}_g$ matrix begins at $p$ in column $1$,~and successively decays by $10\%$ for each subsequent column. The locations of the zeros in each column~are allocated randomly and non-zero elements are drawn from a standard multivariate Gaussian. Again, pairwise scatterplots are shown for $5$ randomly chosen variables for the first~of the five replicate data sets in Figure \ref{Plot:SparsePairs}, to demonstrate the extent of the overlap between~clusters.\vspace{-0.5em}
Results are presented in Figure \ref{Plot:SimulationStudy3}. Performance is comparable to the results of Simulation Study 2, in the sense that, again, the number of clusters is over-estimated. ARI values are nonetheless acceptable. Small clusters are divided into even smaller sub-clusters for which~the model struggles to adaptively shrink the number of factors. The comparability of~the results of these experiments suggests that performance is being driven not by whether the loadings used to generate the data exhibit increasing levels of sparsity across columns, in line with the MGP prior underpinning the model, but by the presence of many small clusters.\clearpage

The over-estimation of $\smash{\widehat{q}_g}$ in the small clusters in simulation studies 2 and 3 suggests that the hyperparameters $\alpha_1$ and $\alpha_2$ related to the MGP column shrinkage parameters may need to be higher in mixture settings to enforce a greater degree of shrinkage as there will be fewer data in each cluster from which local and global shrinkage parameters can be learned, compared to fitting an \textsf{IFA} model on the full data set. Introducing Metropolis-Hastings steps to allow these hyperparameters be cluster-specific and learned from the data, rather than fixed, may also help in this regard.\enlargethispage{\baselineskip}
\begin{figure}[H]
	\setlength{\abovecaptionskip}{1pt plus 0pt minus 1pt}
	\setlength{\belowcaptionskip}{1pt plus 0pt minus 1pt}
	\centering
	\begin{subfigure}[t]{\textwidth}
		\setlength{\abovecaptionskip}{1pt plus 0pt minus 0pt}
		\setlength{\belowcaptionskip}{1pt plus 0pt minus 0pt}
		\centering
		\includegraphics[width=1.015\textwidth, height=1.3125in]{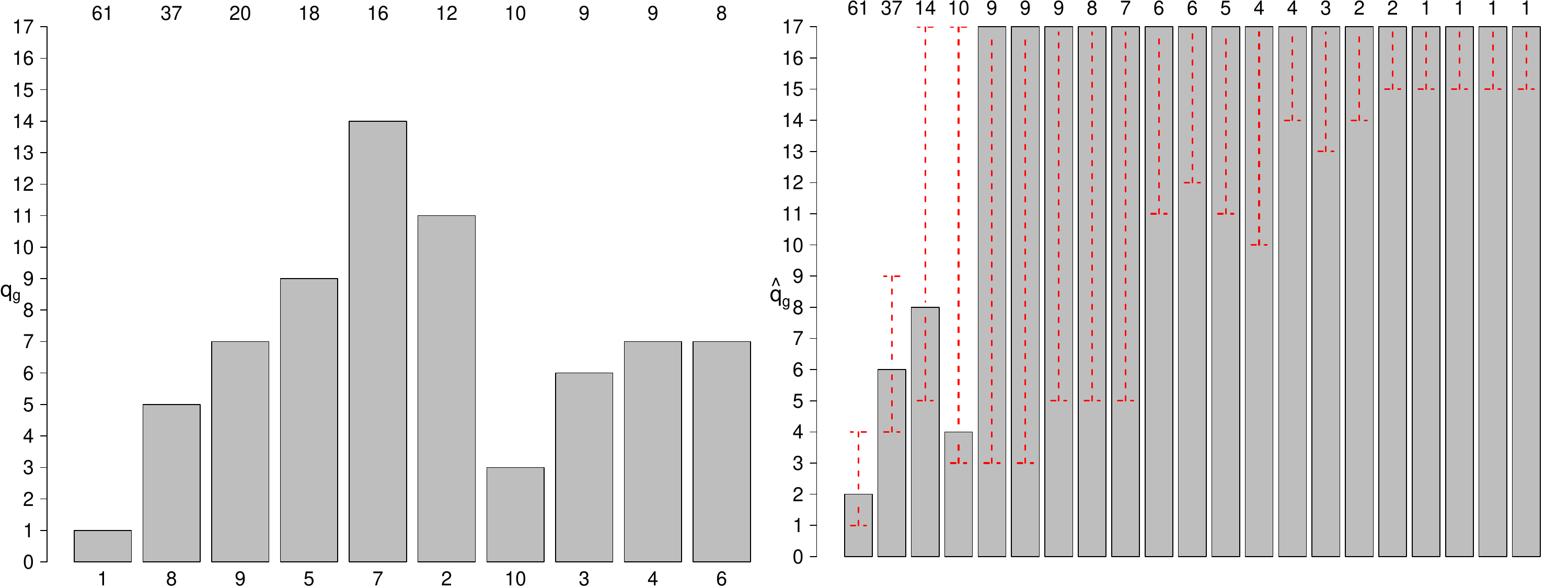}
		\caption{Replicate 1: $\smash{\widehat{G}=21~[21,21]}$, $\mbox{ARI}=0.93$, $\smash{\widehat{\alpha}=0.65}$, $\smash{\widehat{d}=0.31}$.}
		\label{Plot:SimStudy3a}
	\end{subfigure}
	\begin{subfigure}[t]{\textwidth}
		\setlength{\abovecaptionskip}{1pt plus 0pt minus 0pt}
		\setlength{\belowcaptionskip}{1pt plus 0pt minus 0pt}
		\centering
		\includegraphics[width=1.015\textwidth, height=1.3125in]{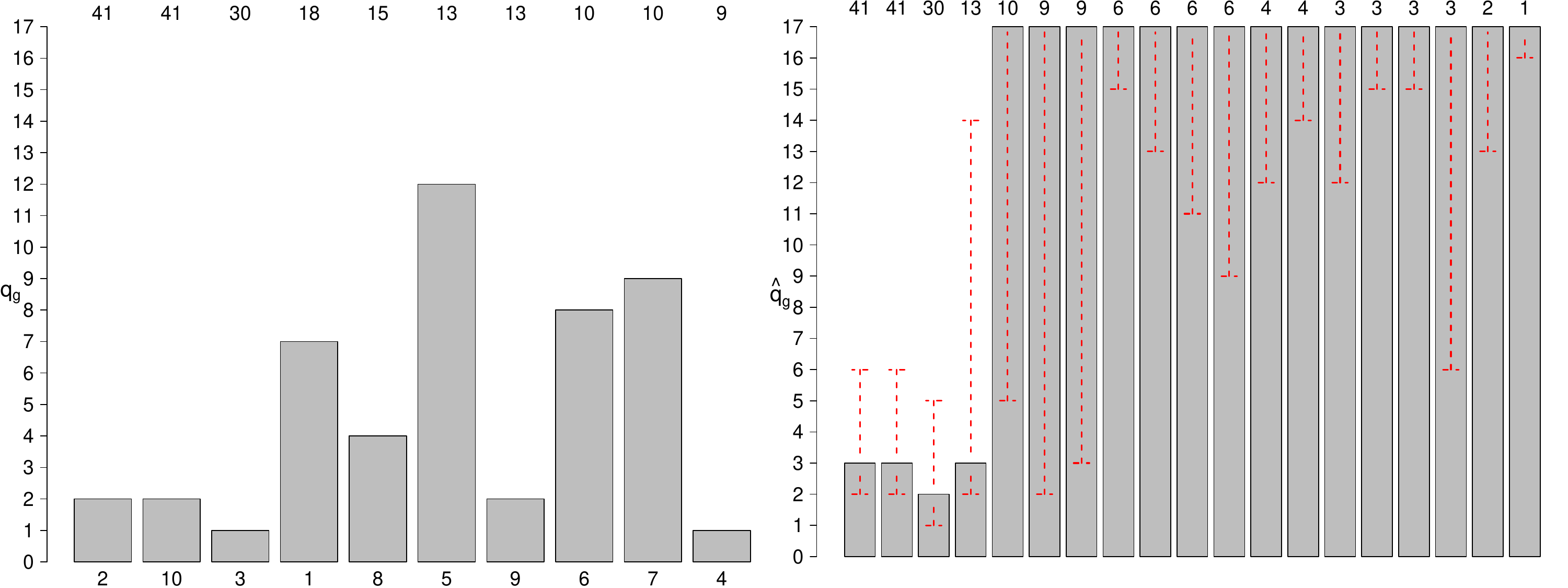}
		\caption{Replicate 2: $\smash{\widehat{G}=19~[18,19]}$, $\mbox{ARI}=0.94$, $\smash{\widehat{\alpha}=0.91}$, $\smash{\widehat{d}=0.23}$.}
		\label{Plot:SimStudy3b}
	\end{subfigure}
	\begin{subfigure}[t]{\textwidth}
		\setlength{\abovecaptionskip}{1pt plus 0pt minus 0pt}
		\setlength{\belowcaptionskip}{1pt plus 0pt minus 0pt}
		\centering
		\includegraphics[width=1.015\textwidth, height=1.3125in]{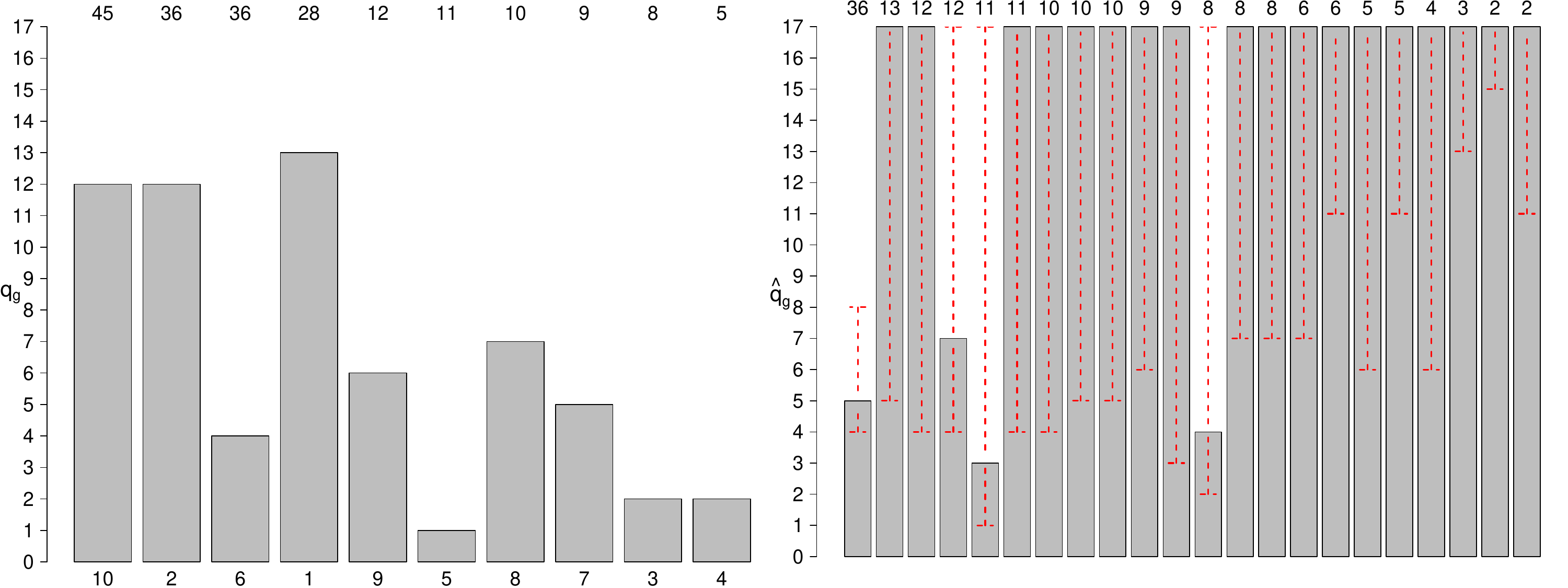}
		\caption{Replicate 3: $\smash{\widehat{G}=22~[22,22]}$, $\mbox{ARI}=0.59$, $\smash{\widehat{\alpha}=1.35}$, $\smash{\widehat{d}=0.18}$.}
		\label{Plot:SimStudy3c}
	\end{subfigure}
	\begin{subfigure}[t]{\textwidth}
		\setlength{\abovecaptionskip}{1pt plus 0pt minus 0pt}
		\setlength{\belowcaptionskip}{1pt plus 0pt minus 0pt}
		\centering
		\includegraphics[width=1.015\textwidth, height=1.3125in]{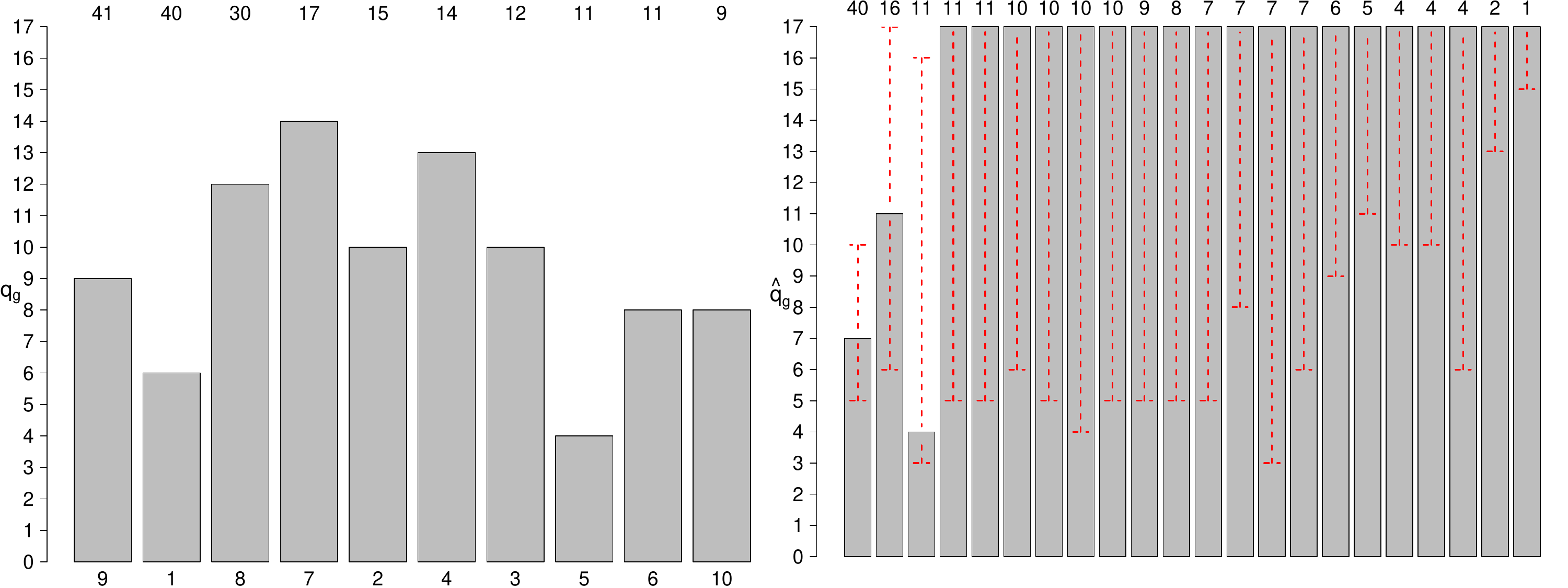}
		\caption{Replicate 4: $\smash{\widehat{G}=22~[22,22]}$, $\mbox{ARI}=0.69$, $\smash{\widehat{\alpha}=1.22}$, $\smash{\widehat{d}=0.21}$.}
		\label{Plot:SimStudy3d}
	\end{subfigure}
	\begin{subfigure}[t]{\textwidth}
		\setlength{\abovecaptionskip}{1pt plus 0pt minus 0pt}
		\setlength{\belowcaptionskip}{1pt plus 0pt minus 0pt}
		\centering
		\includegraphics[width=1.015\textwidth, height=1.3125in]{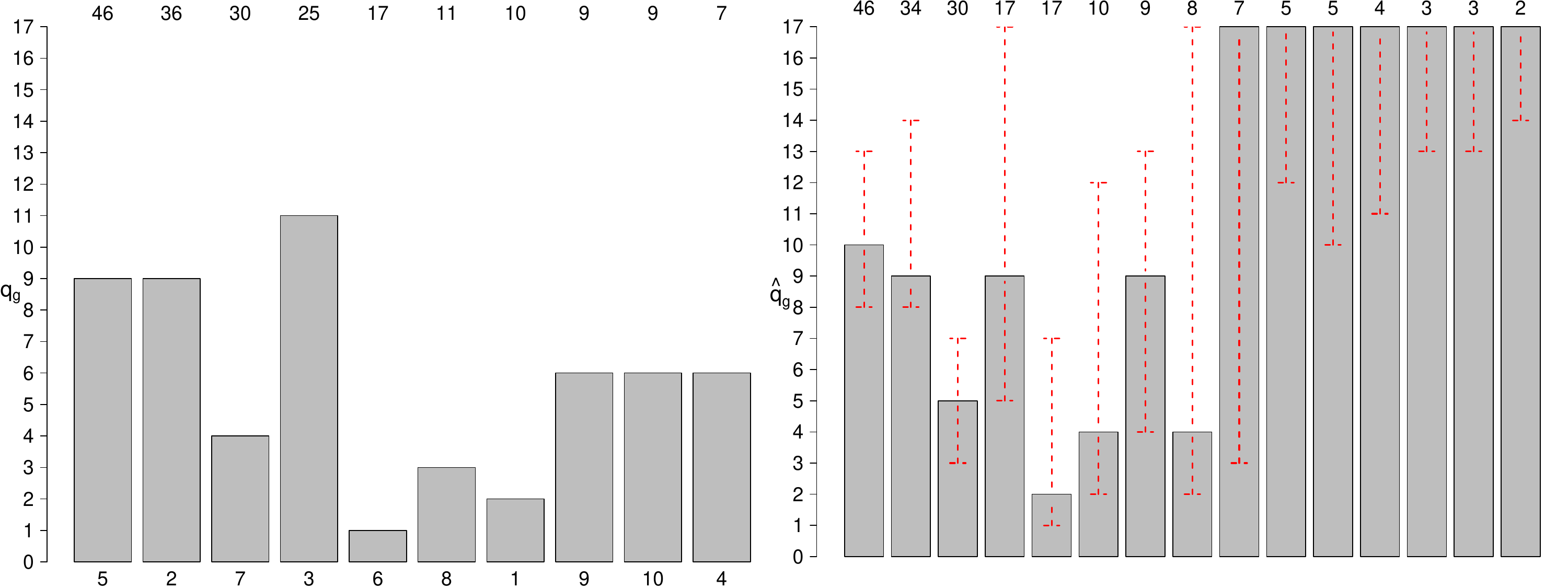}
		\caption{Replicate 5: $\smash{\widehat{G}=15~[14,25]}$, $\mbox{ARI}=0.94$, $\smash{\widehat{\alpha}=1.09}$, $\smash{\widehat{d}=0.18}$.}
		\label{Plot:SimStudy3e}
	\end{subfigure}
	\caption[Barplots of the true and estimated numbers of factors for each replicate data set comprising Simulation Study 3.]{Barplots of the true number of cluster-specific factors $q_g$ (left) and estimates $\smash{\widehat{q}_g}$ (right)~for each replicate data set and corresponding fitted \textsf{IMIFA} model comprising Simulation Study 3.~Bars~are sorted in descending order of $n_g$ and $\smash{\widehat{n}_g}$, respectively, and labelled above with these true and estimated cluster sizes. The plots on the left are also labelled below with the cluster indices. Vertical red lines~in the plots on the right show $95\%$ credible intervals for $\smash{\widehat{q}_g}$. Modal $\smash{\widehat{G}}$ estimates (with $95\%$ credible intervals in brackets), ARI values, and posterior mean estimates $\smash{\widehat{\alpha}}$ and $\smash{\widehat{d}}$ are given for each~replicate.}
	\label{Plot:SimulationStudy3}
\end{figure}

\clearpage

\subsection[Assessing Robustness of the \textrm{IMIFA} Model]{Assessing Robustness of the \textrm{IMIFA} Model}
\label{Subsection:Robustness}
\setcounter{table}{0}
\setcounter{figure}{0}

In order to assess the robustness of the \textsf{IMIFA} model, $\textrm{N}_1\negthinspace\left(0,1\right)$ noise with no clustering information was appended separately to the rows and columns of the olive oil data set. Six new scenarios were generated with $10$, $50$, and $100$ extra variables, and the same numbers of extra observations. Cluster validity is evaluated in Table \ref{Table:OliveRobust} with respect to the $4$ area relabelling in Table \ref{Table:OliveFour}. In the case of extra observations, noise observations are labelled as though they belong to a fifth cluster. Data were mean-centred and unit-scaled only after expansion.

As the number of irrelevant variables increases, the clustering structure can still be uncovered quite well, however mixing becomes slower and there is increasing support for clusters with only one or no factors as the signal-to-noise ratio decreases. As such, variable selection, or at least data pre-processing, may still be required. As rows of noise are appended, \textsf{IMIFA} generally has no difficulty in assigning these observations to a cluster of their own. Interestingly, clusters corresponding to noise observations correctly require no latent factor structure.
\begin{table}[H]
	\setcounter{table}{0}
	\caption[Clustering performance of the IMIFA model on expanded noisy versions of the olive oil data.]{Clustering performance of the \textsf{IMIFA} model on expanded noisy versions of the Italian olive oil data. The run-time relative to running \textsf{IMIFA} on the original data, posterior mean of the PYP parameters $\alpha$ and $d$, modal estimates of $G$ and $\mathbf{Q}$, ARI, and percentage error rate are all given.}
	\label{Table:OliveRobust}
	\centering
	\scriptsize
	\extrarowheight 2.5pt
	
	\begin{tabular}[pos=center]{c | c | c | c | c | c | c | c }
		\multicolumn{1}{c|}{Scenario} & {{\centering}Relative~Time} & {{\centering}$\alpha$} & {{\centering}$d$} & {{\centering}$G$} & {{\centering}{$\mathbf{Q}$}} & {{\centering}ARI} & {{\centering}Error~(\%)}\\\hline
		$N=572, p=18$ & $1.86$ & $0.48$ & $0.01$ & $4$ & $3$, $4$, $4$, $3$ & $0.85$ & $12.59$\\
		$N=572, p=58$ & $3.14$ & $0.47$ & $0.01$ & $4$ & $1$, $2$, $2$, $2$ & $0.74$ & $14.69$\\
		$N=572, p=108$ & $5.64$ & $0.46$ & $0.02$ & $4$ & $0$, $1$, $0$, $2$ & $0.73$ & $17.66$\\
		\hline
		$N=582, p=8$ & $1.10$ & $0.57$ & $0.01$ & $5$ & $6$, $2$, $2$, $2$, $0$ & $0.94$ & $6.87$\\
		$N=622, p=8$ & $1.09$ & $0.56$ & $0.01$ & $5$ & $4$, $1$, $1$, $2$, $0$ & $0.95$ & $6.59$\\
		$N=672, p=8$ & $1.07$ & $0.53$ & $0.01$ & $5$ & $4$, $1$, $2$, $2$, $0$ & $1.00$ & $0.45$
	\end{tabular}
\end{table}

\subsection[Additional Results and Visualisations]{Additional Results and Visualisations}
\label{Subsection:AdditionalResults}
\setcounter{table}{0}
\setcounter{figure}{0}

In this Section, some additional visualisations of the results of the illustrative applications are provided. Specifically, more details are provided on the posterior predictive model fit assessment and the observation-specific cluster membership uncertainties. All plots were produced using the associated \textsf{R} package \texttt{IMIFA} \citep{IMIFAR2021}.

The Posterior Predictive Reconstruction Error (PPRE) has been proposed as a posterior predictive checking strategy for models in the \textsf{IMIFA} family. In short, this involves computing the standardised Frobenius norm of the difference between a matrix of histogram bin counts for the modelled data set and similar matrices constructed using replicate data drawn from the posterior predictive distribution. While the median PPRE value or boxplots of the distribution of PPRE values have been shown to yield useful global measures of model fit in multivariate settings, the histograms themselves can also be studied on a variable-by-variable basis. 

\begin{figure}[H]
	\centering
	\includegraphics[width=0.7\textwidth,keepaspectratio]{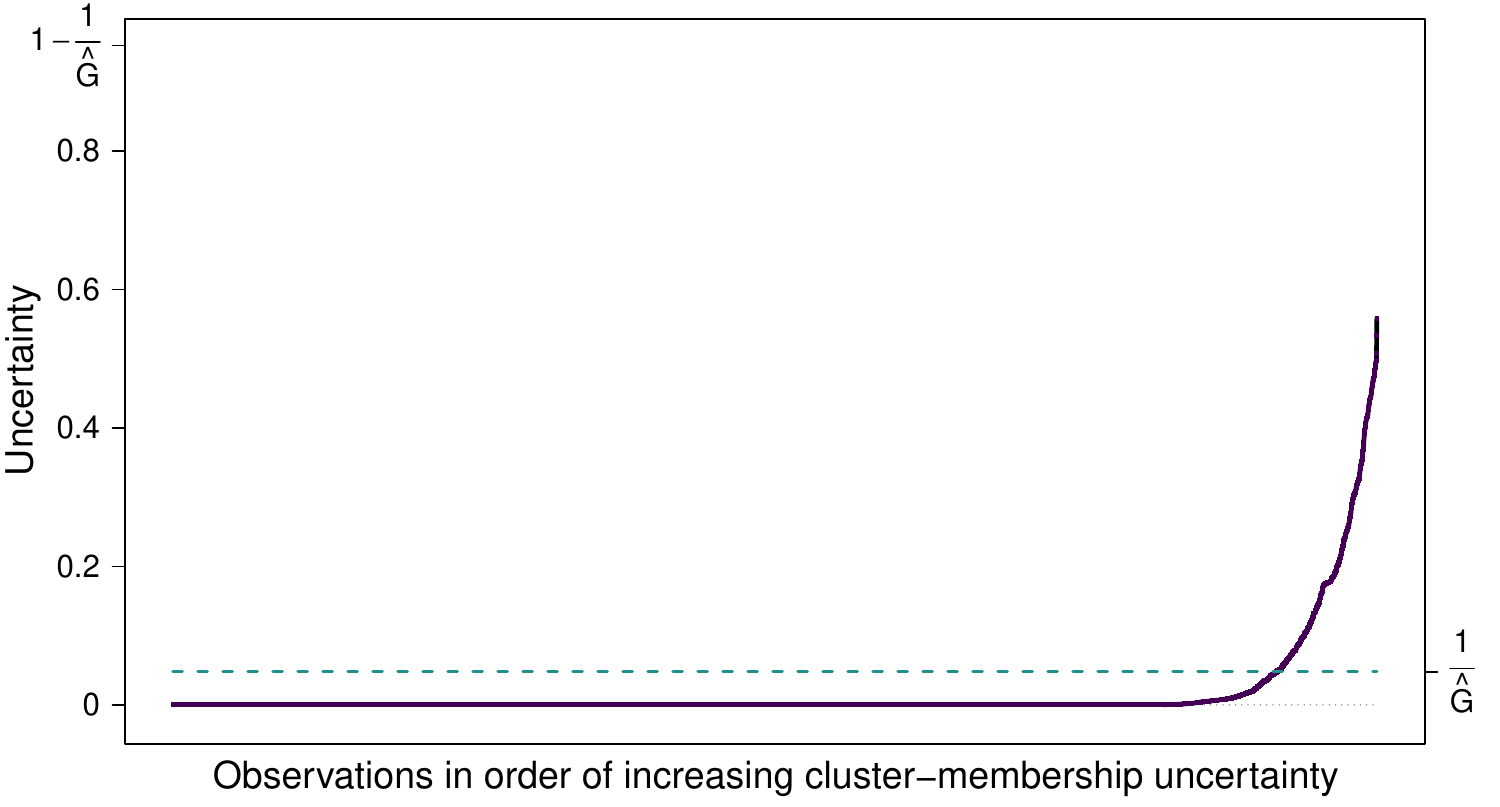}
	\caption[Uncertainty profile plot for the IMIFA model fit to the USPS data.]{Uncertainty profile plot for the $21$-cluster \textsf{IMIFA} model fitted to the USPS digits data, showing observation-specific  uncertainties in increasing order, most of which are below the line at $\nicefrac{1}{\widehat{G}}$.}
	\label{Plot:DigitsUncertainty}
\end{figure}

In high-dimensional settings, such as the spectral metabolomic ($p=189$) and USPS digits ($p=256$) data sets, it is only feasible to examine the histograms for a subset of the variables. Nonetheless, the global median PPRE measures for these data sets are quite good ($0.21$ and $0.05$, respectively). Hence, Figure \ref{Plot:OliveHistograms} shows only the histograms comparing bin counts for the $p=8$ variables in the standardised Italian olive oil data, to which an \textsf{IMIFA} model was fitted, against corresponding counts for the replicate data under the fitted \textsf{IMIFA} model. The true bin counts are within the $95\%$ credible intervals of the replicate data bin counts in the vast majority of cases, indicating good model fit: recall that this \textsf{IMIFA} model achieves a median PPRE of just $0.10$.
\begin{figure}[H]
	\setcounter{figure}{0}
	\centering
	\includegraphics[width=\textwidth, keepaspectratio]{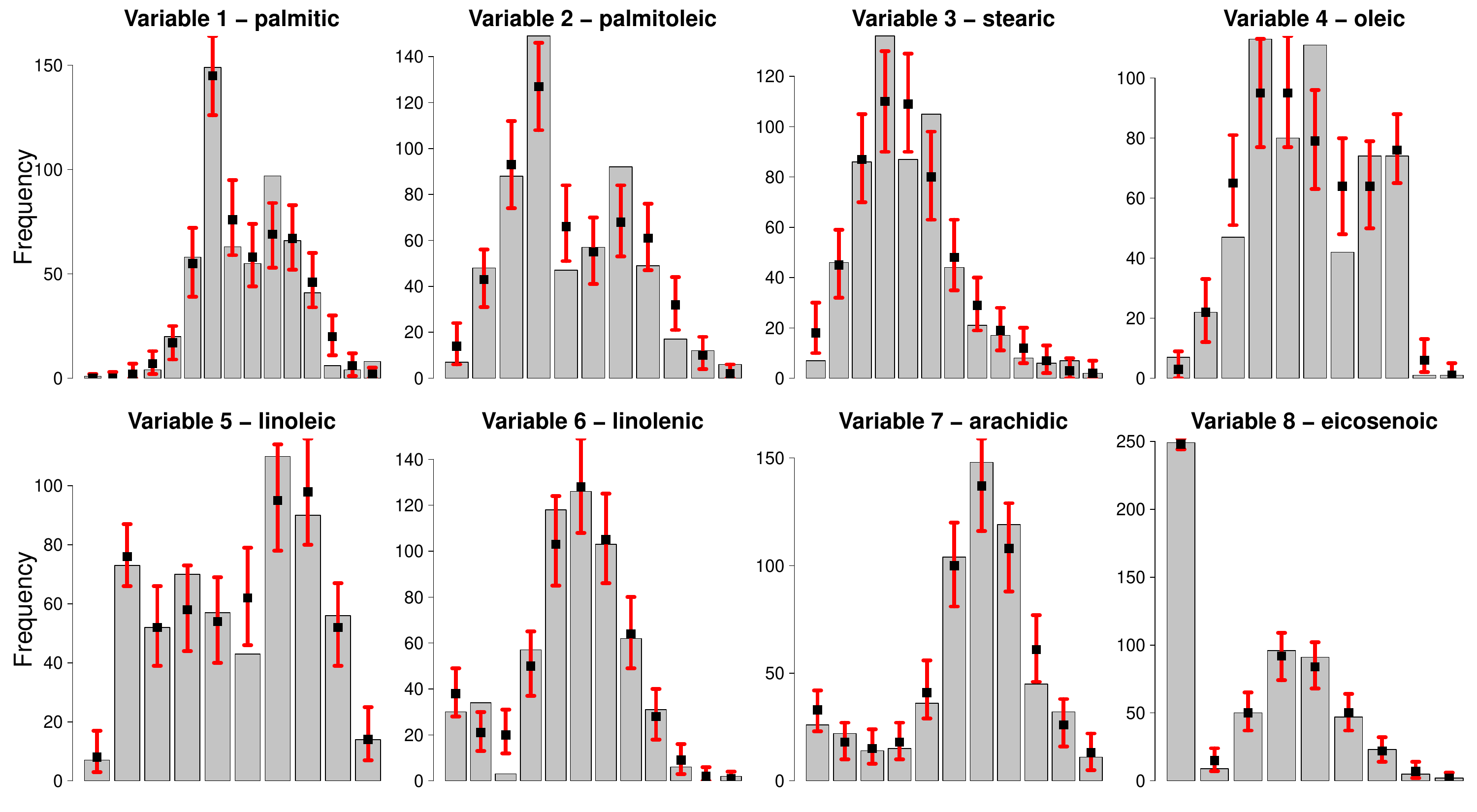}
	\caption[Posterior predictive histograms for the IMIFA model fit to the olive oil data.]{Histograms of the $p=8$ variables in the standardised Italian olive oil data set. The height of each bar corresponds to the modelled data set, while the black squares correspond to the median bin counts of the replicate data sets drawn from the posterior predictive distribution of the fitted \textsf{IMIFA} model (with associated $95\%$ credible intervals given by vertical red lines).}
	\label{Plot:OliveHistograms}
\end{figure}
The \textsf{IMIFA} model fitted to the USPS digits data set uncovers $\smash{\widehat{G}=21}$ clusters. Regarding the uncertainty in the allocations to these clusters, the model-based nature of \textsf{IMIFA} facilitates estimation of the uncertainty with which observation $i$ is assigned to its cluster $g$ via
\[\widehat{U}_i = \minsub{g\,\in\,\left\{1,\ldots,\widehat{G}\right\}}\negmedspace\left(1 - \widehat{z}_{ig}\right),\]
where $\smash{\widehat{z}_{ig}}$ is the estimated probability that observation $i$ belongs to cluster $g$. Figure \ref{Plot:DigitsUncertainty} shows that the observation-specific cluster membership uncertainties are generally quite low, with the mean uncertainty being just $0.02$ and $92$\% of observations being assigned with uncertainty less than $\nicefrac{1}{\widehat{G}}$. A similar plot for the olive oil data is shown in the main text (Figure \ref{Plot:Uncertainty}); uncertainties for the spectral metabolomic data are not shown, as there was no uncertainty in the assignments under the fitted \textsf{IMIFA} model (i.e. $\smash{\widehat{U}_i=0\:\:\forall\:\:i=1,\ldots,N}$). 

\setlength{\bibsep}{6.5pt}
{\fontsize{11.25}{13.25}\selectfont
\bibliographystyle{myplainnat}\interlinepenalty=10000
\renewcommand\bibsection{\subsection*{\refname}}
\bibliography{IMIFA_arXiv_v6_bib}
\end{document}